\documentclass[camera,letterpaper,nomarginnotes,nonarrowgutter]{jpaper}
\newif\ifdraft
\draftfalse

\usepackage{tikz}
\usepackage{mathptmx} 
\usepackage{amsmath,amssymb,amsfonts}
\usepackage{comment}
\usepackage{graphics}
\usepackage{fancyhdr}
\usepackage{booktabs}
\usepackage{pifont}
\usepackage[normalem]{ulem}
\usepackage{cite}
\usepackage{hhline}
\usepackage{xcolor}
\usepackage{setspace}
\usepackage{textcomp}
\usepackage{float}
\usepackage{enumitem}
\usepackage{afterpage}
\usepackage{graphicx}
\usepackage[htt]{hyphenat}
\usepackage{makecell}
\usepackage{xspace}
\usepackage{listings}
\usepackage{multirow}
\usepackage{balance}
\usepackage{glossaries} 
\usepackage{siunitx}
\usepackage{duckuments} 
\usepackage{dblfloatfix} 
\usepackage{lscape}
\usepackage{subcaption}
\usepackage{rotating}
\usepackage[us,12hr]{datetime}
\usepackage[en-GB, useregional=numeric]{datetime2}
\usepackage{bm} 
\usepackage{tabularx}
\usepackage{tablefootnote}

\usepackage[compact]{titlesec}
\usepackage[colorinlistoftodos,prependcaption,textsize=small]{todonotes} 
\usepackage{setspace}
\usepackage{dblfloatfix}    
\usepackage[ruled, linesnumbered]{algorithm2e}
\usepackage{soul}
\usepackage{bm}
\usepackage{etoolbox}
\usepackage{multibib}

\makeatletter
\patchcmd\algocf@Vline{\vrule}{\vrule \kern-0.4pt}{}{}
\patchcmd\algocf@Vsline{\vrule}{\vrule \kern-0.4pt}{}{}
\makeatother
\usepackage[bookmarks=true,breaklinks=true,hidelinks]{hyperref}
\newif\ifsecondbib
\secondbibfalse

\newif\ifarxiv 
\arxivfalse

\newif\ifcameraready
\camerareadyfalse

\newif\ifrev
\revfalse

\newif\ifshepherd
\shepherdfalse

\newif\ifpagenumbers
\pagenumberstrue

\newcounter{version}
\ifcameraready
\else
\fi

\newcommand{\affilETH}[0]{\textsuperscript{\S}}
\newcommand{\affilETU}[0]{\textsuperscript{$\dagger$}}
\newcommand{\affilUOS}[0]{\textsuperscript{$\ddagger$}}
\author{
{Yahya Can Tu\u{g}rul\affilETH\affilETU}\qquad%
{A. Giray Ya\u{g}l{\i}kç{\i}\affilETH}\qquad%
{İsmail Emir Yüksel\affilETH}\qquad%
{Ataberk Olgun\affilETH}\\
{O\u{g}uzhan Canpolat\affilETH\affilETU}\qquad%
{Nisa Bostancı\affilETH}\qquad%
{Mohammad Sadrosadati\affilETH}\qquad%
{O\u{g}uz Ergin\affilUOS\affilETU}\qquad%
{Onur Mutlu\affilETH}\\
{\affilETH \emph{ETH Z{\"u}rich}}\qquad{}\affilETU \emph{TOBB University of Economics and Technology}\qquad{}\affilUOS \emph{University of Sharjah}
}

\newcommand{\tfr}[0]{t_{FCRI}}
\newcommand{\nrh}[0]{N_{RH}}
\newcommand{\nnrh}[0]{N_{RH_{naive}}}
\newcommand{\thpcr}[0]{N_{PCR}}
\newcommand{\thpcrrow}[0]{N_{PCR(row)}}
\newacronym{nrh}{$\nrh{}$}{ the aggressor row needs to be activated more than a certain threshold value, defined as \emph{RowHammer threshold}}
\newacronym{nnrh}{$\nnrh{}$}{the \gls{nrh} value found by using the \emph{naïve} method}
\newacronym{thpcr}{$\thpcr{}$}{the maximum number of consecutive partial charge restorations}
\newacronym{thpcrrow}{$\thpcrrow{}$}{the maximum number of consecutive partial charge restorations for each DRAM row} 
\newacronym{fr}{$FR$}{fully restored bit vector}
\newacronym{tfr}{$\tfr{}$}{full charge restoration interval}
\newacronym{ber}{$BER$}{Bit-Error-Rate}
\newacronym{thammer}{$T_{Hammer}$}{the total time required to hammer the aggressor rows}
\newacronym{tsleep}{$T_{Sleep}$}{the remaining time until $t_{REFW}$}
\newacronym{tlcr}{$T_{LCR}$}{the time passed until the last charge restoration performed with the nominal latency}

\newcommand{\act}[0]{ACT}
\newcommand{\pre}[0]{PRE}
\newcommand{\refresh}[0]{REF}
\newcommand{\wri}[0]{\texttt{WR}}
\newcommand{\rd}[0]{\texttt{RD}}

\usepackage[shortcuts]{extdash}

\newacronym{iqr}{$IQR$}{inter-quartile range}
\newacronym{act}{\act{}}{activate}
\newacronym{pre}{\pre{}}{precharge}
\newacronym{ref}{\refresh{}}{refresh}
\newacronym{wr}{\wri{}}{write}
\newacronym{rd}{\rd{}}{read}
\newacronym{jedec}{JEDEC}{Joint Electron Device Engineering Council}

\newcommand{\tras}[0]{t_{RAS}}
\newcommand{\trp}[0]{t_{RP}}
\newcommand{\trc}[0]{t_{RC}}
\newcommand{\trefi}[0]{t_{REFI}}
\newcommand{\trefw}[0]{t_{REFW}}
\newcommand{\trfc}[0]{t_{RFC}}
\newcommand{\trrd}[0]{t_{RRD}}
\newcommand{\trasred}[0]{t_{RAS(Red)}}

\newacronym{trefw}{$\trefw{}$}{refresh window}
\newacronym{tras}{$\tras{}$}{the latency of sensing the row's data and fully restoring a DRAM cell's charge}
\newacronym{trp}{$\trp{}$}{the time required to set the bitline voltage back to the reference level (e.g., $V_{DD}/2$)}
\newacronym{trefi}{$\trefi{}$}{refresh interval}
\newacronym{trfc}{$\trfc{}$}{refresh latency}
\newacronym{trrd}{$\trrd{}$}{XXXXXXXXXXXXXXXXXXXXXXXXXXXXXXXXXXXXXXXXXXXXXX}
\newacronym{trasred}{$\trasred{}$}{reduced charge restoration latency}

\newcommand{\rhmemisolationrefs}[0]{\cite{fournaris2017exploiting, poddebniak2018attacking, tatar2018throwhammer, carre2018openssl, barenghi2018software, zhang2018triggering, bhattacharya2018advanced, google-project-zero, kim2014flipping, rowhammergithub, seaborn2015exploiting, van2016drammer, gruss2016rowhammer, razavi2016flip, pessl2016drama, xiao2016one, bosman2016dedup, bhattacharya2016curious, burleson2016invited, qiao2016new, brasser2017can, jang2017sgx, aga2017good, mutlu2017rowhammer, tatar2018defeating, gruss2018another, lipp2018nethammer, van2018guardion, frigo2018grand, cojocar2019eccploit,  ji2019pinpoint, mutlu2019retrospective, hong2019terminal, kwong2020rambleed, frigo2020trrespass, cojocar2020rowhammer, weissman2020jackhammer, zhang2020pthammer, yao2020deephammer, deridder2021smash, hassan2021utrr, jattke2022blacksmith, tol2022toward, kogler2022half, orosa2022spyhammer, zhang2022implicit, liu2022generating, cohen2022hammerscope, zheng2022trojvit, fahr2022frodo, tobah2022spechammer, rakin2022deepsteal, park2016statistical, park2016experiments,lim2017active, ryu2017overcoming, yun2018study, yang2019trap, walker2021ondramrowhammer, kim2020revisiting, orosa2021deeper, yaglikci2022understanding, khan2018analysis, agarwal2018rowhammer, li2014write, ni2018write, genssler2022reliability, mutlu2023fundamentally}}

\newcommand{\exploitingRowHammerAllCitations}[0]{\cite{  ji2019pinpoint, aga2017good, aydin2022cyber, barenghi2018software, bhattacharya2016curious, bhattacharya2018advanced, bosman2016dedup, brasser2017can, burleson2016invited, cai2022feasibility, carre2018openssl, cohen2022hammerscope, cojocar2019eccploit, cojocar2020rowhammer, deridder2021smash, fahr2022frodo, frigo2018grand, frigo2020trrespass, google-project-zero, gruss2016rowhammer, gruss2018another, hassan2021utrr, hong2019terminal, islam2022signature, jang2017sgx, jattke2022blacksmith, kaur2022work, kim2014flipping, kim2020revisiting, kogler2022half, kwong2020rambleed, lefforge2023reverse, li2022cyberradar, lipp2018nethammer, liu2022generating, marazzi2022protrr, mus2022jolt, mutlu2017rowhammer, mutlu2019retrospective, orosa2021deeper, orosa2022spyhammer, park2016experiments, park2016statistical, pessl2016drama, poddebniak2018attacking, qiao2016new, rakin2022deepsteal, razavi2016flip, redeker2002investigation, roohi2022efficient, rowhammergithub, saroiu2022price, seaborn2015exploiting, staudigl2022neurohammer, tatar2018defeating, tatar2018throwhammer, tobah2022spechammer, tol2022toward, van2016drammer, van2018guardion, walker2021ondramrowhammer, wang2022research, weissman2020jackhammer, xiao2016one, yang2019trap, yang2022socially, yao2020deephammer, zhang2018triggering, zhang2020pthammer, zhang2022implicit, zheng2022trojvit,fahr2022effects,fournaris2017exploiting,qureshi2021rethinking, luo2023rowpress, luo2024experimental, luo2024rowpress, olgun2023hbm, olgun2024read, jattke2025zenhammer, marazzi2024risch, lang2023blaster, baek2025marionette}}

\newcommand{\understandingRowHammerAllCitations}[0]{\cite{redeker2002investigation, kim2014flipping, park2014active, park2016statistical, yang2016suppression, park2016experiments,lim2017active, ryu2017overcoming, yang2017scanning, lim2018study, yun2018study, yang2019trap, gautam2019row, walker2021ondramrowhammer, kim2020revisiting, orosa2021deeper, jiang2021quantifying, orosa2022spyhammer, cohen2022hammerscope, yaglikci2022understanding, khan2018analysis, agarwal2018rowhammer, li2014write, ni2018write, genssler2022reliability, mutlu2023fundamentally, he2023whistleblower, baeg2022estimation, frigo2020trrespass, mutlu2017rowhammer, mutlu2018rowhammer, mutlu2019retrospective, olgun2023hbm, olgun2023drambender, zhou2023double, luo2023rowpress, olgun2024read, olgun2025variable, lang2023blaster, baek2025marionette, luo2025revisiting}}

\newcommand{\mitigatingRowHammerAllCitations}[0]{\cite{ AppleRefInc, aichinger2015ddr, ajorpaz2022evax, arikan2022processor, aweke2016anvil, bains2015row, bains2016distributed, bains2016row, bennett2021panopticon, brasser2017can, devaux2021method, di2023copy, enomoto2022efficient, fakhrzadehgan2022safeguard, france2022modeling, france2022reducing, frigo2020trrespass, gautam2019row, gomez2016dummy, greenfield2012throttling, gude2023defending, guha2022criticality, han2021surround, hassan2019crow, hassan2021utrr, hong2023dsac, hp2015rowhammer, irazoqui2016mascat, jedec2017ddr4, joardar2022learning, joardar2022machine, juffinger2023csi, kang2020cattwo, kim2014architectural, kim2014flipping, kim2022mithril, kim2023ddr5, konoth2018zebram, lee2019twice, lee2021cryoguard, lenovo2015rowhammer, loughlin2022moesiprime, manzhosov2022revisiting, marazzi2022protrr, marazzi2023rega, mutlu2023fundamentally, naseredini2022alarm, park2020graphene, park2022row, qureshi2021rethinking, qureshi2022hydra, ryu2017overcoming, saileshwar2022randomized, saroiu2022configure, saroiu2022price, saxena2022aqua, saxena2023pt, seyedzadeh2018cbt, sharma2022review, son2017making, tomita2022extracting, van2018guardion, vig2018rapid, wi2023shadow, woo2023scalable, yaglikci2021blockhammer, yaglikci2021security, yaglikci2022hira, yang2016suppression, yang2017scanning, you2019mrloc, zhang2020leveraging, zhang2022softtrr, zhou2022lt, zhou2023dnndefender,loughlin2021stop,nale2021perfect, canpolat2024breakhammer, bostanci2024comet, olgun2024abacus, yaglikci2024spatial, qureshi2024mint, qureshi2024impress, hassan2024selfmanaging, canpolat2025chronus, canpolat2024prac}}

\newcommand{\dramStandardCitations}[0]{\cite{jedec-ddr3, jedec2015hbm, jedec2015lpddr4, jedec2017ddr4, jedec2020ddr5, jedec2020lpddr5, jedec2021gddr6, jedec2024ddr5}}

\newcommand{\refreshBasedRowHammerDefenseCitations}[0]{\cite{lee2019twice, seyedzadeh2017cbt, seyedzadeh2018cbt, kang2020cattwo, park2020graphene, kim2022mithril, kim2014architectural, bains2015row, bains2016distributed, bains2016row, aweke2016anvil, AppleRefInc, kim2014flipping, son2017making, you2019mrloc, yaglikci2021security, frigo2020trrespass, hassan2021utrr, qureshi2022hydra, devaux2021method, lee2021cryoguard, marazzi2022protrr, zhang2022softtrr, joardar2022learning, yaglikci2024spatial, bostanci2024comet, olgun2024abacus, canpolat2024breakhammer, canpolat2024prac, qureshi2024mint, hassan2024selfmanaging, canpolat2025chronus}}

\newcommand{\rowHammerDefenseScalingProblemsCitations}[0]{\cite{kim2020revisiting, yaglikci2021blockhammer, park2020graphene, mutlu2017rowhammer, mutlu2018rowhammer, mutlu2019retrospective, mutlu2023fundamentally, hassan2021utrr, canpolat2024breakhammer, bostanci2024comet, yaglikci2024spatial, qureshi2024mint}}

\definecolor{gfored}{rgb}{0.580, 0.050, 0.211}
\definecolor{ao}{rgb}{0.007, 0.520, 0.867}
\definecolor{moegi}{rgb}{0.357, 0.537, 0.188}
\definecolor{jl}{rgb}{1.0, 0.2, 0.8}
\definecolor{brown(web)}{rgb}{0.65, 0.16, 0.16}
\definecolor{bisque}{rgb}{1.0, 0.89, 0.77}
\definecolor{nbs}{rgb}{0.88, 0.07, 0.37}
\definecolor{yt}{rgb}{0.58, 0.44, 0.86}
\definecolor{iy}{rgb}{0.0, 0.36, 0.05}
\definecolor{burntorange}{rgb}{0.8, 0.33, 0.0}
\definecolor{ouscolor}{rgb}{0.25, 0.41, 0.88}

\newcommand*\circled[1]{\tikz[baseline=(char.base)]{%
  \node[shape=circle,fill,inner sep=0.5pt] (char) {\textcolor{white}{#1}};}%
}%

\newcommand{\ignore}[1]{}

\ifdraft
    \newcommand{\param}[1]{\textcolor{red}{#1}} 
\else
\fi

\definecolor{frenchblue}{rgb}{0.19, 0.55, 0.91}

\usepackage[most]{tcolorbox} 
\tcbset{before skip=1.5pt, after skip=4pt}

\newtcolorbox[auto counter]{obsx}[3][]{%
    colframe = #2!45,
    colback  = #2!10,
    coltitle = #2!20!black, 
    colbacktitle=#2!20,
    coltitle=black,
    fonttitle=\bfseries, 
    title=#3~\thetcbcounter.\ ,
    enhanced,
    attach boxed title to top left={yshift=-2.8mm, xshift=0.15cm},
    bottom=-2.2pt,
    #1%
}

\usepackage[most]{tcolorbox} 
\newtcolorbox[auto counter]{tkx}[2][]{%
    enhanced, breakable, center title,
    colframe = #2!45,
    colback  = #2!10,
    colbacktitle=#2!20,
    left=-0.5pt,
    right=-0.5pt,
    bottom=-2pt,
    top=-0.25pt,
    #1%
}

\newcommand{\secref}[1]{§\ref{#1}}

\newcommand{\algref}[1]{Alg.~\ref{#1}}
\newcommand{\figref}[1]{Fig.~\ref{#1}}

\newcommand{\takeref}[1]{Takeaway~\ref{#1}}
\newcommand{\fnref}[1]{\textsuperscript{\ref{#1}}}

\definecolor{amber}{rgb}{1.0, 0.49, 0.0}
\definecolor{awesome}{rgb}{1.0, 0.13, 0.32}
\definecolor{dollarbill}{rgb}{0.52,0.73,0.4}
\definecolor{moegi}{rgb}{0.357, 0.537, 0.188}
\definecolor{burgundy}{rgb}{0.5, 0.0, 0.13}
\definecolor{ballblue}{rgb}{0.13, 0.67, 0.8}
\definecolor{ups-truck}{rgb}{0.53, 0.28, 0.21}
\definecolor{airforceblue}{rgb}{0.36, 0.54, 0.66}
\definecolor{cadmiumgreen}{rgb}{0.0, 0.42, 0.24}
\definecolor{darkcyan}{rgb}{0.0, 0.55, 0.55}
\definecolor{caribbeangreen}{rgb}{0.0, 0.8, 0.6}
\definecolor{flamingopink}{rgb}{0.99, 0.56, 0.67}
\definecolor{jazzberryjam}{rgb}{0.65, 0.04, 0.37}
\definecolor{mediumpersianblue}{rgb}{0.0, 0.4, 0.65}
\definecolor{coolblack}{rgb}{0.0, 0.18, 0.39}
\definecolor{bleudefrance}{rgb}{0.19, 0.55, 0.91}
\definecolor{ao}{rgb}{0.0, 0.0, 1.0}
\definecolor{babyblueeyes}{rgb}{0.63, 0.79, 0.95}
\definecolor{darkwarmgray}{rgb}{0.2,0,0}
\definecolor{brightpink}{rgb}{1.0, 0.0, 0.5}
\definecolor{iy}{rgb}{0.0, 0.36, 0.05}
\definecolor{nbcolor}{rgb}{1.0, 0.33, 0.64}
\definecolor{lightmauve}{rgb}{0.86, 0.82, 1.0}

\newcommand{\squishlist}{
 \begin{list}{$\circ$}
  { \setlength{\itemsep}{0pt}
     \setlength{\parsep}{0pt}
     \setlength{\topsep}{0pt}
     \setlength{\partopsep}{0pt}
     \setlength{\leftmargin}{1em}
     \setlength{\labelwidth}{1em}
     \setlength{\labelsep}{0.5em} } }

\newcommand{\squishsublist}{
\begin{list}{$\rightarrow$}
 { \setlength{\itemsep}{0pt}
    \setlength{\parsep}{0pt}
    \setlength{\topsep}{-10em}
    \setlength{\partopsep}{-3pt}
    \setlength{\leftmargin}{1em}
    \setlength{\labelwidth}{1em}
    \setlength{\labelsep}{0.5em} } }

\newcommand{\squishend}{
  \end{list}  }

\newcommand{\head}[1]{\noindent\textbf{#1.}}

\ifrev

    \newcommand{\cqrev}[1]{{#1}}
    \newcommand{\iqrev}[1]{{#1}}
    
    \newcommand{\cql}[1]{}
    \newcommand{\iql}[1]{}
\else
    \newcommand{\cqrev}[1]{{#1}}
    \newcommand{\iqrev}[1]{{#1}}
    
    \newcommand{\cql}[1]{}
    \newcommand{\iql}[1]{}
\fi

\ifshepherd

    \newcommand{\srev}[1]{{#1}}
    \newcommand{\ssrev}[1]{{#1}}
    \newcommand{\sql}[1]{}
\else
    \newcommand{\srev}[1]{{#1}}
    \newcommand{\ssrev}[1]{{#1}}
    \newcommand{\sql}[1]{}
\fi

\ifdraft
\fi
\newcommand{\gf}[2]{\ifnum#1=\value{version}{#2}\else{#2}\fi}
\newcommand{\agy}[2]{\ifnum#1=\value{version}{#2}\else{#2}\fi}
\newcommand{\atb}[2]{\ifnum#1=\value{version}{#2}\else{#2}\fi}
\newcommand{\yct}[2]{\ifnum#1=\value{version}{#2}\else{#2}\fi}
\newcommand{\ous}[2]{\ifnum#1=\value{version}{#2}\else{#2}\fi}
\newcommand{\iey}[2]{\ifnum#1=\value{version}{#2}\else{#2}\fi}
\newcommand{\nb}[2]{\ifnum#1=\value{version}{#2}\else{#2}\fi}
\newcommand{\hluo}[2]{\ifnum#1=\value{version}{#2}\else{#2}\fi}
\newcommand{\om}[2]{\ifnum#1=\value{version}{#2}\else{#2}\fi}

\newcommand{\agytodo}[2]{\ifnum#1=\value{version}\todo[size=\scriptsize, linecolor=orange, bordercolor=orange, backgroundcolor=white]{\textcolor{blue}{TODO:~#2}}\else{}\fi}
\newcommand{\ycttodo}[2]{\ifnum#1=\value{version}\todo[size=\scriptsize, linecolor=orange, bordercolor=orange, backgroundcolor=white]{\textcolor{yt}{TODO:~#2}}\else{}\fi}
\newcommand{\ieytodo}[2]{\ifnum#1=\value{version}\todo[size=\scriptsize, linecolor=orange, bordercolor=orange, backgroundcolor=white]{\textcolor{iey}{TODO:~#2}}\else{}\fi}

\newcommand{\agycomment}[2]{\ifnum#1=\value{version}\todo[size=\scriptsize, linecolor=orange, bordercolor=orange, backgroundcolor=white]{\textcolor{blue}{Giray: #2}}\else{}\fi}
\newcommand{\atbcomment}[2]{\ifnum#1=\value{version}\todo[size=\scriptsize, linecolor=orange, bordercolor=orange, backgroundcolor=white]{\textcolor{orange}{Atb: #2}}\else{}\fi}
\newcommand{\yctcomment}[2]{\ifnum#1=\value{version}\todo[size=\scriptsize, linecolor=orange, bordercolor=orange, backgroundcolor=white]{\textcolor{yt}{Yahya: #2}}\else{}\fi}
\newcommand{\ouscomment}[2]{\ifnum#1=\value{version}\todo[size=\scriptsize, linecolor=orange, bordercolor=orange, backgroundcolor=white]{\textcolor{ouscolor}{Oguzhan: #2}}\else{}\fi}
\newcommand{\gfcomment}[2]{\ifnum#1=\value{version}\todo[size=\scriptsize, linecolor=orange, bordercolor=orange, backgroundcolor=white]{\textcolor{purple}{Geraldo: #2}}\else{}\fi}
\newcommand{\ieycomment}[2]{\ifnum#1=\value{version}\todo[size=\scriptsize, linecolor=orange, bordercolor=orange, backgroundcolor=white]{\textcolor{iy}{Ismail: #2}}\else{}\fi}
\newcommand{\omcomment}[2]{\ifnum#1=\value{version}\todo[size=\scriptsize, linecolor=orange, bordercolor=orange, backgroundcolor=white]{\textcolor{gfored}{Onur: #2}}\else{}\fi}
\newcommand{\nbcomment}[2]{\ifnum#1=\value{version}\todo[size=\scriptsize, linecolor=orange, bordercolor=orange, backgroundcolor=white]{\textcolor{nbcolor}{Nisa: #2}}\else{}\fi}
\newcommand{\hluocomment}[2]{\ifnum#1=\value{version}\todo[size=\scriptsize, linecolor=orange, bordercolor=orange, backgroundcolor=white]{\textcolor{moegi}{Haocong: #2}}\else{}\fi}
\newcommand{\versionedparam}[2]{\ifnum#1=\value{version}{#2}\else{#2}\fi}
\providecommand{\param}[1]{\versionedparam{\value{version}}{#1}}

\newcommand{\X}[0]{PaCRAM}
\newcommand{\Xh}[0]{\mbox{PaCRAM-H}}
\newcommand{\Xm}[0]{\mbox{PaCRAM-M}}
\newcommand{\Xs}[0]{\mbox{PaCRAM-S}}
\newcommand{\Xlong}[0]{Partial Charge Restoration for Aggressive Mitigation}

\newcommand*\nCHIPS{388}

\newcommand{\revtag}[1]{}

\newcounter{obs}
\newcounter{tkw}
\newcommand\takeaway[1]{
\refstepcounter{tkw}
\begin{tkx}{lightmauve}
\small
\noindent\textbf{Takeaway~\thetkw.} \textit{#1}
\end{tkx}
}
   
\newcounter{take}
\titlespacing\section{0pt}{5pt plus 2pt minus 2pt}{0pt plus 2pt minus 2pt}
\titlespacing\subsection{0pt}{5pt plus 2pt minus 2pt}{0pt plus 2pt minus 2pt}
\titlespacing\subsubsection{0pt}{5pt plus 2pt minus 2pt}{0pt plus 2pt minus 2pt}

 \makeatletter
 \g@addto@macro{\normalsize}{%
   \setlength{\abovedisplayskip}{2pt plus 0.5pt minus 1pt}
   \setlength{\belowdisplayskip}{2pt plus 0.5pt minus 1pt}
   \setlength{\abovedisplayshortskip}{0pt}
   \setlength{\belowdisplayshortskip}{0pt}
   \setlength{\intextsep}{2pt plus 1pt minus 1pt}
   \setlength{\textfloatsep}{2pt plus 1pt minus 1pt}
   \setlength{\floatsep}{2pt plus 1pt minus 1pt}
   \setlength{\skip\footins}{3.5pt plus 1pt minus 1pt}}
\makeatother

\title{\Large Understanding RowHammer Under Reduced Refresh Latency:\\Experimental Analysis of Real DRAM Chips and Implications on Future Solutions\vspace{-1em}}

\newcounter{passversion}
\newcounter{subpassversion}
\pagenumberstrue
\fancypagestyle{firstpage}
{
    \fancyhead{}

    \fancyhead[C]{
        \raggedleft
            \includegraphics[width=12mm,height=12mm]{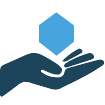}
            \includegraphics[width=12mm,height=12mm]{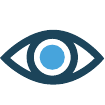}
            \includegraphics[width=12mm,height=12mm]{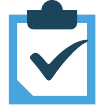}
    }

    \fancyfoot[C]{\thepage}
}
\newcites{module}{Module References}

\makeatletter
\def\bstctlcite{\@ifnextchar[{\@bstctlcite}{\@bstctlcite[@auxout]}}
\def\@bstctlcite[#1]#2{\@bsphack
  \@for\@citeb:=#2\do{%
    \edef\@citeb{\expandafter\@firstofone\@citeb}%
    \if@filesw\immediate\write\csname #1\endcsname{\string\citation{\@citeb}}\fi}%
  \@esphack}
\makeatother

\begin{document}
\maketitle
\thispagestyle{firstpage}
\pagestyle{plain}
\bstctlcite{IEEEexample:BSTcontrol}

\ifcameraready
    \setcounter{version}{99}
\else
    \setcounter{version}{1200}
\fi

\setstretch{0.9305}

\begin{abstract}

\agy{2}{Read disturbance in modern DRAM chips is a widespread weakness that is used for breaking memory isolation, one of the fundamental building blocks of \om{1}{system} security and privacy.}
RowHammer is a prime example of read disturbance in DRAM where repeatedly accessing (hammering) a row of DRAM cells (DRAM row) induces bitflips in physically nearby DRAM rows {(victim rows)}. 
Unfortunately, shrinking technology node size exacerbates \agy{2}{RowHammer}
{and} as such, significantly fewer accesses can induce bitflips \om{1}{in} newer DRAM chip generations. To ensure \om{1}{robust} DRAM operation, state-of-the-art mitigation mechanisms \agy{2}{restore the charge in potential victim rows (i.e., \om{2}{they perform} preventive refresh or charge restoration).} 
\srev{With newer DRAM chip \sql{R2.1}generations, these mechanisms perform preventive refresh more aggressively and cause larger performance, energy, or area overheads.}
\yct{20}{Therefore, it is essential to develop a better understanding and in-depth insights into the preventive refresh to secure real DRAM chips at low cost.}

\yct{20}{In this paper, our goal is to \om{2}{mitigate RowHammer at low cost by understanding} the preventive refresh latency and the impact of reduced refresh latency on RowHammer. To this end, we present the first rigorous experimental study on the interactions between refresh latency and RowHammer characteristics in real DRAM chips. Our experimental characterization \iey{20}{using} 388 real DDR4 DRAM chips from three major manufacturers demonstrates that a preventive refresh latency can be significantly reduced (by 64\%) at the expense of requiring slightly more (by 0.54\%) preventive refreshes.
\srev{To \sql{R2.2}investigate the impact of reduced \yct{9}{preventive} refresh latency on system performance and energy efficiency, 
we reduce the \yct{9}{preventive} refresh latency and adjust the aggressiveness of existing RowHammer solutions \om{1}{{by developing a new}} mechanism, \Xlong{} (\X{}).}
\ssrev{Our results show that by reducing the \yct{9}{preventive} refresh latency, \X{} reduces the performance and energy overheads induced by \om{1}{five} state-of-the-art RowHammer \yct{9}{mitigation mechanisms} with \yct{7}{small} additional area overhead.
Thus, \X{} introduces a novel perspective \om{1}{into} addressing RowHammer vulnerability at low cost by leveraging our experimental observations.}} To aid future research, we open-source our \X{} implementation at \url{https://github.com/CMU-SAFARI/PaCRAM}.\yctcomment{2}{Repo is ready, can be made public.}

\end{abstract}
\section{Introduction}
\label{sec:introduction}

{To ensure system \om{1}{robustness (i.e., reliability, security, and safety)}, it is critical to {maintain} {memory isolation: accessing a memory address should \emph{not} cause unintended side-effects on data stored on other addresses. {Unfortunately}, with aggressive technology node scaling, dynamic random access memory (DRAM)~\cite{dennard1968dram}, the prevalent {main} memory technology}, suffers from increased read disturbance: accessing (reading) a DRAM cell degrades the data integrity of other physically close \yct{0}{and} \emph{unaccessed} DRAM cells\om{1}{~\cite{kim2014flipping, kim2023flipping, mutlu2017rowhammer, mutlu2023fundamentally, mutlu2019retrospective}}. \om{1}{This is because reading a DRAM cell causes physically nearby DRAM cells to lose their charge (i.e., charge leakage) and exhibit bitflips.}
\emph{RowHammer} \agy{0}{is a} prime example of such {DRAM read {disturbance} \om{2}{phenomena} {where} a row of DRAM cells (i.e., a DRAM row) can experience bitflips when another physically nearby DRAM row (i.e., aggressor row) is} repeatedly opened (i.e., hammered)~\rhmemisolationrefs{}.}
\om{1}{\emph{RowPress} is another example of DRAM read disturbance \om{3}{phenomenon} where a DRAM row can experience bitflips when another physically nearby DRAM row is kept open for a long time \om{2}{period} (i.e., pressed)~\cite{luo2023rowpress, luo2024rowpress}. \agy{1}{Prior work~\cite{luo2024experimental} shows that combining RowHammer and RowPress significantly exacerbates disturbance on a victim row and significantly reduces the number and duration of row activations needed to induce the first read disturbance bitflip.}}

{Many prior works demonstrate \om{1}{security} attacks on a wide range of systems \om{2}{that} exploit read disturbance {to escalate {privilege}, leak private data, and manipulate critical application outputs~\om{1}{\exploitingRowHammerAllCitations{}}. \yctcomment{2}{includes RH-RP work}
To make matters worse, recent experimental studies\om{1}{~\understandingRowHammerAllCitations{}}\yctcomment{1}{HBM papers} have found that {newer DRAM chip generations are more susceptible to read disturbance}. {For example,} DRAM chips manufactured in {2018-2020} can experience RowHammer {bitflips} after an order of magnitude fewer row activations compared to the chips manufactured in {2012-2013}~\cite{kim2020revisiting}. \om{1}{Recent studies~\cite{olgun2023hbm, olgun2024read} also show that \agy{1}{modern} HBM2 DRAM chips are \agy{1}{as} vulnerable to both RowHammer and RowPress \agy{1}{as modern DDR4 and LPDDR4 DRAM chips}.}

\om{1}{To ensure robust DRAM operation, several prior works~\refreshBasedRowHammerDefenseCitations{} \agy{1}{propose to} perform an operation called \emph{preventive refresh}. Preventive refresh restores the \agy{1}{electrical} charge \agy{1}{stored in potential victim cells} (i.e., \om{2}{performs} charge restoration). This operation is performed before 
\agy{1}{the victim row gets disturbed significantly enough to experience bitflips due to RowHammer or RowPress. For precise timing of preventive refreshes, these works define a threshold called \emph{RowHammer threshold}, which represents the minimum hammer count needed to induce the first bitflip in a victim row.}
Existing RowHammer mitigation \yct{9}{mechanisms} can also prevent RowPress bitflips when they are configured to be more aggressive, which is practically equivalent to configuring them for smaller RowHammer thresholds~\cite{luo2023rowpress, luo2024rowpress, luo2024experimental}.}

{Charge restoration of a DRAM cell has a non-negligible latency\om{1}{, denoted by the (LP)DDRx timing parameter $t_{RAS}$}, which is typically $32-35 ns$~\dramStandardCitations{}. 
During charge restoration of cells in a DRAM row, the DRAM bank that contains the row becomes unavailable, and thus refreshing a DRAM row can \om{1}{delay} memory accesses and induce performance overheads.}
\om{1}{As read \om{2}{disturbance} in DRAM chips \om{3}{worsens}, RowHammer mitigation \om{1}{mechanisms} face a trade-off between system performance and area overhead. Some solutions rely on \emph{more aggressive} preventive refreshes, resulting in higher performance overheads, but lower area overheads (i.e., \emph{high-performance-overhead mitigations}), while other solutions introduce larger area overheads to detect a RowHammer attack more precisely, but incur lower performance overheads (i.e., \emph{high-area-overhead mitigations})~\rowHammerDefenseScalingProblemsCitations{}.}

\srev{To mitigate read disturbance at \emph{both} low performance and area overheads, it is {critical} to understand
{the \om{1}{effects of the }refresh 
on} {read disturbance in DRAM chips}. 
\om{2}{This understanding can lead to} more \om{2}{robust solutions for} current and future DRAM-based {memory} systems.}} 
\agy{3}{Many prior works~\understandingRowHammerAllCitations{} study the characteristics of DRAM read disturbance in various aspects.} \yct{1}{However, }\agy{0}{even though charge restoration and read disturbance \om{1}{both greatly affect} charge leakage in DRAM cells,
\emph{no} prior work rigorously studies the interaction between them and \om{2}{the} implications \om{2}{of this interaction} on RowHammer \yct{9}{mitigation mechanisms} \om{1}{and attacks}.}

\om{1}{\textbf{Our goal} is to enable RowHammer mitigation at \emph{both} low performance and area overheads by understanding RowHammer under reduced refresh latency \om{1}{in real DRAM chips}.}
\yct{20}{To this end}, we present \yct{1}{i)} the first rigorous experimental study on the interactions between refresh latency and read disturbance characteristics of real modern DRAM chips~\yct{20}{and ii) demonstrate the potential benefits of our empirical observations by \om{1}{developing} a mechanism, \om{1}{\emph{\Xlong{} (\X{})}}. 
\om{2}{The key idea of \X{} is to reliably reduce} the latency of the preventive refreshes \om{1}{in} both high-performance-overhead and high-area-overhead RowHammer mitigation \om{1}{mechanisms}\om{2}{, thereby improving system performance and energy efficiency.} By doing so, \X{} introduces a novel perspective \om{1}{into} addressing the RowHammer vulnerability while incurring \yct{7}{small} additional area overhead over existing RowHammer mitigation \om{1}{mechanisms}.
}

\head{\om{1}{Experimental} characterization} We \agy{1}{present }the first rigorous experimental characterization \agy{1}{study} \om{1}{that examines the effects of refresh latency on RowHammer vulnerability, on \nCHIPS{} real DDR4 DRAM \agy{1}{chips}.} 
\om{2}{Our experimental characterization demonstrates that the latency of preventive refreshes can be safely reduced with either \emph{no} change or a relatively small reduction in the RowHammer threshold, and such reduced refresh latency does \emph{not} further exacerbate RowHammer vulnerability beyond the initial degradation, even after thousands of preventive refreshes.}
\om{2}{Therefore, the latency of a vast majority of preventive refreshes can be reduced reliably without jeopardizing the data integrity of a DRAM chip.}
}

\head{\yct{20}{Potential benefits of reduced refresh latency}} 
\yct{20}{We showcase the benefits of our empirical observations \om{1}{by developing} a mechanism, \om{1}{\emph{\Xlong{} (\X{})}}}.
\X{} is implemented in the memory controller together with existing RowHammer \yct{9}{mitigation mechanisms}. 
\om{1}{\X{} carefully \om{2}{and reliably} reduces the latency of preventive refreshes, restoring the charge of the potential victim cells partially (i.e., partial charge restoration).
To ensure secure operation, \X{} adjusts the aggressiveness of \om{2}{a} RowHammer \yct{9}{mitigation mechanism} (i.e., configures the existing \yct{9}{mechanism} with the reduced RowHammer threshold) \om{2}{as some DRAM chips exhibit an increased vulnerability to RowHammer when they are refreshed using reduced refresh latency}.} 
\om{1}{To configure \X{}, we use our experimental data (i.e., reduced charge restoration latency, RowHammer threshold) from real DRAM characterization.}
We demonstrate that \X{} significantly increases the system performance~\yct{20}{(energy efficiency)} by reducing the overheads induced by \yct{20}{five} state-of-the-art RowHammer \yct{9}{mitigation mechanisms},
PARA~\cite{kim2014flipping}, 
RFM~\cite{jedec2020ddr5},
PRAC~\cite{jedec2024ddr5},
Hydra~\cite{qureshi2022hydra}, and 
Graphene~\cite{park2020graphene}
by \yct{20}{18.95\% (14.59\%), 12.28\% (11.56\%), 2.07\% (1.15\%), 2.56\% (2.18\%), and 5.37\% (4.50\%)} on average \yctcomment{7}{checked all numbers}across \param{62} workloads \om{2}{while introducing \om{3}{small} additional area overhead \om{3}{in the memory controller} (0.09\% of the area of a high-end Intel Xeon processor~\cite{wikichipcascade})}.

We make the following contributions:

\begin{itemize}

    \item We present the first rigorous characterization of the \om{1}{effects of reduced refresh latency on} RowHammer vulnerability \om{1}{in real DRAM chips}. Our experimental results on \param{\nCHIPS{}} real DRAM chips \om{1}{from three major manufacturers show that \om{2}{latency of a vast majority of preventive refreshes can be reduced reliably}.}
    
    \item \om{1}{To demonstrate the potential benefits of our experimental observations, we propose \om{1}{\emph{\Xlong{} (\X{})}}, a \om{1}{new} mechanism} that \om{2}{reliably} reduces the latency of preventive refreshes \om{3}{performed by} RowHammer mitigation mechanisms\om{3}{, and \om{6}{accordingly} adjusts the aggressiveness of RowHammer mitigation mechanisms}.

    \item \ssrev{We showcase \X{}'s \om{1}{effectiveness by evaluating its} impact using five state-of-the-art RowHammer mitigation mechanisms. \om{1}{Our results} show that \X{} significantly reduces \agy{20}{their} performance \yct{20}{and energy} overheads while introducing \om{3}{small additional area overhead in the memory controller.}}

\end{itemize}

\section{Background}
\label{sec:background}
\subsection{DRAM Organization and Operation}
\label{sec:dram_organization}

\begin{figure}[ht]
    \centering
    \includegraphics[width=\linewidth]{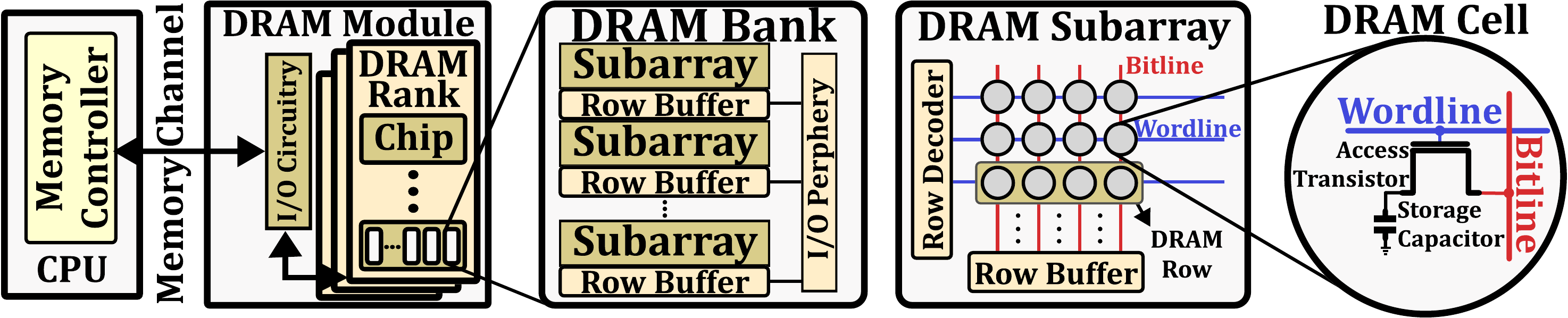}
    \caption{\srev{DRAM \sql{R4}organization}}
    \label{fig:dram_organization}
\end{figure}

\head{\om{2}{DRAM} organization}~\figref{fig:dram_organization} {shows the hierarchical organization of a modern DRAM-based main memory. The memory controller connects to a DRAM module over a memory channel. A module contains one or multiple DRAM ranks that time-share the memory channel. A rank consists of multiple DRAM chips.
Each DRAM chip \agy{0}{has} multiple DRAM banks \agy{0}{each of which contains multiple subarrays}. A DRAM bank is organized as a two-dimensional array of DRAM cells, where a row of cells is called a DRAM row. A DRAM cell consists of i) a storage capacitor, which stores one bit of information in the form of electrical charge, and ii) an access transistor.}

\begin{figure}[ht]
\centering
\includegraphics[width=0.8\linewidth]{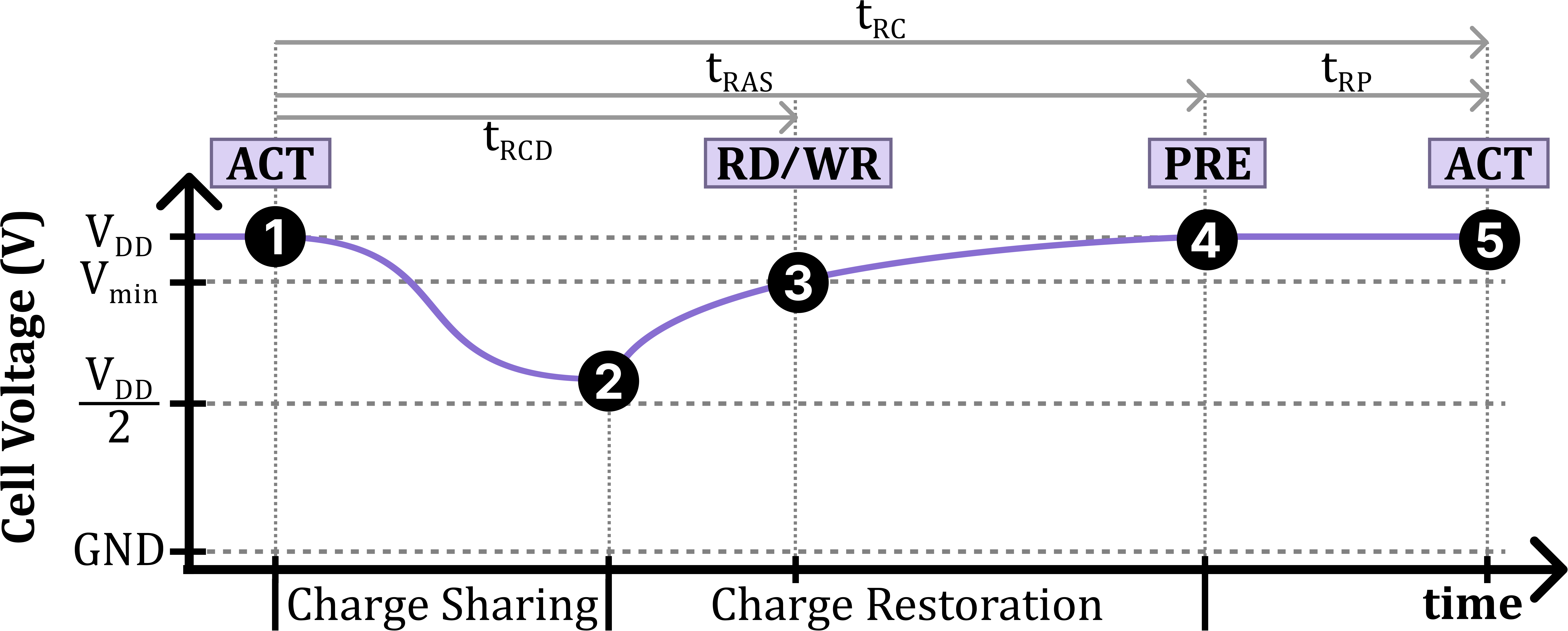}
\caption{\om{3}{Timeline of a DRAM cell's activation process}}
\label{fig:pr_timeline}
\end{figure}

\head{DRAM operation and timing}
\iey{0}{There are four \yct{1}{main} DRAM commands to access a DRAM row: $ACT$, $RD$, $WR$, and $PRE$. \figref{fig:pr_timeline} illustrates the relationship between the commands issued to access a DRAM cell and their governing timing parameters in \om{3}{five} steps. Initially, the cell capacitor stores $V_{DD}$ and is precharged.~\circled{1}~\om{3}{The} memory controller issues the $ACT$ command. 
\om{3}{$ACT$ asserts the corresponding \om{2}{row's} wordline, which} consequently enables all the access transistors in the DRAM row, \om{3}{conducting} cell capacitors to corresponding bitlines, and \om{3}{initiating} an analog \om{3}{\emph{charge sharing}} process between each cell capacitor and its corresponding bitline.~\circled{2}~\om{3}{Once} the cell and bitline voltages equalize due to charge sharing, \om{3}{\emph{charge restoration}} starts. During charge restoration, the sense amplifiers are enabled to first detect the bitline voltage shift and later restore the bitline to a full $V_{DD}$ or $GND$ depending on the direction of the shift.~\circled{3}~\om{3}{Once} the cell is restored to a voltage level that is ready to access \yct{3}{($V_{min}$)}, which requires waiting for \om{3}{the} $t_{RCD}$ timing parameter (i.e., the timing parameter between $ACT$ and $RD$/$WR$ commands), $RD$ or $WR$ commands can be issued to the bank.~\circled{4}~\om{3}{The} memory controller can follow an $ACT$ with $PRE$ to the same bank after at least the time interval for \gls{tras}. $t_{RAS}$ ensures that enough time has passed to fully restore the DRAM cells of the activated row to a ready-to-precharge voltage. The latency of $PRE$ is governed by \yct{6}{\gls{trp}}.~\circled{5}~After $t_{RP}$, the memory controller can issue an $ACT$ to open a new row in the same bank.}
\yct{7}{Thus, an $ACT$ following another $ACT$ to the same bank can be issued only after $t_{RAS} + t_{RP}$ \om{8}{(i.e., $t_{RC}$)}.}
\yctcomment{6}{I will add tRC to fig.}

\head{{DRAM refresh}}
\agy{3}{DRAM is a volatile memory technology: a DRAM cell leaks charge over time and can lose its data (i.e., charge leakage).}
If charge leakage exceeds a level, the data of the cell \emph{cannot} be sensed by the sense amplifier correctly. These errors are known as data retention failures. To prevent \yct{9}{these} failures, 
the memory controller issues $REF$ commands to refresh DRAM cells. A $REF$ command \agy{3}{internally} activates DRAM rows to sense and restore the charge \om{3}{levels} of the cells\om{3}{;} \agy{3}{which takes around \om{3}{$32-35 ns$}~\dramStandardCitations{}}. A single refresh can be thought of as opening (i.e., activating) a row and closing (i.e., precharging) the bank, as shown in~\figref{fig:pr_timeline}.
{To maintain data integrity, the memory controller periodically refreshes each row in a time interval called \gls{trefw} ($32 ms$ for DDR5~\cite{jedec2020ddr5} and $64 ms$ for DDR4~\cite{jedec2017ddr4}). To ensure all rows are refreshed every \gls{trefw}, the memory controller issues $REF$ commands with a time interval called \gls{trefi} ($3.9 \mu s$ for DDR5~\cite{jedec2020ddr5} and $7.8 \mu s$ for DDR4~\cite{jedec2017ddr4}).} \agy{0}{Each $REF$ command blocks the whole DRAM bank or rank for a time window called \gls{trfc}. \yct{6}{\agy{3}{\gls{trfc} is} a \om{3}{function of the} charge restoration {latency} and the number of rows refreshed with a $REF$ command. \om{7}{\gls{trfc} is} \yct{6}{\om{3}{$195ns$ and $350ns$ for 8Gb DDR5~\cite{jedec2020ddr5} \agy{3}{and \om{7}{8Gb} DDR4~\cite{jedec2017ddr4} DRAM} chips}}\yct{7}{, respectively}.}}
\subsection{DRAM Read Disturbance}
\label{sec:background_dram_read_disturbance}
\yct{3}{As DRAM manufacturing technology
\agy{3}{node size scales down, DRAM cell \om{6}{sizes} and cell-to-cell \om{6}{distances} shrink. As a result, accessing a DRAM cell can \om{6}{more easily} disturb data stored in another physically nearby cell. \om{8}{This phenomenon} is called \emph{read disturbance}}.
Two prime examples of read disturbance in modern DRAM chips are \emph{RowHammer}~\cite{kim2014flipping} and \emph{RowPress}~\cite{luo2023rowpress}, where repeatedly accessing (\emph{hammering}) or keeping active (\emph{pressing}) a DRAM row induces bitflips in physically nearby DRAM rows, respectively. In RowHammer and RowPress terminology, the row that is hammered or pressed is called the \emph{aggressor} row, and the row that experiences bitflips \iey{7}{is} the \emph{victim} row. For read disturbance bitflips to occur\agy{3}{,} \gls{nrh}.
Several characterization studies~\understandingRowHammerAllCitations{} show that as DRAM technology continues to scale to smaller nodes, DRAM chips are becoming increasingly vulnerable to RowHammer (i.e., newer chips have lower \gls{nrh} values).}\footnote{For example, one can induce RowHammer bitflips with 4.8K activations in the chips manufactured in 2020 while a row needs to be activated 69.2K times in older chips manufactured in 2013\om{6}{~\cite{kim2020revisiting}}.} 
\om{2}{Recent studies~\cite{olgun2023hbm, olgun2024read} also show that modern HBM2 DRAM chips are as \om{3}{vulnerable} to both RowHammer and RowPress as modern DDR4 and LPDDR4 DRAM chips.}
\agy{0}{To make matters worse, \gls{nrh} reduces significantly by crafting access patterns that exploit both RowHammer and RowPress~\cite{luo2023rowpress}.}

\head{DRAM read disturbance mitigation \om{0}{mechanisms}}
Many prior works propose mitigation mechanisms~\mitigatingRowHammerAllCitations{} to protect DRAM chips against RowHammer bitflips\om{3}{,} leveraging different approaches. \agy{0}{These \om{0}{mechanisms} \om{3}{usually} perform two \om{3}{main} tasks: i)~execute a \om{3}{\emph{trigger algorithm}} and ii)~perform a \om{3}{\emph{preventive action}}. The trigger algorithm observes the memory access patterns and triggers a preventive action based on the result of a probabilistic or a deterministic process. 
The preventive action is \om{3}{usually} one of i)~refreshing victim rows \yct{3}{(i.e., \emph{preventive refresh})}, ii)~dynamically remapping aggressor rows, and iii)~throttling unsafe accesses.}
\yct{20}{The DDR5 standard introduces a new command called \emph{refresh management} (RFM)~\cite{jedec2020ddr5} to protect DRAM chips against RowHammer attacks. The memory controller tracks the activation count of each DRAM bank and issues an RFM command when the count exceeds a threshold value. When the DRAM module receives an RFM command, it internally takes preventive actions within the given time window. 
\agy{3}{RFM tracks the row activation counts \om{6}{at} bank granularity rather than row granularity. Therefore, RFM commands are issued based on the total activation count across thousands of rows in a bank with \emph{no} notion of the row-level temporal locality of those activations. 
As such, the memory controller can trigger an excessive \om{3}{number} of RFM commands\om{3}{,} causing high performance overheads~\cite{canpolat2024prac}.}
To \om{3}{reduce} the number of RFM commands, the latest DDR5 standard introduces a new \om{0}{mechanism} called \emph{Per Row Activation Counting} (PRAC)~\cite{jedec2024ddr5}. PRAC enables fine-grained tracking by implementing \om{3}{per-row} activation counters inside the DRAM chip. When a row's counter exceeds a threshold value, the DRAM chip requests an RFM command from the memory controller with a new \emph{back-off} signal. Once the memory controller receives this signal, it issues an RFM command and allows the DRAM chip to take preventive actions.}
\agy{0}{Existing RowHammer mitigations can also \om{3}{be used to} prevent \om{3}{\emph{RowPress}} bitflips when their trigger algorithms are configured to be more aggressive, which is practically equivalent to configuring them for sub-1K \gls{nrh} values~\cite{luo2023rowpress, luo2024experimental, luo2024rowpress}.}

{RowHammer mitigation \om{0}{mechanisms} face an important trade-off between two \om{3}{sets of metrics:} i)~system performance \om{3}{(and energy efficiency)} and ii)~area cost. \om{3}{We} categorize RowHammer mitigation \om{1}{mechanisms} depending on their dominant overheads, i)~\emph{high-performance-overhead mitigations} and ii)~\emph{high-area-overhead mitigations}.
High-performance-overhead mitigations \om{3}{(e.g., PARA~\cite{kim2014flipping} and RFM~\cite{jedec2020ddr5})} employ basic detection algorithms that introduce lower area overhead but falsely detect many aggressor rows, causing many unnecessary preventive refreshes. Due to the high number of preventive refreshes, these mitigations introduce higher performance and energy \om{3}{overheads than high-area-overhead mechanisms}.
High-area-overhead mitigations \om{3}{(e.g., PRAC~\cite{jedec2024ddr5}, Hydra~\cite{qureshi2022hydra}, and Graphene~\cite{park2020graphene})} utilize \om{3}{more sophisticated} trigger algorithms that accurately detect aggressor rows and avoid performing unnecessary preventive refreshes. Therefore, these mitigations introduce lower performance and energy overheads but higher area overheads.} \om{1}{As DRAM read disturbance worsens with shrinking technology, \gls{nrh} values are expected to reduce even more\om{3}{~\understandingRowHammerAllCitations{}}.}
Unfortunately, existing RowHammer mitigation \om{6}{mechanisms} incur either high performance \om{3}{(and energy)} overheads or high area overheads. \om{7}{Both high-performance-overhead and high-area-overhead mitigation \om{6}{mechanisms} are not scalable with increasing RowHammer vulnerability.}
Therefore, reducing the overheads of existing RowHammer \yct{9}{mitigation mechanisms} is critical \om{3}{to enable efficient scaling of DRAM-based systems into the future}.


\section{Motivation}
\label{sec:motivation}

\begin{figure}[ht]
\centering
\includegraphics[width=0.85\linewidth]{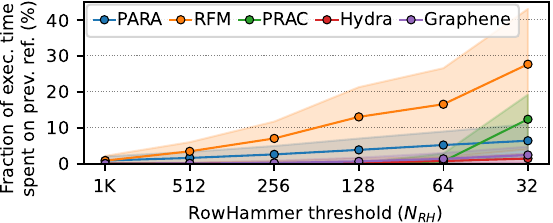}
\caption{\om{3}{Preventive refresh overhead of five RowHammer mitigation mechanisms as RowHammer vulnerability worsens}}
\label{fig:performance_motivation}
\end{figure}

\head{\yct{3}{Preventive refresh} overheads of RowHammer mitigation \om{1}{mechanisms}}
\srev{Prior works~\rowHammerDefenseScalingProblemsCitations{} already show that existing RowHammer mitigations \sql{R2.1}incur prohibitively high performance or area overheads as \gls{nrh} reduces \om{3}{with DRAM technology scaling}.} \agy{3}{This is because blocking a DRAM bank for performing a preventive refresh
might delay memory accesses and thus degrade system performance. We analyze} 
\yct{3}{the fraction of execution time \agy{3}{during which} a DRAM bank is \agy{3}{busy performing} preventive refreshes,}
following the methodology described in \secref{sec:eval_methodology} for \agy{3}{\param{five} state-of-the-art RowHammer mitigation \om{1}{mechanisms} and} \om{3}{six} different \gls{nrh} values \om{3}{on} \param{60} multi-programmed 4-core workload mixes.
\figref{fig:performance_motivation} shows a line plot where each curve represents a different RowHammer mitigation \om{1}{mechanism}. The x-axis shows \om{7}{the} tested \gls{nrh} values and the y-axis shows
the fraction
of \om{3}{execution} time spent on preventive refreshes.
The shaded areas mark the minimum and maximum values across all workload mixes.
\yctcomment{7}{fixed colors}
\vspace{10pt}


\cqrev{
We make \param{four} observations from \figref{fig:performance_motivation}.
First, all tested RowHammer mitigation \om{1}{mechanisms} spend a larger fraction of their time performing preventive refreshes 
as \om{3}{\gls{nrh} reduces (i.e., DRAM chips become more vulnerable to read disturbance)}. 
Second, RFM's, PRAC's, and PARA's \om{3}{overheads} reach up to 43.05\%, 19.19\%, and 10.95\%, respectively, in the worst-case across our tests.
\agy{3}{Third, \param{RFM} and \param{PARA} exhibit the highest time spent on preventive refreshes for all tested \gls{nrh} values larger than 32 (e.g., 19.30\% and 6.12\% of total execution time, on average across tested workload mixes at an \gls{nrh} of 64, respectively), while they introduce almost zero area overhead~\cite{kim2014flipping, jedec2020ddr5} (not shown in the figure).}
%
Fourth, among the tested RowHammer mitigations, Graphene and Hydra \om{3}{spend the least} time \om{3}{on} preventive refreshes, \om{3}{e.g., only}
\param{2.68\%} and \param{1.54\%} \om{3}{of total execution time} for an \gls{nrh} value of \param{32} on average across all tested workload mixes. Hydra \om{6}{wastes} the least \agy{3}{fraction of execution time} \om{6}{on} preventive refreshes\om{7}{,} but incurs significant \om{6}{system-level} slowdowns \om{7}{(not shown here; see~\secref{sec:eval_perf})} because it maintains its \om{3}{counter} metadata in DRAM and occupies \om{3}{the} memory channel to retrieve and update \om{3}{its metadata}~\cite{canpolat2024breakhammer, bostanci2024comet, olgun2024abacus}. 
\param{Graphene} \agy{3}{spends} \iey{6}{\agy{6}{a} higher fraction of execution time on preventive refreshes than Hydra, but incurs the lowest system-level slowdowns across all tested RowHammer mitigations (not shown here; see~\secref{sec:eval_perf}). Graphene achieves the lowest overall performance overhead}
at the expense of increasingly large chip area overhead that reaches \iey{0}{\param{$10.38mm^2$} (\param{$4.45\%$} of the chip area of an Intel Xeon processor~\cite{wikichipcascade})
when configured with an \gls{nrh} of 32 for a \om{6}{dual-rank} system with 16 banks \om{6}{per} rank.}
With these \param{four} observations, we \yct{1}{demonstrate} that existing RowHammer mitigations incur significant \yct{3}{preventive refresh overheads} or area overheads as \gls{nrh} decreases. Therefore, we conclude that reducing the overheads of such mechanisms at low cost is critical.
}

\head{Reducing the time \agy{0}{and energy} spent for preventive refreshes}\yctcomment{3}{Fig 3 provides the time cost with the nominal tRAS, here we will provide the cost of preventive refs with reduced tRAS.}
Prior works already show that DRAM timing constraints include large guardbands\om{6}{~\cite{lee2015adaptive, liu2013experimental, chang2016understanding, chang2017understanding, chang2017understandingphd, kim2018solar, yaglikci2022understanding, mathew2017using, lee2017design, chandrasekar2014exploiting, das2018vrldram}}\yctcomment{6}{added DATE2014, SIGMETRICS2017, VRL-DRAM}\yctcomment{7}{fixed refs}. To reduce the performance overhead of RowHammer mitigation \om{1}{mechanisms}, we explore the idea of reducing the time spent on \om{3}{each} preventive refresh.
To this end, we 
experimentally evaluate the effect of reducing \gls{tras}
\agy{3}{on the} i)~reliability of real DDR4 DRAM chips using the methodology described in \agy{3}{\secref{sec:methodology_dramchips}}, and ii)~\agy{3}{time and energy costs} of preventive refreshes.

\begin{figure}[ht]
\centering
\includegraphics[width=0.92\linewidth]{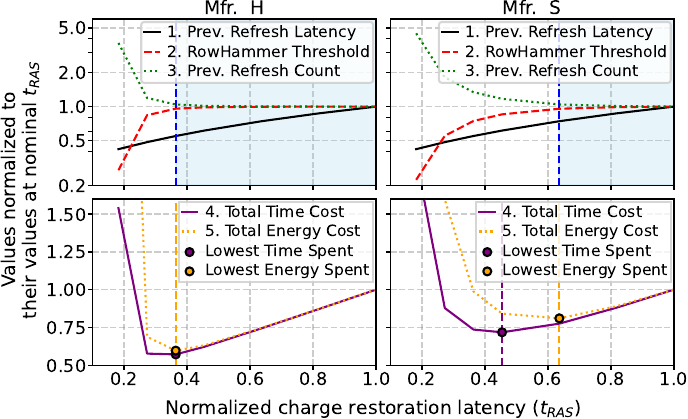}
\caption{\srev{Effect of reducing charge restoration latency on \sql{R4}the time \om{3}{and energy spent on} preventive refreshes}}
\label{fig:m_plot}
\end{figure}

\figref{fig:m_plot} shows the results of our motivational analysis for two representative DDR4 DRAM modules \om{1}{one from \agy{3}{Mfrs.~H and~S} \om{3}{(see \secref{sec:methodology_dramchips})}}. We reduce the charge restoration latency from right to left on the x-axis. $x=1.0$ marks the nominal \gls{tras} value \om{3}{(i.e., $33ns$)}. The y-axis shows the values of five different curves normalized to their respective values at the nominal \gls{tras} value, three of which are \om{3}{in} the top two subplots\iey{7}{,} and the remaining two are in the bottom two subplots. The enumerated five curves show \yct{3}{the following\om{6}{:}} 
1)~\om{3}{\emph{Prev\om{6}{entive} Refresh Latency}:} the time spent to perform a single preventive refresh \yct{3}{(solid \om{6}{black} line in top subplots)}. \yct{3}{DRAM standards~\dramStandardCitations{} do \emph{not} report how the preventive refresh latency is calculated. \om{6}{We} compute it as the sum of \gls{tras} and $t_{RP}$ timing parameters\om{6}{, since} a preventive refresh is functionally equivalent to opening and closing a DRAM row.}
2)~\om{3}{\emph{RowHammer Threshold}:} the minimum activation count needed to induce the first RowHammer bitflip, i.e., \gls{nrh}, observed on real DRAM chips~\yct{0}{when reduced charge restoration latency is used} \yct{3}{(dashed \om{6}{red} line in top subplots)}.
3)~\om{3}{\emph{Prev\om{6}{entive} Refresh Count}:} the number of preventive refreshes, \agy{3}{computed as 1/\gls{nrh} based on the insight that a preventive refresh is performed every \gls{nrh} activations \yct{3}{(dotted \om{6}{green} line in top subplots)}.}
4)~\om{3}{\emph{Total Time Cost}:} the time spent for all preventive refreshes as \agy{3}{the} product of \yct{7}{Preventive Refresh Count} and \yct{7}{Preventive Refresh Latency} \yct{3}{(solid purple line in bottom subplots)}.
5)~\om{3}{\emph{Total Energy Cost}:} energy consumption \om{6}{of} all preventive \agy{20}{refreshes} as the product of \yct{7}{Preventive Refresh Count} and the total time spent doing so \yct{3}{(dotted \om{6}{yellow} line in bottom subplots)}.

From \figref{fig:m_plot}, we make \param{four} observations. 
\yct{3}{First, \om{7}{Preventive Refresh Latency} reduces proportionally as charge restoration latency decreases.}
\yct{3}{Second, \om{6}{reducing the} charge restoration latency i)~\om{7}{reduces RowHammer Threshold} (\gls{nrh}) by less than \om{6}{only} $5\%$ \yct{6}{when} \gls{tras} is reduced by $64\%$ and $36\%$} for \om{1}{the \om{6}{two} \agy{3}{modules} from \agy{3}{Mfrs.~H and~S}}, respectively (marked with dashed blue lines), and ii)~\om{6}{reduces} \gls{nrh} \om{6}{much} more significantly if \gls{tras} is reduced \om{7}{more} than these values. 
\yct{3}{Third, \om{6}{Total Time Cost} \om{6}{of preventive refreshes}
\agy{3}{i)~reduces} with reducing \gls{tras} until an inflection point (marked with a dark purple circle \om{6}{in \figref{fig:m_plot} (bottom)} at 36\% and 45\% of the nominal \gls{tras} value for modules from \agy{3}{Mfrs.~H and~S}, respectively)\om{6}{,} \agy{3}{and}
\agy{3}{ii)~increases beyond the respective inflection points.} 
\agy{3}{This is because} the \om{7}{higher Preventive Refresh Count} \om{6}{at lower \gls{tras}} overwhelms the reduction in \om{7}{Preventive Refresh Latency}. Hence, the lowest \om{7}{Total Time Costs} are observed at these inflection points with a \om{7}{Total Time Cost} reduction of $43\%$ and $28\%$ for the \om{6}{two} \agy{3}{modules} from \agy{3}{Mfrs.~H and~S}, respectively.} 
\yct{3}{Fourth, \om{6}{Total Energy Cost \om{6}{of preventive refreshes} also has \om{6}{similar} inflection points}
(marked with a dark \om{6}{yellow} circle \om{6}{in \figref{fig:m_plot} (bottom)})
at $36\%$ and $64\%$ of the nominal \gls{tras} value where the total energy cost can be reduced by $40\%$ and $19\%$ for the \agy{3}{modules} from \agy{3}{Mfrs.~H and~S}, respectively.}\yctcomment{7}{yes they are consistent with the figure}
Based on this motivational analysis, we conclude that reducing \agy{3}{charge} restoration latency is a promising approach to reduce the performance and energy overheads of RowHammer mitigations. 

\head{{Characterizing the Impact of {Charge} Restoration Latency}}
{To reliably reduce the {charge} restoration latency of preventive refreshes, it is critical to understand the limits of a DRAM chip. 
\yct{3}{Therefore, it is \om{6}{important} to investigate i)~how RowHammer \om{6}{T}hreshold and the number of RowHammer bitflips
change with reduced charge restoration latency, ii)~whether \agy{3}{using reduced charge restoration latency for many consecutive \yct{3}{preventive} refreshes further worsens RowHammer threshold or number of RowHammer bitflips},
and iii)~how the data retention time of a DRAM row changes with the reduced charge restoration latency.}
Although many prior works~\understandingRowHammerAllCitations{} study \om{7}{various aspects of} the \om{3}{read disturbance} characteristics of DRAM (e.g., \om{3}{access and data patterns}, voltage, temperature), \emph{no} prior work characterizes the effect of \agy{3}{charge} restoration latency on RowHammer.
}

\yct{7}{In this paper, we conduct a rigorous experimental characterization study to understand how reducing charge restoration latency affects RowHammer vulnerability in real DDR4 DRAM chips, along with its effects on system performance and energy efficiency.}

\section{Characterization Methodology}
\label{sec:methodology}
\om{3}{We} describe our DRAM testing infrastructure, the real DDR4 DRAM chips tested, and our testing methodology.

\subsection{DRAM Testing Infrastructure}
\label{subsec:methodology_infra}

\begin{figure}[ht]
\centering
\includegraphics[width=0.9\linewidth]{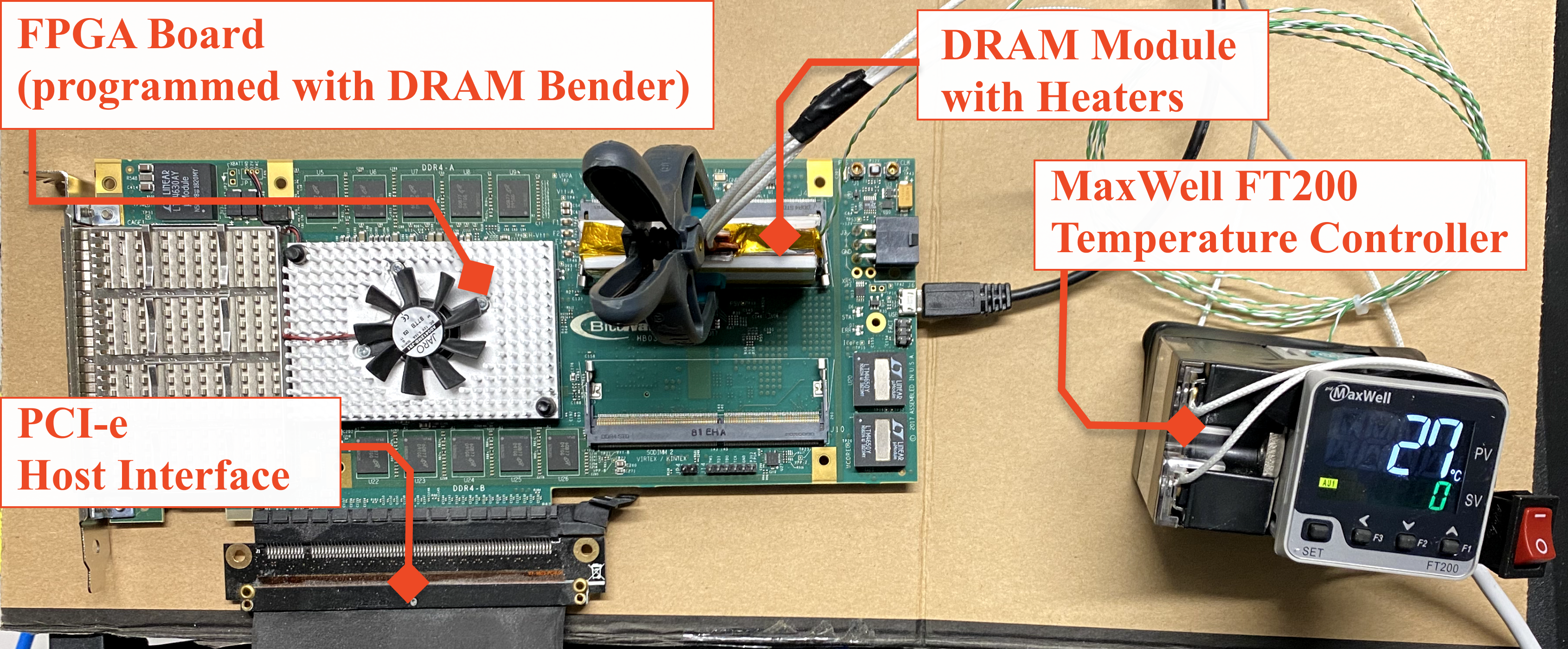}
\caption{\srev{\om{3}{Photograph of our} DRAM \sql{R4}Bender infrastructure}}
\label{fig:infrastructure}
\end{figure}

We test {commodity} DDR4 DRAM chips using an FPGA-based DRAM testing infrastructure that consists of four main components {(as Fig.~\ref{fig:infrastructure} illustrates)}: i)~a host machine that generates the test {program} and {collects experiment results}, ii)~an FPGA development board (Xilinx Alveo U200~\cite{alveo_u200}), programmed with DRAM Bender~\cite{olgun2023drambender, safari-drambender} {(based on SoftMC~\cite{hassan2017softmc, softmcgithub})}, {to execute our test programs}, iii)~a thermocouple temperature sensor and a pair of heater pads pressed against the DRAM chips {to maintain a {target temperature level}}, and iv)~a PID temperature controller (MaxWell FT200~\cite{maxwellFT200}) that controls the heaters and keeps the temperature at the desired level \agy{0}{with a precision of $\pm$\SI{0.5}{\celsius}}.\footnote{\yct{3}{To test the reliability of our temperature controller during RowHammer tests, we repeat RowHammer tests on all rows in round-robin fashion at nine different hammer counts \yct{3}{for 24 hours}. We sample the temperature of three modules (one from each manufacturer) every 5 seconds and observe that the variation in temperature is less than \SI{0.5}{\celsius}.
}}

\head{Eliminating interference sources} 
To observe \agy{0}{RowHammer bitflips} \yct{0}{without any circuit-/system-level interference},
we \yct{0}{take \param{three} measures }\agy{0}{ to eliminate} potential sources of interference, \yct{0}{as \om{3}{done} in} prior works~\cite{kim2020revisiting, orosa2021deeper, yaglikci2022understanding, hassan2021utrr, luo2023rowpress}.
First, we disable periodic refresh during the execution of our test programs to \agy{0}{prevent \yct{0}{any unintended charge restoration by periodic refreshes or }potential on-DRAM-die \gf{0}{TRR} mechanisms~\cite{frigo2020trrespass, hassan2021utrr}}.
Second, we \yct{0}{make sure the runtime of any program does \emph{not} exceed the refresh window such that \agy{6}{\emph{no}} data retention failures are observed.}
Third, we \yct{0}{confirm} that the tested DRAM modules and chips have neither rank-level nor on-die ECC.
\agy{0}{{\agy{6}{By doing so,}} we directly observe {and analyze} all {circuit-level} bitflips {without interference} {from architecture-level correction and mitigation mechanisms}\om{3}{~\cite{patel2020beer, patel2019understanding}}.}

\subsection{Tested DDR4 DRAM Chips}
\label{sec:methodology_dramchips}

{Table~\ref{tab:dram_chip_list} \om{3}{lists} the \agy{0}{\nCHIPS{} real DDR4 DRAM chips (\om{3}{\agy{3}{belonging} to 10 U-DIMM, 10 R-DIMM, and 10 SO-DIMM modules})} that we test from all three major DRAM manufacturers, which cumulatively hold a \SI{97}{\percent} share of the DRAM market~\cite{statista2024marketshare}.}
{\agy{0}{To investigate whether our analysis applies to different DRAM technologies, designs, and manufacturing processes,
we test various} DRAM chips with different die densities and die revisions from each DRAM chip manufacturer.}\footnote{\agy{0}{A DRAM chip's} technology node is \agy{0}{\emph{not} always} publicly available. 
A die revision code of X indicates that there is \emph{no} public information available about the die revision. We provide more detail on all tested DRAM modules, their RowHammer characteristics under partial charge restoration, and their \X{} configuration parameters in Appendix \ref{sec:ext_tables}.
}
{To maintain a reasonable experiment time, we perform our characterization using a total of 3K rows \om{3}{in a randomly selected} bank from each module we tested; 1K from the beginning, 1K from the middle, and 1K from the end of the DRAM bank\om{3}{, similarly} to prior \om{3}{works}~\cite{luo2023rowpress, yaglikci2022hira, kim2014flipping, orosa2021deeper, yaglikci2022understanding, olgun2023hbm, gao2022frac, gao2019computedram}}.
\yctcomment{6}{I will prepare the extended table for appendix}
\vspace{5pt}
\begin{table}[ht!]
  \renewcommand{\arraystretch}{0.7}
  \setlength{\tabcolsep}{2pt}
  \centering
  \footnotesize

  {
  \caption{Tested DDR4 DRAM chips}
    \begin{tabular}{l|ccccccc}
    \multirow{2}{*} {{\bf Chip Mfr.}} & \textbf{Module} & \textbf{\#Chips} & \textbf{Form} & {{\bf Die}} & {{\bf Die}} &{{\bf Chip}} & {{\bf Date}} \\ 
                            & \textbf{IDs} & \textbf{(\#Modules)} & \textbf{Factor} & {{\bf Density}} & {{\bf Rev.}} & {{\bf Org.}} & {{\bf Code}} \\   
        \hline                 
               & H0   & 8 (1) &   SO-DIMM &      4Gb  & M   &   x8   &   N/A   \\  
               & H1   & 8 (1) &   SO-DIMM &      4Gb  & X   &   x8   &   N/A   \\  
               & H2   & 8 (1) &   SO-DIMM &      4Gb  & A   &   x8   &   N/A   \\  
    Mfr. H     & H3   & 32 (1) &   R-DIMM &      8Gb  & M   &   x4   &   N/A   \\  
    (SK Hynix) & H4-5   & 32 (2) &   R-DIMM &      8Gb  & D   &   x8   &   2048   \\  
               & H6   & 32 (1) &   R-DIMM &      8Gb  & A   &   x4   &   N/A   \\  
               & H7-8   & 32 (2) &   U-DIMM &      16Gb  & C   &   x8   &   2136   \\  
        \hline                 
               & M0-1-2   & 48 (3) &     R-DIMM      & 8Gb  & B   &   x4   &   N/A   \\  
               & M3   & 16 (1) &     SO-DIMM      & 16Gb  & F   &   x8   &   2237   \\  
    Mfr. M     & M4   & 4 (1) &     SO-DIMM      & 16Gb  & E   &   x16   &   2046   \\  
    (Micron)   & M5   & 32 (1) &     R-DIMM      & 16Gb  & E   &   x4   &   2014   \\  
               & M6   & 4 (1) &     SO-DIMM      & 16Gb  & B   &   x16   &   2126   \\  
        \hline                 
               & S0-1   & 32 (2) &        U-DIMM       &      4Gb  & F   &   x8   &   N/A   \\  
               & S2-3-4   & 24 (3) &        SO-DIMM       &      4Gb  & E   &   x8   &   1708   \\  
               & S5   & 4 (1) &        SO-DIMM       &      4Gb  & C   &   x16   &   N/A   \\  
    Mfr. S     & S6-7-8-9   & 32 (4) &        U-DIMM       &      8Gb  & D   &   x8   &   2110   \\  
    (Samsung)  & S10   & 16 (1) &        R-DIMM       &      8Gb  & C   &   x8   &   1809   \\  
               & S11   & 8 (1) &        R-DIMM       &      8Gb  & B   &   x8   &   2052   \\  
               & S12   & 8 (1) &        U-DIMM       &      8Gb  & A   &   x8   &   2212   \\  
               & S13   & 8 (1) &        U-DIMM       &      16Gb  & B   &   x8   &   2315   \\                
        \hline
    \end{tabular}
    \label{tab:dram_chip_list}
    }
\end{table}


\subsection{Testing Methodology}
\label{subsec:testing_methodology}

\head{Metrics}
{To characterize \agy{20}{a DRAM row's RowHammer} vulnerability, we examine \agy{0}{two metrics:  {i)~\gls{nrh}, the minimum number of activations per\yctcomment{3}{Yes, we always use double-sided.} aggressor \om{3}{row} to \om{3}{observe} at least one RowHammer bitflip and ii)~\gls{ber}, the fraction of DRAM cells in a row that experience a RowHammer bitflip}.
A higher \gls{nrh} (\gls{ber}) indicates lower (higher)} vulnerability.}

\SetAlFnt{\scriptsize}
\RestyleAlgo{ruled}
\begin{algorithm}
{\setstretch{0.8}
\caption{Test for profiling the effect of reduced \gls{tras} on RowHammer vulnerability}\label{alg:test_alg}
    \DontPrintSemicolon
    \SetAlgoLined
    \SetKwFunction{FMain}{main\_test\_loop}
    \SetKwFunction{FHammer}{perform\_RH}
    \SetKwFunction{initialize}{init\_rows}
    \SetKwFunction{measureber}{perform\_RH}
    \SetKwFunction{check}{check\_for\_bitflips}
    \SetKwFunction{Hammer}{hammer\_doublesided}
    \SetKwFunction{InitAggr}{initialize\_aggressor\_rows}
    \SetKwFunction{InitVictim}{initialize\_victim\_row}
    \SetKwFunction{PartRestoration}{partial\_restoration}
    \SetKwFunction{Sleep}{sleep\_until\_tREFW}
    \SetKwProg{Fn}{Function}{:}{}
    
    \tcp{$t_{RAS(Nom)}$: nominal charge restoration latency}
    \tcp{$t_{RAS(Red)}$: reduced charge restoration latency}
    \tcp{$RA_{vic}$: victim row address}
    \tcp{$N_{PR}$: number of consecutive partial restorations}
    \tcp{$HC$: hammer count per aggressor row}
    \tcp{$DP$: data Pattern}
    \Fn{\PartRestoration{$RA_{vic}$, $t_{RAS(Red)}$, $N_{PR}$}}{
        \tcp{perform $N_{PR}$ partial charge restorations with $t_{RAS(Red)}$}
        \For{$i = 0; \ i < N_{PR}; \ i\text{++}$}{
            ACT({$RA_{vic}$}, wait=$t_{RAS(Red)}$)\;
            PRE({wait=$t_{RP}$})\;
        }
    }
    \Fn{\FHammer{$RA_{vic}$, $DP$, $HC$, $t_{RAS(Red)}$, $N_{PR}$}}{
        \tcp{perform double-sided RH and return the bitflip count}
        \initialize($RA_{vic}$, $DP$) \tcp{initialize aggressor and victim rows}
        \PartRestoration{$RA_{vic}$, $t_{RAS(Red)}$, $N_{PR}$}\;
        \Hammer({$RA_{vic}$}, $HC$) \tcp{hammer aggressor rows}
        \Sleep{} \tcp{wait until the end of $t_{REFW}$}
        \KwRet \check($RA_{vic}$) \tcp{count the bitflips}
    }
    
    \Fn{\FMain{}}{
        \tcp{measure $N_{RH}$ for each $t_{RAS(Red)}$, $N_{PR}$, and $RA_{vic}$}
        \ForEach{$t_{RAS(Red)}$ in from $t_{RAS(Nom)}$ to $6ns$ at steps of $3ns$}{
            \ForEach{$N_{PR}$ in from $TestedN_{PR}$}{
                \ForEach{$RA_{vic}$ in $TestedRows$}{
                    \tcp{{find the worst-case data pattern (WCDP)}}
                    
                    \ForEach{$DP$ in [RS, RSI, CS, CSI, CB, CBI]}{
                        \FHammer{$RA_{vic}$, $DP$, 100K, $t_{RAS(Red)}$, $N_{PR}$}\;
                        $WCDP$ = $DP$ that causes the most bitflip\;
                    }
                    \tcp{measure $BER$ with 100K hammers}
                    $BER$ = \FHammer{$RA_{vic}$,\ $WCDP$,\ $100K$,\ $t_{RAS(Red)}$,\ $N_{PR}$}\;
                    \tcp{check for retention bitflips without hammering, return $N_{RH}=0$ if there are retention bitflips}
                    $ret\_bfs$ = \FHammer{$RA_{vic}$,\ $WCDP$,\ $0$,\ $t_{RAS(Red)}$,\ $N_{PR}$}\;
                    \If{$ret\_bfs\ >\ 0$}{
                        \Return $0$, $BER$\;
                    }
                    \tcp{measure final $N_{RH}$ using bi-section search}
                    $HC_{high} = 100K$;
                    $HC_{low} = 0$;
                    $HC_{step} = 1K$;
                    $N_{RH} = 100K$;\;
                    \While{$HC_{high} - HC_{low} > HC_{step}$}{
                        $HC_{cur} = (HC_{high} + HC_{low})/2$\;
                        $rh\_bfs$ = \FHammer{$RA_{vic}$,\ $WCDP$,\ $HC_{cur}$,\ $t_{RAS(Red)}$,\ $N_{PR}$}\;
                        $HC_{low} = (rh\_bfs\ ==\ 0)\ ?\ HC_{cur}\ :\ HC_{low}$\;
                        $HC_{high} = (rh\_bfs\ !=\ 0)\ ?\ HC_{cur}\ :\ HC_{high}$\;
                        $N_{RH} = (rh\_bfs\ ==\ 0)\ ?\ N_{RH}\ :\ HC_{cur}$\;
                    }
                    \Return $N_{RH}$, $BER$\;
                }
            }
        }
    }
}
\end{algorithm}

\yctcomment{8}{Double-checked the line numbers.}
\head{Tests}
\yct{3}{Alg.~\ref{alg:test_alg} describes our main test loop and helper functions.}
\yct{3}{Our main test loop (lines 12-36\agy{3}{)}
measures \gls{nrh} and \gls{ber} of each tested row for different charge restoration latencies and different numbers of restorations.}
Our RowHammer testing function (lines 6-11\agy{3}{)} uses \om{6}{the} double-sided hammering pattern~\cite{kim2014flipping, kim2020revisiting, seaborn2015exploiting, orosa2021deeper, luo2023rowpress}, where we hammer two physically adjacent (aggressor) rows to a victim row in an alternating manner. We perform double-sided {hammering} with the maximum activation rate possible within DDR4 command timing {specifications}~\cite{jedec2017ddr4} as this access pattern is \om{3}{demonstrated to be} the most effective RowHammer access pattern on DRAM chips when RowHammer mitigation mechanisms are disabled~\cite{kim2014flipping, kim2020revisiting, frigo2020trrespass, cojocar2020rowhammer, seaborn2015exploiting, orosa2021deeper, olgun2023hbm}.
\yct{3}{After initializing the aggressor and victim rows \om{6}{(line 7)}, we perform \om{3}{partial charge} restoration on \yct{3}{the sandwiched victim row (i.e., the row between two aggressor rows) (line~8\agy{3}{)} \yct{6}{using \gls{trasred}} \emph{before}} we perform double-sided hammering (line~9\agy{3}{)}.
{\yct{6}{To guarantee that reducing charge restoration latency does \emph{not} cause any failures}, we keep the victim rows unaccessed for \agy{3}{a time window as large as} \gls{trefw} (i.e.\agy{3}{,} until \agy{3}{the partially restored row would be} refreshed by periodic refresh)
(line~10\agy{3}{)}}. \yct{6}{By doing so, we} ensure that we can detect if reduced charge restoration latency causes a data retention failure. An \gls{nrh} value of 0 indicates that the victim row experiences a bitflip \emph{without} \om{6}{hammering} due to data retention failure (lines~21-24\agy{3}{)}.}
\yct{3}{To analyze the effect of partial charge restoration, we sweep the charge restoration latency and the number of consecutive charge restorations (lines 1-5).}
\yct{2}{We run all our experiments for five iterations and record the lowest (highest) observed \gls{nrh} (\gls{ber}) in our analyses similar to prior works~\cite{luo2023rowpress, olgun2023hbm, yaglikci2022hira, yaglikci2022understanding, orosa2021deeper, kim2020revisiting, lee2017design, kim2014flipping}.}

\head{{Data patterns}}
{We use six commonly used data patterns~\cite{chang2016understanding,chang2017understanding,khan2014efficacy,khan2016parbor,khan2016case,kim2020revisiting,lee2017design,mukhanov2020dstress,orosa2021deeper, kim2014flipping, liu2013experimental}: row stripe (0xFF/0x00),
checkerboard (0xAA/0x55), column stripe (0xAA/0xAA) and their inverses. Before measuring a row's \gls{nrh} \om{3}{and \gls{ber}}, we identify the worst-case data pattern (WCDP) (lines 16-19\agy{3}{)} for each row among six data patterns. We use the data pattern that causes the most bitflips to measure \gls{nrh} \om{3}{and \gls{ber}} of that row.
}

\head{Finding $\bm{N}_{\bm{RH}}$ and $\bm{BER}$}
\yct{1}{We perform a bi-section search\om{3}{~\cite{bisectionsearch}} to measure a row's \gls{nrh} (lines \param{25-32}\agy{3}{)}. The search algorithm has three parameters: $HC_{high}$, $HC_{low}$, and $HC_{step}$. \om{3}{$HC_{high}$ ($HC_{low}$) specify the upper (lower)} bound for \gls{nrh} and $HC_{step}$ determines the resolution of the \gls{nrh} measurement. In our tests,
$HC_{high}$, $HC_{low}$, and $HC_{step}$ are 100K, 0, and 1K, respectively. \yct{3}{To measure a row's \gls{ber}, we perform double-sided hammering with a hammer count of 100K and check for the bitflip \om{6}{count} (line~20\agy{3}{)}.}}

\head{Finding physically adjacent rows}
DRAM manufacturers use internal address mapping schemes~\cite{cojocar2020rowhammer, salp} to translate \emph{logical} DRAM addresses (e.g., row, bank, column) into physical DRAM addresses (e.g., the physical location of a row). These schemes enable post-manufacturing row repair by remapping faulty rows~\cite{hassan2019crow} and organize internal DRAM structures cost-efficiently~\cite{khan2016parbor, vandegoor2002address}. Mapping schemes vary significantly across DRAM chips~\cite{barenghi2018software, cojocar2020rowhammer, horiguchi1997redundancy, itoh2013vlsi, khan2016parbor, kim2014flipping, patel2020beer, orosa2021deeper}, requiring reverse-engineering to analyze physical row organization. For each victim row, we carefully identify the two neighboring aggressor rows for double-sided RowHammer using reverse-engineering techniques described in prior works~\cite{kim2020revisiting, orosa2021deeper, luo2023rowpress}.

\head{{Temperature}}
{
We conduct all our tests at three temperature levels \SI{50}{\celsius}, \SI{65}{\celsius}, and \SI{80}{\celsius}, which is the highest possible temperature in our infrastructure that leaves a safe margin of \SI{5}{\celsius} to \SI{85}{\celsius}, where the refresh rate is doubled~\cite{jedec2017ddr4}. We report our results for \SI{80}{\celsius} unless stated otherwise.}

\section{\om{3}{Effect of Charge Restoration Latency\\on RowHammer}}
\label{sec:rh_cr}
\agy{3}{We present} the first rigorous characterization of the effect of charge restoration latency on RowHammer vulnerability in real DDR4 DRAM chips.
\subsection{Effect on RowHammer Threshold}
\label{sec:effect_nrh}

\begin{figure}[ht]
\centering
\includegraphics[width=1\linewidth]{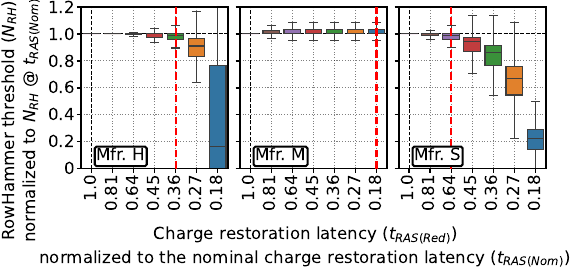}
\caption{\yct{3}{$\bm{N}_{\bm{RH}}$ versus charge restoration latency in all tested DRAM rows}}
\label{fig:nRH_single_pcr}
\end{figure}

We 
measure \gls{nrh} of a DRAM row \agy{1}{while sweeping the charge restoration latency used for restoring the charge of \om{6}{each} victim row.}
\figref{fig:nRH_single_pcr} demonstrates \gls{nrh} \om{3}{as} charge restoration \agy{1}{latency \om{3}{reduces} in a box-and-whiskers plot.}\footnote{\label{fn:boxplot}{\agy{1}{The box is lower-bounded by the first quartile (i.e., the median of the first half of the ordered set of data points) and upper-bounded by the third quartile (i.e., the median of the second half of the ordered set of data points).
The \gls{iqr} is the distance between the first and third quartiles (i.e., box size).
Whiskers show the minimum and maximum values.}}}
The x-axis shows the normalized charge restoration latency used \agy{1}{for restoring} victim rows \agy{1}{before hammering} and the y-axis shows the \gls{nrh} value \yct{3}{for each tested victim row, normalized to \agy{1}{the} \gls{nrh}} \agy{1}{value when the victim row is refreshed before hammering using the nominal charge restoration latency \om{3}{(i.e., $t_{RAS}=33ns$}). \agy{3}{Each subplot shows data from a different manufacturer, \om{6}{depicting \gls{nrh} values of all tested rows} across all tested chips from that manufacturer.}}
\yct{7}{Dashed red lines represent the lowest charge restoration latency that does \emph{not} significantly \yct{9}{($<3\%$)} decrease \gls{nrh} for each manufacturer.}

We make \param{three} observations from \figref{fig:nRH_single_pcr}. First, \gls{nrh} of the tested rows \agy{1}{significantly} decreases \agy{1}{as} charge restoration latency \agy{1}{reduces for Mfrs.~H and~S.} 
\agy{1}{Second, at very low \gls{tras} values (e.g., $x=0.18t_{RAS}$), charge restoration is \emph{not} strong enough for a subset of DRAM cells to retain data for $64ms$, and thus they exhibit an \gls{nrh} value of 0, meaning that they experience \om{3}{data retention} bitflips \om{3}{(without being hammered)}.} 
\agy{1}{Third}, reducing charge restoration latency \agy{1}{by \param{64\%}, \param{82\%}}\om{6}{, and 36\%} \yct{6}{(shown with dashed red lines)} \om{3}{negligibly affects} \yct{7}{($<3\%$)} the \gls{nrh} of the \agy{1}{tested} rows \agy{1}{from} Mfrs.~H, M\om{6}{, and~S}, respectively. 
\agy{1}{We hypothesize that Mfrs.~H and~M implement} large \om{7}{safety margins (i.e., guardbands)} \agy{1}{in their respective nominal} charge restoration latencies\agy{1}{, such that \om{6}{these latencies}} can be \agy{1}{significantly} reduced without \agy{1}{exacerbating} \om{3}{the RowHammer vulnerability}. 
\om{3}{Based on these observations, we derive:}

\takeaway{Charge restoration latency can be reduced \agy{3}{down to a safe minimum value without affecting} the \om{3}{RowHammer threshold} of real DRAM chips.\label{take:nrh_cr}}

\begin{figure}[ht]
\centering
\includegraphics[width=1\linewidth]{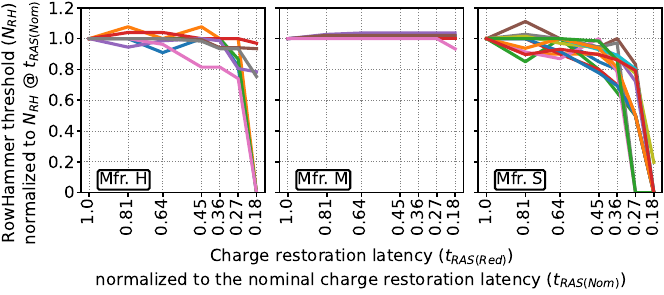}
\caption{\om{3}{Lowest observed $\bm{N}_{\bm{RH}}$ versus charge restoration latency in all tested DRAM modules}}
\label{fig:nrh_changemodule}
\end{figure}

\noindent
\iqrev{\textbf{\om{3}{Lowest} $\bm{N}_{\bm{RH}}$ across rows \yct{6}{with} reduced $\bm{t}_{\bm{RAS}}$.} \figref{fig:nrh_changemodule} demonstrates how the \om{3}{\om{6}{lowest} observed} \gls{nrh} across tested DRAM rows \om{6}{changes with} reduced \gls{tras} in a line plot, where the DRAM row that exhibits the \om{3}{lowest} \gls{nrh} may or may \emph{not} be the same row \om{7}{across different} \gls{tras} values. The x-axis shows \om{3}{normalized} charge restoration \om{6}{latency} and the y-axis shows the \om{3}{lowest} \gls{nrh} value in a DRAM module under reduced \gls{tras}, normalized to the \om{3}{lowest} \gls{nrh} under nominal \gls{tras}.
\yct{10}{Each subplot represents modules from a different manufacturer and each curve represents a different module.}
}

\yctcomment{6}{absolute nrh values changes significantly between modules, but we can figure out something for the extended version}
\yctcomment{7}{fixed colors}

\yctcomment{10}{updated the figure}
\iqrev{From \figref{fig:nrh_changemodule}, we make two observations. 
\om{3}{First, for all tested modules from Mfr.~M, the lowest \gls{nrh} value does \emph{not} \yct{3}{significantly} change
even at $x=0.27t_{RAS}$ \om{7}{(73\% reduction in \gls{tras})}.}
\om{3}{Second, the modules from Mfrs.~H and~S are resilient to reduced charge restoration latency. Their lowest \gls{nrh} values decrease by less than 3\%, on average across tested modules, even when the charge restoration latency is reduced by 36\% and 19\% for Mfrs.~H and~S, respectively.}\yctcomment{6}{We used the average nRH of the modules.}
\om{3}{Based on these observations, we conclude that reducing \om{6}{the} charge restoration latency \om{6}{can have} a \om{6}{small ($<3\%$) impact on the} lowest observed \gls{nrh} values in the tested DRAM modules.}}
\omcomment{7}{36 and 19 are small reductions 3 percent seems high? Is the data in This figure consistent with the data in figure 6?}
\yctcomment{7}{I checked the numbers and they are correct, we use 5 iterations for each test. Therefore, I think 3\% is logical due to VRD. Regarding figure 6, they represent each row's nrh change while this plot represents the nrh of the module.}

\begin{figure}[ht]
\centering
\includegraphics[width=0.9\linewidth]{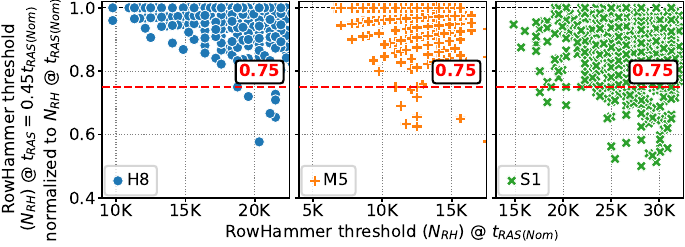}
\caption{\yct{3}{$\bm{N}_{\bm{RH}}$ at $\bm{0.45t}_{\bm{RAS}}$ versus $\bm{N}_{\bm{RH}}$ at nominal $\bm{t}_{\bm{RAS}}$ for \agy{3}{tested rows from \yct{10}{three} representative modules \yct{10}{H8, M5, and S1}}}}
\label{fig:nrh_single_pcr_per}
\end{figure}

To analyze \yct{3}{how \gls{nrh} values of individual rows change, we compare the normalized \gls{nrh} values to their \gls{nrh} values under nominal \gls{tras}}.
\agy{3}{\yct{3}{\figref{fig:nrh_single_pcr_per} shows} a scatter plot where each data point represents a DRAM row from \yct{10}{three} representative modules: \yct{10}{H8, M5, and S1}. The x-\yct{3}{axis shows} the row's \yct{3}{\gls{nrh}} under nominal charge restoration latency, \yct{3}{and the y-axis shows the row's \gls{nrh} under reduced charge restoration latency (\om{6}{to} $0.45t_{RAS}$) normalized to the \gls{nrh} under nominal charge restoration latency}.
\yct{10}{Each subplot indicates \om{11}{DRAM} rows from different manufacturers.}}
\yct{6}{The dashed red line represents $y=0.75$ where the \gls{nrh} value decreases by 25\% with $0.45t_{RAS}$.}
\yctcomment{10}{Added Micron}

{We make two observations from~\figref{fig:nrh_single_pcr_per}.
First, a small fraction of rows are significantly more sensitive to partial charge restoration than the vast majority of the rows. 
\yct{3}{For example, fewer than 0.45\%\yct{10}{, 0.66\%,} and 10.34\% of the rows from Mfrs.~H\yct{10}{, M,} and~S experience an \gls{nrh} reduction of more than 25\% \yct{3}{(i.e., below the red dashed line)} when they are refreshed using $0.45t_{RAS}$.}
\yct{3}{Second, rows with the lowest \gls{nrh} values under nominal charge restoration latency do \emph{not} exhibit the largest reductions in \gls{nrh} under reduced \gls{tras}. To illustrate, the rows with low \gls{nrh} values under nominal \gls{tras} (e.g., $x=10K$ at the left subplot) \agy{3}{exhibit values close to $1.0$ on the y-axis,}
while the rows with high \gls{nrh} values (e.g., $x=20K$ at the left subplot) can experience \om{6}{larger} \gls{nrh} reductions.}
\yct{3}{Based on these observations, we derive:}

\takeaway{\yct{7}{Charge restoration latency can be reduced down to a safe minimum value without significantly affecting the lowest observed \gls{nrh} of the tested DRAM modules.\label{take:nrh_dist}}}

\subsection{Effect on RowHammer Bit-Error-Rate}
\label{sec:effect_ber}

{\om{3}{We next} study the impact \om{3}{of} reduced charge restoration latency on the RowHammer \gls{ber}\yctcomment{6}{defined in sec 4.3}.} \figref{fig:ber_single_pcr} demonstrates \om{3}{how} RowHammer \gls{ber} \om{3}{changes as} charge restoration \agy{1}{latency \om{3}{reduces}} \yct{9}{(using a similar style as \figref{fig:nRH_single_pcr}).}
\begin{figure}[ht]
\centering
\includegraphics[width=1\linewidth]{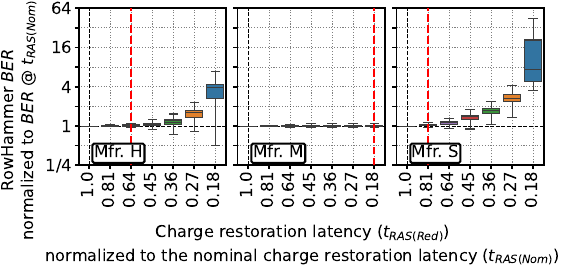}
\caption{\yct{3}{$\bm{BER}$ versus charge restoration latency in all tested DRAM rows}}
\label{fig:ber_single_pcr}
\end{figure}
We make \param{two} observations from \figref{fig:ber_single_pcr}.
First, \gls{ber} of the tested rows \agy{1}{increases} \om{3}{superlinearly} \agy{1}{as} charge restoration latency \agy{1}{reduces for Mfrs.~H and~S.} 
\agy{1}{Second}, reducing charge restoration latency \agy{1}{by \param{36\%}, \param{82\%}}\om{6}{, and 19\%} \yct{6}{(shown with dashed red lines)} does \emph{not} significantly increase \yct{6}{($<3\%$)} the \gls{ber} of the \agy{1}{tested} rows \agy{1}{from} Mfrs.~H, M, and~S, respectively. \yct{3}{From these observation, we derive:}

\takeaway{\yct{3}{Charge restoration latency can be reduced down to a safe minimum value without \yct{7}{significantly} increasing the RowHammer \gls{ber}.}\label{take:ber_cr}}


\subsection{\om{6}{Combined} Effect of Temperature\\and Charge Restoration Latency on RowHammer}
\label{sec:effect_temp}

\srev{\yct{3}{We} study the \om{7}{combined} impact of temperature and charge restoration latency on RowHammer vulnerability in terms of \gls{nrh} and \gls{ber}.
\figref{fig:nrh_single_pcr_temp} demonstrates \yct{3}{how \om{6}{\gls{nrh} and} \gls{ber} change as charge restoration latency reduces using} a box-and-whiskers plot.\fnref{fn:boxplot}
The x-axis shows the normalized restoration latency used \agy{1}{for refreshing} victim rows \agy{1}{before hammering} and the y-axis shows the \gls{nrh} (top \yct{3}{subplots}) and \gls{ber} (bottom \yct{3}{subplots}) \agy{1}{values \yct{3}{for each tested victim row},} normalized to \agy{1}{the} \gls{nrh} and \gls{ber} \agy{1}{values when the victim row is refreshed using the nominal charge restoration latency. Different colors represent different temperature values.} 
}

\yctcomment{10}{Added Micron}
\begin{figure}[ht]
\centering
\includegraphics[width=1\linewidth]{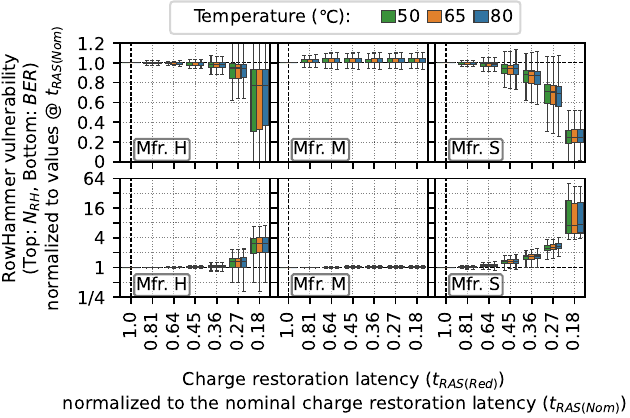}
\caption{\yct{3}{RowHammer vulnerability in terms of $\bm{N}_{\bm{RH}}$ (top subplots) and $\bm{BER}$ (bottom subplots) versus charge restoration latency for each tested DRAM row at \om{6}{three different temperatures}}}
\label{fig:nrh_single_pcr_temp}
\end{figure}

\agy{3}{From \figref{fig:nrh_single_pcr_temp}, we observe that temperature does \emph{not} significantly \om{6}{impact} the effect of reducing charge restoration latency on RowHammer vulnerability.}
\yct{10}{For example, when the temperature is raised from \SI{50}{\celsius} to \SI{80}{\celsius}, the normalized \gls{nrh} (\gls{ber}) changes by only \param{0.31\% (1\%), 0.20\% (0.02\%), and 0.08\% (9\%)} for the chips from Mfrs.~H, M, and S, respectively.}}
\yct{3}{Based on this observation, we derive:}

\takeaway{\yct{3}{Our experimental results show no significant impact of temperature on the effect of charge restoration latency on RowHammer vulnerability.\label{take:nrh_temp}}}

\subsection{Effect of Repeated Partial Charge Restoration\\on RowHammer}
\label{sec:rh_rep}

\agy{1}{Reducing charge restoration latency can lead to partially restoring \agy{20}{DRAM cell} charge. Repeating partial charge restoration can eventually degrade the stored charge to a level that is \emph{not} strong enough \om{6}{to maintain data integrity \om{7}{in} the presence of charge leakage}. \yct{3}{Therefore, we} investigate the effect of repeated partial charge restoration on the RowHammer vulnerability.}
\yct{3}{To do so, we perform different numbers of consecutive preventive refreshes on the victim row with reduced charge restoration \om{6}{latency} before hammering the aggressor rows.}

\begin{figure}[ht]
\centering
\includegraphics[width=1\linewidth]{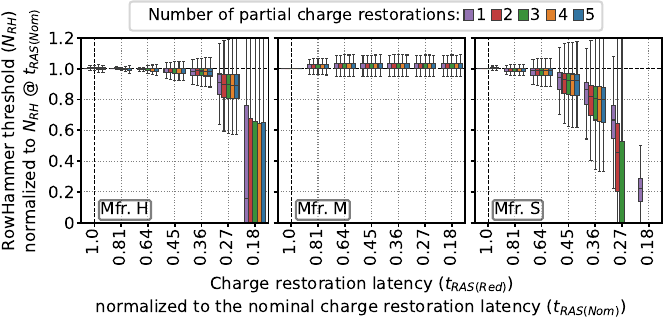}
\caption{\yct{3}{$\bm{N}_{\bm{RH}}$ versus repeated partial charge restoration in all tested DRAM rows}}
\label{fig:nrh_multiple_pcr}
\end{figure}

\figref{fig:nrh_multiple_pcr} demonstrates \yct{3}{how \gls{nrh} changes} when the victim row is repeatedly \om{6}{preventively} refreshed with reduced charge restoration latency 
using a box-and-whiskers plot.\fnref{fn:boxplot}
The x-axis shows the normalized charge restoration latency used \agy{1}{for refreshing} victim rows \agy{1}{\yct{3}{before hammering (see \secref{subsec:testing_methodology})}} and the y-axis shows the \gls{nrh} \agy{1}{value \yct{3}{of each tested row},} normalized to \agy{1}{the} \gls{nrh} \agy{1}{value when the victim row is refreshed using the nominal charge restoration latency.} \agy{1}{Different colors denote the number of restorations performed on the victim row with reduced charge restoration latency.} \yct{3}{Each subplot shows data from a different manufacturer, aggregated across all tested chips from that manufacturer.}

\yctcomment{10}{Added Micron}

\agy{1}{We make \param{three} observations from \figref{fig:nrh_multiple_pcr}. 
First, \gls{nrh} of tested rows from \yct{10}{Mfrs.~H and M} is \emph{not} significantly affected by the number of restorations as the \gls{nrh} values are similar across different numbers of restorations. 
Second, \gls{nrh} of the tested rows from Mfr. S decreases as the number of restorations increases. For example, under $0.36t_{RAS}$, a clear downward trend is observed in the \om{6}{\gls{nrh}} values. 
Third, performing partial charge restoration multiple times can lead to data retention bitflips at very low \gls{tras} values. For example, a single partial charge restoration using $0.27t_{RAS}$ does \emph{not} lead to data retention bitflips for tested rows from Mfr. S, while repeating \om{6}{such partial} restoration twice \om{6}{across two preventive refreshes} can lead to data retention bitflips (i.e., \om{6}{some cells} experience bitflips with \emph{no} hammering\om{6}{, as induced by $N_{RH}=0$)}.}
\yct{3}{Based on these observations, we conclude that \agy{3}{using reduced charge restoration latency \om{7}{repeatedly for} many consecutive preventive refreshes} might lead to data retention bitflips at very low \gls{tras} values.}

\begin{figure}[ht]
\centering
\includegraphics[width=0.9\linewidth]{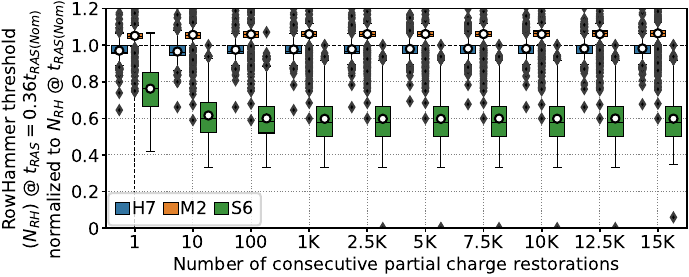}
\caption{\yct{3}{$\bm{N}_{\bm{RH}}$ versus \om{7}{number of} repeated partial \om{6}{charge} restorations in all tested rows from \yct{10}{three} modules}}
\label{fig:nrh_ext_pcr}
\end{figure}

\agy{1}{To investigate the effect of repeated \om{7}{partial} charge restoration more rigorously, we scale up the number of \agy{3}{repeated partial charge} restorations up to 15K \agy{3}{at the beginning of each test for \yct{10}{three} representative modules: \yct{10}{H7, M2, and S6}}.
\figref{fig:nrh_ext_pcr} demonstrates how \gls{nrh} \agy{1}{changes when the victim row is repeatedly refreshed before hammering with} \yct{6}{charge restoration latency of $0.36t_{RAS}$} \agy{1}{\agy{3}{using} a box-and-whiskers plot.}\fnref{fn:boxplot}
The x-axis shows the number of \yct{6}{partial charge restorations} performed on the victim row with $0.36t_{RAS}$. The y-axis shows the \gls{nrh} \agy{1}{values} \agy{3}{when the victim rows are repeatedly refreshed with reduced charge restoration latency\yct{6}{,} normalized \yct{6}{to their \gls{nrh} values with the nominal \gls{tras}}}.
\agy{1}{Different colors denote different modules.}}


\yctcomment{10}{Added Micron}

We make \param{\agy{3}{three}} observations from \figref{fig:nrh_ext_pcr}.
First, \gls{nrh} \om{6}{values} of the rows from \yct{10}{Mfrs.~H and M} are \emph{not} significantly affected by the number of partial charge restorations. For example\yct{1}{, \gls{nrh} values \yct{10}{of the tested rows from Mfr.~H} \yct{3}{change only by $1.24\%$} \om{6}{if rows} are refreshed 15K times using $0.36t_{RAS}$ \om{6}{versus only one time}.}
Second, performing restorations using reduced charge restoration latency many times can lead to data retention bitflips in the rows from Mfr.~S. 
\yctcomment{7}{We checked the data and they are correct, for high latencies, it does not matter the number of refreshes since after each refresh the voltage level will be the same.}
For example, \agy{3}{using reduced charge restoration latency for} 2.5K \agy{3}{consecutive \om{7}{preventive} refreshes}
causes some cells to lose their charge \emph{without} hammering (\om{6}{as seen by} $N_{RH}=0$).
\yct{6}{Third, to prevent bitflips due to repeated partial charge restorations, there should be a limit on the number of consecutive refreshes with reduced charge restoration latency.}
For example, rows from \yct{10}{Mfrs.~H, M, and S} experience \emph{no} bitflips when their rows are
refreshed \yct{10}{15K, 15K, and 1K} times with \yct{6}{$0.36t_{RAS}$}, respectively. 
After the rows from \yct{10}{Mfrs.~H, M, and S} are refreshed \yct{10}{15K, 15K, and 1K} times, respectively, we need to perform a charge restoration using nominal \gls{tras} to fully restore the charges of the cells.
\agy{3}{Based on our analysis of the repeated partial charge restoration, we derive:}

\takeaway{\yct{6}{Reduced charge restoration latency can be used safely for many consecutive refreshes without causing any failures.}\label{take:rh_rep}}

\section{\yct{3}{Effect of Charge Restoration Latency on \om{7}{the} Half-Double \om{7}{Access Pattern}}}
\label{sec:halfdobule_chargerestoration}
\iey{7}{An aggressor row can cause bitflips in victim rows that are not physically adjacent~\cite{kim2014flipping,kim2020revisiting}. The impact of RowHammer on a victim row decreases and eventually disappears as the physical distance between the victim and the aggressor rows increases. To account for this characteristic, prior works define \emph{blast radius} as the distance between an aggressor row and its furthest victim row~\cite{kim2014flipping,kim2020revisiting, frigo2020trrespass,orosa2021deeper,cojocar2020rowhammer,deridder2021smash,hassan2021utrr,jattke2022blacksmith,kogler2022half,park2020graphene, yaglikci2021security, yaglikci2021blockhammer, hassan2024selfmanaging, devaux2021method, patel2022case, patel2024rethinking}. A recent RowHammer attack, Half-Double~\cite{kogler2022half}, introduces an access pattern (i.e., Half-Double access pattern) that exploits blast
radius to induce bitflips with a significantly lower activation count. Half-Double demonstrates that}
\yct{7}{hammering a \emph{far} aggressor row (i.e., physically two rows away from the victim row) many times and then hammering the \emph{near} aggressor row (i.e., physically \om{8}{adjacent to} the victim row) a much smaller number of times lead\iey{7}{s} to bitflips in some DDR4 DRAM chips.}
Half-Double \yct{7}{access pattern is} based on exploiting two weaknesses of a poorly designed RowHammer mitigation mechanism i)~underestimation of blast radius (i.e., the \agy{7}{maximum distance between an aggressor and a victim} row) and ii)~ignoring the hammering effect of the victim row preventive refreshes. To mitigate Half-Double bitflips, existing RowHammer mitigation mechanisms perform preventive refreshes on \agy{7}{all four rows in} $\pm2$\agy{7}{-row distance} of the aggressor row~\cite{olgun2024abacus, canpolat2024prac, qureshi2024mint, qureshi2024impress, hassan2021utrr, yaglikci2021blockhammer}. For completeness, we analyze the impact of reducing charge restoration latency on \iey{7}{the} Half-Double \om{7}{access pattern}. To do so, we modify \algref{alg:test_alg} such that the hammering function performs \iey{7}{the} Half-Double access pattern instead of double-sided RowHammer on randomly selected \param{\agy{7}{two}} modules from \agy{7}{each of Mfrs.~H and~S (four modules in total).}

\begin{figure}[!h]
\centering
\includegraphics[width=1\linewidth]{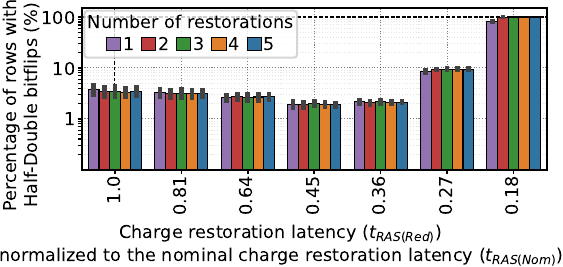}
\caption{\srev{Percentage of rows with Half-Double bitflips \om{8}{versus} reduced charge restoration latency \om{8}{for preventive refreshes}}}
\label{fig:half_cr}
\end{figure}

\srev{
\figref{fig:half_cr} demonstrates the impact of reduced \gls{tras} on the Half-Double access pattern in terms of the percentage of rows with Half-Double bitflips for the modules from Mfr.~H \yct{7}{\om{7}{(}we do \emph{not} observe any Half-Double bitflips in the modules from Mfr.~S\om{7}{)}}.\yctcomment{7}{Yes, we should test Mfr. M as well, added to TODO.}
The x-axis shows the normalized charge restoration latency used for refreshing victim rows before Half-Double and the y-axis shows the percentage of rows with Half-Double bitflips. Different colors represent different number\om{7}{s} of partial charge restorations.
}\yct{7}{Error bars represent the minimum and maximum values across each tested module from Mfr.~H.}

\srev{
We make four observations.
First, the \om{7}{tested} modules from Mfr.~S do \emph{not} exhibit Half-Double bitflips within a \gls{trefw} (i.e., 64ms) interval\om{7}{, even though they do exhibit double-sided RowHammer bitflips (\secref{sec:effect_nrh})}. 
Second, the total number of Half-Double bitflips is significantly lower than the total number of double-sided RowHammer bitflips (not shown in the figure).
Third, \yct{7}{when charge restoration latency is reduced by 64\%,} the \om{8}{percentage} of rows with bitflips decreases \yct{7}{(by 39.31\%)}\yctcomment{8}{checked the precision for all numbers}.
\yct{7}{As charge restoration latency is reduced from $0.36t_{RAS}$ to $0.18t_{RAS}$, the \yct{8}{percentage} of rows with bitflips significantly increases.}
\yctcomment{7}{we do not have a hypothesis for now and we are trying.}
Fourth, the number of charge restorations does \emph{not} significantly affect the \yct{8}{percentage} of rows with bitlips. \yct{7}{For example, the \yct{8}{percentage} of rows with bitflips varies by 1.50\% when \yct{8}{each tested row is} refreshed one time using $0.36t_{RAS}$ versus refreshed five times.}
We conclude that i)~the double-sided RowHammer pattern is significantly more effective than \om{7}{the} Half-Double access pattern and ii)~charge restoration latency can be reduced to a safe minimum value without affecting the Half-Double \yct{8}{access pattern} vulnerability of a DRAM chip.
}

\section{\yct{3}{Effect of Charge Restoration Latency\\on Data Retention Failures}}
\label{sec:retention_chargerestoration}

\srev{We analyze \yct{7}{how the data retention time of a DRAM row changes as charge restoration latency reduces}\om{7}{.}\yctcomment{7}{We will test more modules and data pattern. added to TODO.}
To do so, we select \yct{10}{2, 1, and 4} modules from \yct{10}{Mfrs.~H, M, and~S}, respectively, and test all rows in a \yct{7}{randomly selected} bank from each module at \SI{80}{\celsius} using \param{two} data patterns \yct{7}{(all 1s and 0s)}. \yct{7}{To analyze the effect of reduced charge restoration latency on data retention time,} we \yct{7}{perform charge restoration on} the \yct{7}{tested} row \yct{7}{using} reduced charge restoration latency \param{once or ten} times. \yct{7}{Then,} we wait for the data retention time we want to test and check for bitflips. To analyze the change in the data retention time of DRAM cells, we test the rows with \yct{7}{different} data retention times ($64ms,\ 96ms,\ 128ms,\ 256ms,\ 512ms,\ 1024ms$).}

\begin{figure}[ht]
\centering
\includegraphics[width=1\linewidth]{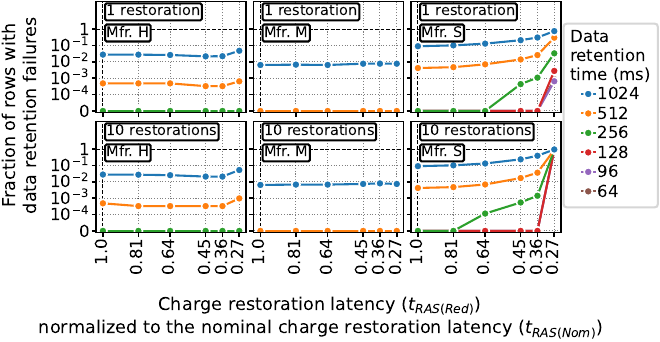}
\caption{\srev{Fraction of rows with data retention failures under repeated partial charge restoration}}
\label{fig:ret_cr}
\end{figure}

\srev{~\figref{fig:ret_cr} demonstrates the \yct{7}{effect} of reduced \yct{7}{charge} restoration latency on data retention failures \yct{10}{for different manufacturers (\om{11}{each represented in different} columns of subplots)}.
The x-axis shows the normalized charge restoration latency used \agy{1}{for refreshing} rows. The y-axis shows the fraction of rows with data retention failures \yct{10}{when the tested row is refreshed using reduced latency one time (top subplots) or ten times (bottom subplots)}.
\yct{10}{Each curve represents a different data retention time \yct{7}{and overlapping lines use the color corresponding to the higher data retention time.}}}
\yct{10}{Each data point presents the average values across all tested modules from a manufacturer.}
\yctcomment{7}{Replotted the figure with distinct colors and added this explanation.}
\vspace{10pt}

\srev{We make \yct{10}{seven} observations from \figref{fig:ret_cr}.
\yct{10}{First, the tested rows from Mfrs.~H and M do \emph{not} experience data retention bitflips at $256ms$ and $512ms$ even when the rows are refreshed ten times using $0.27t_{RAS}$, respectively.}
\yct{10}{Second, for Mfr.~H, the fraction of rows with bitflips slightly increases when $t_{RAS}$ is reduced from  $0.36t_{RAS}$ to $0.27t_{RAS}$, while the number of \agy{7}{consecutive} partial charge restorations does \emph{not} affect the data retention failures.}
\yct{10}{Third, for Mfr.~M, reducing charge restoration latency does \emph{not} affect the data retention failures.}
\yct{10}{Fourth}, the tested rows \yct{10}{from Mfr.~S} do \emph{not} experience data retention bitflips for $256ms$ even when they are refreshed ten times using $0.36t_{RAS}$.\yctcomment{7}{the data was correct but plot was hard to separate, fixed it.}
\yct{10}{Fifth}, the data retention time of some rows \yct{10}{from Mfr.~S} decreases as charge restoration latency reduces. For example, while \emph{no} data retention failures occur when the rows are refreshed using $0.36t_{RAS}$ even with a data retention time of $256ms$, some rows experience data retention bitflips when they are refreshed using $0.27t_{RAS}$.
{\yct{10}{Sixth}, \yct{10}{for Mfr.~S}, the number of \agy{7}{consecutive} partial charge restorations significantly affects the number of rows with data retention bitflips. \yct{8}{For example, \param{472$\times$} more rows experience data retention bitflips at $256ms$ when they are refreshed ten times instead of one time using $0.27t_{RAS}$.}}}
\yct{7}{\yct{10}{Seventh}, our observations are aligned with \agy{7}{a} prior work~\cite{das2018vrldram} that develop\om{7}{s} a detailed circuit-level analytical model and analyze\om{7}{s} the effect of reduced charge restoration latency on the data retention times (not shown in the figure).}
\yct{7}{From these observations, we derive:}

\takeaway{Charge restoration latency can be reduced to a safe minimum value without causing data retention failures.\label{take:retention}}

\section{\yct{4}{\X{}}}
\label{sec:mechanism}
\yct{20}{To demonstrate the potential benefits of our experimental observations, \om{1}{we propose a new memory controller-based mechanism, \emph{\Xlong{} (\X{})}}. \X{} \yct{4}{works with an existing RowHammer mitigation mechanism and} leverages our experimental observations to safely reduce the \agy{4}{latency of preventive refreshes} that the existing mitigation \agy{4}{mechanism} perform\om{7}{s}. \srev{By doing so, \X{} 
introduces a novel perspective \om{1}{into} addressing RowHammer vulnerability at low cost.}}

\subsection{Overview of \X{}}
\label{sec:mech_overview}
\begin{figure}[ht]
\centering
\includegraphics[width=0.9\linewidth]{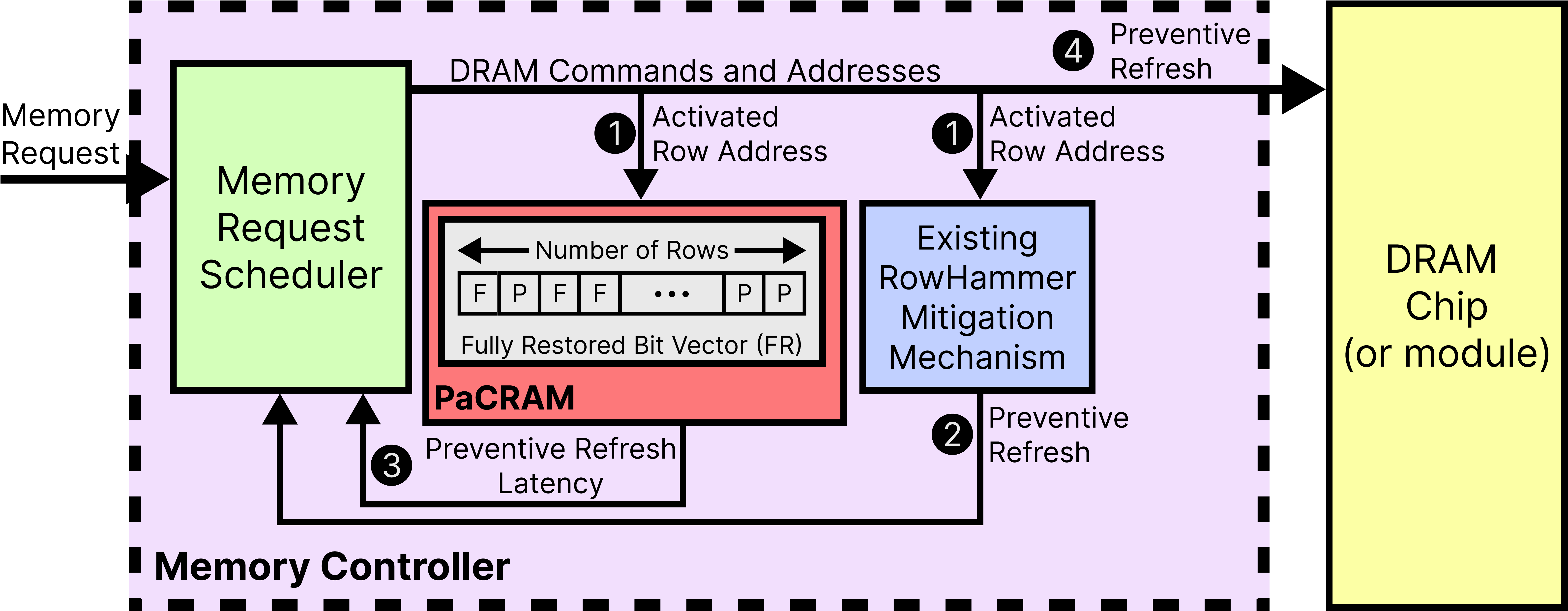}
\caption{\yct{4}{Overview of \X{}}}
\label{fig:pacram}
\end{figure}

\yct{4}{\figref{fig:pacram} illustrates \om{5}{a} high-level overview of \X{}.}
\yct{4}{When a DRAM row is activated~\circled{1}, \X{} and the existing RowHammer mitigation mechanism
\om{5}{are} provided with the activated row address.}
The \om{5}{mitigation} mechanism decides whether to perform a preventive refresh
and issues a preventive refresh if it detects a potential aggressor \om{5}{row}~\circled{2}.
\yct{7}{Simultaneously, \X{} determines whether preventive refresh latency can be reduced and provides \agy{7}{the memory controller with reduced or nominal preventive refresh latency~\circled{3}.}
\agy{7}{The memory request scheduler schedules a preventive refresh (generated by the existing mitigation mechanism) using the preventive refresh latency (determined by \X{})~\circled{4}.}
}

\subsection{Reducing Preventive Refresh Latency}
\label{sec:mech_reducing}

\yct{7}{\X{} uses two preventive refresh latencies: i)~\emph{nominal latency} \agy{7}{for full charge restoration}
and ii)~\emph{reduced latency} \agy{7}{for partial charge restoration}.\footnote{\om{8}{\X{} does \emph{not} affect the latency of periodic refreshes.\label{fn:onlyprev}}}}\yctcomment{8}{added the footnote here and refer it on the evaluation}
\agy{7}{To safely perform partial charge restoration, \X{} overcomes two challenges.}
\agy{4}{First, reducing preventive refresh latency can reduce the row's \gls{nrh} (}\takeref{take:nrh_cr}\agy{4}{).} 
\agy{7}{To address this challenge,}
\X{} configures the existing RowHammer mitigation mechanism with \om{8}{a} reduced \gls{nrh}, \om{5}{based on experimental characterization of the DRAM chips that are being controlled.}
\yct{7}{Second, repeated consecutive partial charge \agy{7}{restoration} might cause failures (\takeref{take:rh_rep}). \agy{7}{To address this challenge,}}
\yct{7}{\X{} limits the number of consecutive partial charge restorations.}
\agy{7}{To this end, \X{} assigns each DRAM row \om{8}{into one} of two states: i)~$F$: the row has to be refreshed using \emph{full} charge restoration and ii)~$P$:~the row can be refreshed using \emph{partial} charge restoration. Initially, all rows are in the $F$-state 
and \X{} periodically pulls each row's state to $F$.
When a row is refreshed using \emph{full} charge restoration, \X{} switches the row's state to $P$, so that future \yct{7}{preventive} refreshes can be performed using \emph{partial} charge restoration.
The \om{8}{time} interval \om{8}{for} resetting rows to $F$-state is the smallest time window in which the maximum number of refreshes with partial charge restoration can be safely performed (\secref{sec:mech_implementation}).}
\yct{8}{By doing so, \X{} performs full charge restoration for the first preventive refresh targeting a DRAM row and partial charge restoration for the rest of the preventive refreshes until the row's state is periodically reset into $F$.}

\head{\agy{7}{Security}}
\yct{8}{\X{} safely reduces charge restoration latency of preventive refreshes by applying two modifications to the existing RowHammer mitigation mechanism, based on experimental characterization data}: i)~reducing \gls{nrh} and ii)~limiting the maximum number of consecutive refreshes using partial charge restoration. Therefore, \X{} guarantees that the existing RowHammer mitigation mechanism is correctly configured based on the DRAM chip's vulnerability level. An existing RowHammer mitigation mechanism's security against DRAM read disturbance-based exploits is identical to that of the mechanism integrated with \X{}.



\subsection{\agy{7}{Detailed} Implementation \agy{7}{and Configuration}}
\label{sec:mech_implementation}

\head{\agy{7}{Full charge restoration interval ($\bm{\tfr{}}$)}}
\agy{7}{To ensure robust execution, we calculate the smallest time interval at which a DRAM row should be refreshed using full charge restoration. We assume the worst case, where a DRAM row is hammered as frequently as possible (i.e., activated every $t_{RC}$).
\yct{8}{The latency of a preventive refresh that performs partial charge restoration is computed as the sum of the reduced charge restoration latency and \iey{8}{the nominal} precharge latency (\gls{trasred} + \gls{trp}).} \agy{7}{The RowHammer mitigation mechanism performs a preventive refresh \yct{7}{after \gls{nrh} row activations}.} Therefore, 
\agy{7}{the smallest interval of such refreshes is the cumulative latency of performing i)~\gls{nrh} activations ($N_{RH} \times t_{RC}$) and ii)~the preventive refresh (\gls{trasred} + \gls{trp}), resulting in (\gls{nrh}$t_{RC}$ + \gls{trasred} + \gls{trp}).} 
From experimental characterization data (\om{8}{\figref{fig:nrh_ext_pcr}}), we obtain \gls{thpcr}.
The smallest time window that can contain \gls{thpcr} of such preventive refreshes is $\thpcr{} (\nrh{}\trc{} + \trasred{} + \trp{})$, which we denote as the \gls{tfr}.}\footnote{\yct{0}{If \gls{tfr} is larger than \gls{trefw}, \X{} uses the reduced charge restoration latency for \emph{all} preventive refreshes \om{5}{since} periodic refresh performs full charge restoration before a row can \om{5}{receive} \gls{thpcr} partial charge restorations.}}
\agy{7}{For example, the tested module S6 exhibits the \gls{nrh} and \gls{thpcr} values of 3.9K and 2K, respectively, \yct{7}{at $t_{RAS(Red)} = 0.36t_{RAS}$} and thus it requires rows to be refreshed with full charge restoration at an interval of \param{$374ms$}}.

\head{\agy{7}{Maintaining row states}}
\agy{7}{To maintain each row's state at low cost, \X{} implements a bit vector, called \gls{fr}. \gls{fr} stores a single bit for each row in the DRAM module. Each bit in \gls{fr} indicates whether a row \om{8}{should be} refreshed using nominal latency within a \gls{tfr}. \gls{fr} can be implemented as an SRAM array in the memory controller.}

\subsection{Hardware Complexity}
\label{sec:eval_area}
{\X{}'s metadata storage \yct{7}{(i.e., \gls{fr})} can be implemented \om{5}{using} SRAM in the memory controller. \om{5}{The} metadata size is independent of \gls{nrh}, making \X{} scalable with the increasing RowHammer vulnerability.}
\X{} stores one bit of information per DRAM row \om{5}{on the fully restored bit vector (\gls{fr})}. We evaluate \X{}'s chip area and access latency \agy{4}{overheads} using CACTI~\cite{cacti}. Our results show that \X{} has an area cost of $0.0069mm^2$ per DRAM bank \yct{4}{where a DRAM bank consists of} 64K DRAM rows.
When configured for a dual rank system with 16 banks at each rank, \X{} requires storage of $8KB$ per bank,
\agy{4}{corresponding to a $0.09\%$ overhead on} a high-end Intel Xeon processor~\cite{wikichipcascade} \agy{4}{($1.35\%$ \om{7}{of} the memory controller area~\cite{skylakedieshot})}.
The metadata access latency \om{7}{in SRAM} \agy{4}{is} $0.27ns$, which is \agy{4}{significantly smaller than the latency of a DRAM row activation (e.g., $14ns$~\cite{datasheetM393A1K43BB1}), and can be} hidden by the row activation latency.


\subsection{{\X{} with \agy{4}{on-DRAM-die RowHammer\\Mitigation Mechanisms}}}
\label{sec:ondiemitigations}
\agy{4}{\om{8}{Various} prior works propose on-DRAM-die RowHammer mitigation mechanisms~\cite{bennett2021panopticon, canpolat2024prac, marazzi2022protrr, marazzi2023rega, hassan2024selfmanaging, qureshi2024mint, son2017making, jedec2024ddr5, canpolat2025chronus}, including the PRAC mechanism (\secref{sec:background_dram_read_disturbance}) in \om{7}{the recent} DDR5 standard~\cite{jedec2024ddr5, canpolat2025chronus}}.
\iqrev{\agy{4}{To support these mechanisms, \X{} can also} be \yct{4}{implemented in \om{5}{a} DRAM chip} \agy{4}{with an on-DRAM-die RowHammer mitigation mechanism.}
\agy{7}{When a preventive refresh is scheduled,}
\yct{7}{\X{} determines whether the refresh latency can be safely reduced \yct{8}{by checking \gls{fr}} and stores the preventive refresh latency information in the mode registers (MR)~\cite{jedec2020ddr5}.}
\om{5}{When} \agy{7}{the} memory controller
\agy{4}{performs a preventive refresh (e.g., by issuing an RFM command~\cite{canpolat2024prac}), it uses the latency specified in the MR.}}
\yct{7}{\agy{7}{\om{8}{We believe} better memory interfaces}
\om{8}{can facilitate} easier \agy{7}{and more efficient integration of} on-DRAM-die RowHammer mitigation mechanisms and \agy{7}{\om{8}{\X-like} mechanisms that make RowHammer mitigation more efficient and scalable}. 
\agy{7}{For example,} Self-Managing DRAM~\cite{hassan2024selfmanaging} \agy{7}{enables} DRAM chips to autonomously \agy{7}{perform maintenance operations. In a memory system that implements Self-Managing DRAM, \X{} can easily be implemented in the DRAM chip with \emph{no} modifications to the memory controller or the memory interface.}}

\section{Evaluation}
\label{sec:evaluation}

\begin{figure*}[b]
\centering
\includegraphics[width=1\linewidth]{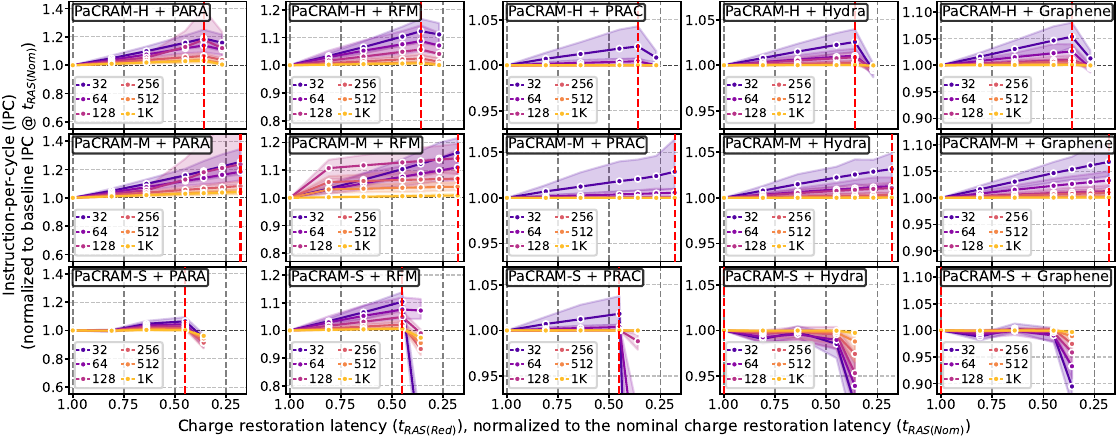}
\caption{\yct{6}{\om{8}{S}ystem performance versus different preventive refresh latencies for different manufacturers, RowHammer mitigation mechanisms, and $\bm{N}_{\bm{RH}}$ values, \om{7}{averaged across 62 single-core workloads}}}
\label{fig:lrr_analysis}
\end{figure*}

\subsection{Methodology}
\label{sec:eval_methodology}
\yct{20}{We showcase \X{}'s impact on system performance and energy efficiency, using Ramulator 2.0~\cite{ramulator2github, luo2023ramulator2} (based on Ramulator~\cite{kim2016ramulator, ramulatorgithub}).}
Table~\ref{table:configs} lists the simulated system configuration, \agy{20}{assuming}
a realistic system with a single \om{7}{core} \yct{1}{or four} \agy{4}{cores}, connected to \yct{1}{DDR5 memory.}
\vspace{5pt}
\begin{table}[ht]
\renewcommand{\arraystretch}{0.85}
\centering
\footnotesize
\caption{Simulated system configuration}
\label{table:configs}
\resizebox{\linewidth}{!}{
\begin{tabular}{ll}
\hline
\textbf{Processor}                                                   & \begin{tabular}[c]{@{}l@{}} \yct{7}{1 or 4 cores}, 3.2GHz clock frequency,\\ 4-wide issue, 128-entry instruction window\end{tabular}  \\ \hline
\textbf{DRAM}                                                        & \begin{tabular}[c]{@{}l@{}}DDR5, 1 channel, 2 rank, 8 bank groups, \\2 banks/bank group, \yct{7}{64K} rows/bank\end{tabular}  \\ \hline
\begin{tabular}[c]{@{}l@{}}\textbf{Memory Ctrl.}\end{tabular} & \begin{tabular}[c]{@{}l@{}}64-entry read and write requests queues,\\Scheduling policy: FR-FCFS~\cite{rixner2000memory,zuravleff1997controller} \\Address mapping: MOP\cite{kaseridis2011minimalistic}\end{tabular}   \\ \hline
\textbf{Last-Level Cache}& \begin{tabular}[c]{@{}l@{}} 2MB per core \end{tabular}  \\ \hline
\end{tabular}}
\end{table}

\head{Workloads}
{We \yct{1}{evaluate \X{} using}
five benchmark suites: {SPEC CPU2006~\cite{spec2006}}, SPEC CPU2017~\cite{spec2017}, {TPC~\cite{tpcweb}, MediaBench~\cite{fritts2009media}, and YCSB~\cite{ycsb}}. 
\yct{7}{From these benchmark suites, }\yct{1}{we randomly select \param{62} \om{5}{single-core} workloads and \param{60}
multi-programmed 4-core workload mixes.}
{For every workload, we generate memory traces corresponding to 100M instructions using SimPoint~\cite{simpoint}.}
We simulate these traces until \yct{1}{each} core executes \param{100M}
instructions with a warmup period of \param{10M} instructions, similarly to prior \agy{4}{works}~\cite{kim2020revisiting, yaglikci2021blockhammer, yaglikci2022hira}.}


\head{Metrics}
{{We evaluate \X{}'s impact on \emph{system performance} (in terms of \yct{0}{instructions-per-cycle, IPC \yct{1}{and weighted speedup\om{5}{~\cite{eyerman2008systemlevel, snavely2000symbiotic}}})
and \emph{DRAM energy consumption}.}}}\yctcomment{5}{we use DRAMPower which only calculates DRAM energy.}


\head{\yct{4}{Comparison points and configuring \X{}}}
We evaluate \X{} with \yct{20}{five} state-of-the-art preventive refresh based RowHammer mitigation \om{7}{mechanisms:}
PARA~\cite{kim2014flipping}, 
RFM~\cite{jedec2020ddr5}, 
PRAC~\cite{jedec2024ddr5}, 
Hydra~\cite{qureshi2022hydra}, and 
Graphene~\cite{park2020graphene}.
We use the implementations that are publicly available in \om{8}{the} Ramulator 2 \om{7}{code repository}~\cite{ramulator2github}. To showcase \X's improvements, we implement \X{} as a plugin that works with other mechanisms \yct{7}{and reduces the latency of preventive refreshes issued by RowHammer mitigation mechanisms}.\fnref{fn:onlyprev}
We use a blast radius of 2 (i.e., a preventive refresh performs charge restoration \om{7}{of victim rows that are within distance} $\pm2$ of an aggressor row) to account for \om{7}{the} Half-Double \om{5}{access pattern~\cite{kogler2022half}}, similarly to prior works~\cite{olgun2024abacus, canpolat2024prac, qureshi2024mint, qureshi2024impress, hassan2021utrr, yaglikci2021blockhammer, canpolat2024breakhammer}.

\yct{4}{We use the following three configurations in our simulations\om{5}{:} 
i)~\emph{No mitigation}: \yct{6}{the} baseline that does \emph{not} implement any RowHammer mitigation mechanism\iey{7}{,}\yctcomment{10}{added PaCRAM-M to fig16-17-18}
ii)~\emph{No \X{}}: \yct{6}{the configuration that implements a RowHammer mitigation mechanism \emph{without} employing \X{}\iey{7}{, and}}
\yct{10}{iii)~\emph{\Xh{}}, \emph{\Xm{}}, and \emph{\Xs{}}: \om{7}{three specific} \yct{6}{configurations that implement a RowHammer mitigation mechanism \agy{7}{and} \X{}\agy{7}{,}
\omcomment{7}{Do we show these modules’ data specifically in earlier sections?}\yctcomment{7}{we added an example to sec8.3 and will add an appendix table.}
\om{7}{configured for three different modules from Mfrs.~H \agy{7}{(H5)}, M~(M2), and~S \agy{7}{(S6)}}, \agy{7}{respectively}.}}} \iqrev{To configure \Xh{}\yct{10}{, \Xm{},} and \Xs{}, for each module, \iql{IQA2}we first \om{7}{experimentally characterize} the \gls{nrh} of the module for \agy{7}{the} charge restoration latencies \om{8}{of $33ns$ (nominal $t_{RAS}$), $27ns$ ($0.81t_{RAS}$), $21ns$ ($0.64t_{RAS}$), $15ns$ ($0.45t_{RAS}$), $12ns$ ($0.36t_{RAS}$), $9ns$ ($0.27t_{RAS}$)\yct{11}{, and $6ns$ ($0.18t_{RAS}$)}.}\yctcomment{7}{We will further improve the placement of this figure.}
\yct{4}{We then calculate \emph{\gls{nrh} reduction ratio} as the ratio of \gls{nrh} value with reduced \om{7}{charge restoration} latency and \gls{nrh} value with nominal \om{7}{charge restoration} latency.} 
\agy{7}{We configure the RowHammer mitigation mechanisms
for the \gls{nrh} values of 1024, 512, 256, 128, 64, and 32 row activations per aggressor row and when the RowHammer mitigation mechanism is integrated with \X{}, we reduce \gls{nrh} according to the reduction ratios of H5\yct{10}{, M2,} and S6 \om{5}{obtained from experimental characterization (\secref{sec:effect_nrh})}.}
For example, \yct{6}{when H5 (i.e., the module used for \Xh{}) is refreshed 300 times using $0.27t_{RAS}$, the lowest observed \gls{nrh} of the module decreases by 8\% (i.e., reduces from 10.2K to 9.4K).}
\yct{6}{\agy{7}{Therefore,} we
\agy{7}{use the scaled} down \gls{nrh} values of 942, 471, 235, 117, 58, and 29 \agy{7}{to simulate the RowHammer mitigation mechanisms, integrated with \Xh{}}.
\yct{8}{\agy{7}{To configure \X{} correctly in the presence of the need for refreshing rows using full charge restoration before a row can receive \gls{thpcr} partial charge restorations (\takeref{take:rh_rep}),} we compute \agy{7}{the interval of full charge restoration (\gls{tfr})} as explained in \secref{sec:mech_implementation}.}
\yctcomment{7}{We will add the appendix table reporting these values for all modules.}
}
}

\subsection{Performance and Energy Analysis}
\label{sec:eval_perf}

\head{Effect of reducing \agy{7}{the charge restoration} latency \agy{7}{of preventive refreshes} on system performance}
\figref{fig:lrr_analysis} demonstrates how \agy{7}{reducing charge restoration latency for} preventive refreshes affects \agy{7}{system} performance \agy{7}{for} five RowHammer mitigation mechanisms \agy{7}{(columns of subplots)} \agy{7}{when they are integrated with \Xh{}\yct{10}{, \Xm{},} and \Xs{} (rows of subplots)}. 
The x-axis shows the charge restoration latency
normalized to \om{7}{the} nominal restoration latency. 
\yct{7}{The y-axis shows the IPC of the \om{8}{single-core} \iey{7}{workloads} normalized to the baseline where the mitigation mechanisms do \emph{not} \yct{7}{use}
\X{} (i.e., the nominal charge restoration latency is used for preventive refreshes).}
Each curve represents \om{7}{a} different \gls{nrh} value.
\yct{7}{The shades show the variation across all tested workloads.} \yct{8}{The dashed red line in each plot represents the best-observed charge restoration latency, which provides the highest performance improvement across tested charge restoration latencies.}

We make \param{\agy{7}{five}} observations from \figref{fig:lrr_analysis}. 
First, \Xh{} \yct{10}{and \Xm{}} \yct{7}{improve system performance by reducing preventive refresh overheads} \yct{7}{of each tested} RowHammer mitigation mechanism for each \iey{7}{tested} \gls{tras} and \gls{nrh} value. 
Second, as we decrease the preventive refresh latency, the performance of \Xh{} and \Xs{} first increases until a point, then decreases since RowHammer vulnerability increases (i.e., \gls{nrh} of the DRAM chip decreases) with the reduced charge restoration latency. 
Third, \Xs{} reduces the performance of Hydra and Graphene as reducing charge restoration latency increases the RowHammer vulnerability and leads to more preventive refreshes.
Fourth, \X{} can reduce the charge restoration latency for \yct{10}{Mfrs.~H and M} more than Mfr.~S as the DRAM chips from \yct{10}{Mfrs.~H and M} have larger \om{7}{\gls{tras}} guardbands than the chips from Mfr. S and \om{7}{thus} are more resilient to reduced charge restoration latencies.
\agy{7}{Fifth, the} \yct{6}{best-observed}
charge restoration latency for \Xh{} \yct{10}{and \Xm{}} are $0.36t_{RAS}$ \yct{11}{and $0.18t_{RAS}$}\yctcomment{11}{Before PaCRAM-M, 0.27 was the lowest value we test, for M, I tested 0.18 as well and updated the numbers and the fig 16 accordingly.}, across all tested RowHammer mitigation mechanisms\yct{10}{, respectively.} The best-observed charge restoration latency for \Xs{} is $0.45t_{RAS}$ when used with PARA, RFM, and PRAC.\footnote{\agy{7}{For the rest of our analysis, \Xh{}\yct{10}{, \Xm{},} and \Xs{} use \yct{1}{the \yct{7}{best-observed}} charge restoration latencies.
}}
\agy{1}{From these observations, we}
\agy{7}{conclude that \X{} increasingly improves system performance by reducing \iey{7}{the} charge restoration latency of preventive refreshes until an inflection point, where the latency reduction is overwhelmed by the increasing number of preventive refreshes.}



\begin{figure}[ht]
\centering
\includegraphics[width=1\linewidth]{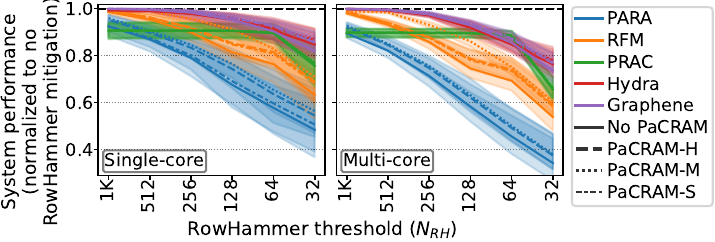}
\caption{\yct{6}{System performance of \X{}}}
\label{fig:perf_imp}
\end{figure}

\noindent
\textbf{\X's impact on system performance.}
\figref{fig:perf_imp} demonstrates the \yct{1}{performance of \X{}} normalized to performance when \emph{no} mitigation mechanism is employed (y-axis) for six different \gls{nrh} values (x-axis) and \yct{20}{five} different RowHammer mitigation mechanisms. \yct{1}{Left (right) subplot represents the performance as IPC (weighted speedup)
for \yct{7}{
single-core (multi-core) systems, averaged across 62 single-core workloads (60 multi-programmed workload mixes).}}
Each color represents \agy{6}{a} different \agy{6}{RowHammer mitigation mechanism} and the \agy{6}{marker} style represents \om{7}{different configurations}. \yct{7}{The shades show the variation across \iey{7}{all} tested workloads.}

We make \param{four} observations from \figref{fig:perf_imp}. 
First, \Xh{} \yct{10}{and \Xm{}} increase system performance for all \gls{nrh} values by reducing the performance overheads of all tested RowHammer mitigation mechanisms. For example, \Xh{} improves \yct{7}{single-core (multi-core)} system performance on average by \yct{7}{18.95\% (10.81\%), 12.28\% (10.84\%), 2.07\% (0.76\%), 2.56\% (2.00\%), and 5.37\% (4.31\%) \yct{7}{when used with}
PARA, RFM, PRAC, Hydra, and Graphene
\yct{7}{at an \gls{nrh} of 32, across all tested workloads.}}
Second, \agy{7}{the} performance improvement of \yct{7}{\X{}} increases as \gls{nrh} decreases for all \iey{7}{tested RowHammer} mitigation mechanisms. For example, \Xh{} increases \yct{7}{single-core} system performance by \yct{20}{0.69\% (12.28\%)} \om{7}{when used with} RFM at an \gls{nrh} of \yct{20}{1K (32)}, on average.
Third, the performance improvements of \Xh{} \yct{10}{and \Xm{}} \yct{1}{are higher than \agy{7}{that} of \Xs{} \yct{7}{when used with any tested mitigation mechanism}.
\yct{10}{For example, \Xh{} and \Xm's \yct{7}{multi-core} performance improvements are \yct{7}{$0.46\%$ and $4.48\%$ higher\om{11}{, on average,} than \Xs's improvement, respectively, when used with RFM}.}
\srev{Fourth, \X{} provides larger performance improvements \om{7}{when used with} high-performance-overhead mitigations (PARA and RFM) compared to high-area-overhead mitigations (PRAC, Hydra, and Graphene). PRAC, Hydra, and Graphene perform fewer preventive refreshes and cause smaller performance overheads at the cost of higher hardware complexity compared to PARA and RFM\om{7}{;} hence \X{} can reduce the latency of more preventive refreshes when \yct{7}{used} with PARA or RFM.}}
\agy{1}{Based on these observations, we derive:}

\takeaway{\X{} significantly improves \yct{7}{both single-core and multi-core} system performance.\label{take:perf_attack}}

\begin{figure}[ht]
\centering
\includegraphics[width=\linewidth]{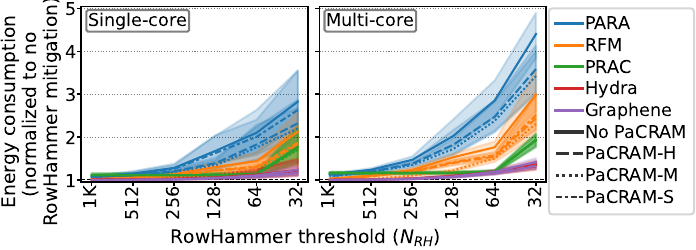}
\caption{\srev{Energy \sql{R4}consumption of \X{}}}
\label{fig:edp}
\end{figure}

\head{\X's impact on \agy{7}{DRAM} energy consumption}
\figref{fig:edp} shows \X{}'s impact on DRAM energy consumption \om{8}{(using a similar style as~\figref{fig:perf_imp}).}
We make \param{three} observations. 
First, \Xh{} \yct{10}{and \Xm{}} reduce energy consumption when used with any tested mitigation mechanism for all \gls{nrh} values \om{8}{(due to lower energy consumed by lower latency preventive refreshes and reduced execution time).}
\yct{7}{For example, \Xh{} reduces \om{8}{the DRAM} energy consumption of \om{8}{the} single-core (multi-core) system by 14.59\% (17.75\%), 11.56\% (18.05\%), 1.15\% (0.26\%), 2.18\% (2.55\%), and 4.50\% (5.76\%)
when used with
PARA, RFM, PRAC, Hydra, and Graphene, respectively, on average at an \gls{nrh} of 32.}
\yct{7}{Second, \Xs's energy consumption is higher than \Xh{} \yct{10}{and \Xm's} energy consumptions. For example, \Xs's single-core energy consumption is 8.69\% \yct{10}{and 11.57\%} higher than \Xh{} \yct{10}{and \Xm's} energy consumption\yct{10}{, respectively, when used with PARA} on average \agy{7}{across 62 single-core workloads}.
\yct{7}{Third, all configurations consume more energy as \gls{nrh} reduces.}
From these observations, we derive:}

\takeaway{\X{} significantly reduces DRAM energy consumption.\label{take:edp}}

\noindent
\srev{\textbf{\X's impact on high-performance-overhead \sql{R2.3}and high-area-overhead mitigations.}}
\srev{\X{} significantly reduces the performance and energy overheads of high-performance-overhead RowHammer mitigation \yct{7}{mechanisms}, e.g., PARA \agy{7}{by \param{18.95\%}} and RFM \agy{7}{by \param{12.28\%} for \gls{nrh}=32} \om{7}{(\figref{fig:perf_imp})}.
\agy{7}{This is}
\om{7}{especially} important because PARA and RFM are significantly more scalable to lower \gls{nrh} values and larger bank counts {due to their low area cost}, unlike the best-performing mechanisms (e.g., Graphene). 

When \yct{7}{used} with \X{}, high-area-overhead mitigation \yct{7}{mechanisms} that already incur high area overheads (e.g., Graphene, Hydra, and PRAC) experience substantial performance and energy overhead reductions at the cost of small additional area overhead. \yct{7}{For example, \X{} reduces the performance overhead of Graphene by 5.37\% \om{7}{(at $N_{RH}=32$)} \agy{7}{at the cost of an additional 0.09\% on top of Graphene's 4.45\% chip area overhead (only 2.02\% increase in Graphene's area).}}
\yct{7}{Therefore, \X{} introduces new system design points for RowHammer mitigation, balancing i)~performance and energy efficiency and ii)~area overhead.}
}

\section{\om{7}{Profiling Overhead}}
\label{sec:limitations}
\noindent
\iqrev{\textbf{{Characterization of DRAM chips.}}
{For \X{} \om{8}{to work robustly}, one needs to correctly identify the \om{7}{appropriate} charge restoration reduction. Doing so requires profiling the DRAM chips in use.
This profiling can be performed in different ways at low cost as described in our methodology \om{8}{(\secref{subsec:testing_methodology})} and several prior works (e.g.,\yct{7}{~\cite{chang2017understanding,liu2013experimental, lee2017design, yaglikci2024spatial, patel2017reaper, liu2013experimental, choi2020reducing, orosa2021deeper, orosa2021codic}}). 
First, the system can \om{8}{perform} profiling the very first time DRAM is initialized and configure \X{}. 
Second, DRAM vendors can \om{8}{perform} profiling at manufacturing time and embed configuration data in the Serial Presence Detect (SPD) circuitry~\cite{jedec-spd}. The memory controller can read the configuration data from the SPD circuitry and configure \X{}.
Third, the system can perform \om{8}{\emph{online profiling}} to configure \X{}\om{8}{~\cite{patel2017reaper, qureshi2015avatar, liu2013experimental, khan2014efficacy, lee2017design}}\yctcomment{9}{cited DIVA-DRAM}.
\yct{20}{To quantify the profiling cost, we demonstrate an optimized profiling methodology.}
}
}

\noindent
\head{Profiling cost} 
As \secref{subsec:testing_methodology} describes, our profiling methodology tests a DRAM row for five different $t_{RAS}$ values with ten different \yct{7}{consecutive} partial charge restorations \yct{7}{counts} at five different hammer counts for five iterations. Given that each test includes a wait time of \gls{trefw} ($64ms$), we concurrently test and, by doing so, overlap the testing time of 1270 different DRAM rows in a DRAM bank. 
Therefore, we test 1270 DRAM rows within an $80s$ time window ($64ms$ $\times$ $5$ $t_{RAS}$ values $\times$ $10$ different numbers of partial charge restorations $\times$ $5$ hammer counts $\times$ $5$ iterations).
As each DRAM row contains 8KB of data, our methodology's profiling throughput is 127 KB/s (1270 $\times$ 8KB $/$ 80s).
Following this methodology, profiling a DRAM bank with 64K
rows takes 68.8 minutes. However, this profiling can be performed in chunks of \om{7}{80 seconds} spread across time by blocking only $9.9MB$ of data at a time. We leave further optimizations of profiling to future work.

\noindent
\iqrev{\textbf{{System reliability.}}
{\X{} is limited to systems that employ \iql{IQF1}DRAM chips that implement guardbands in \om{7}{the} $t_{RAS}$ timing parameter. We experimentally demonstrate on real DDR4 DRAM chips that there exist significantly large guardbands in \om{8}{the} $t_{RAS}$ timing parameter. Our findings are in line with prior works\yct{7}{~\cite{liu2013experimental, lee2015adaptive, chang2016understanding, chang2017understanding, chang2017understandingphd, kim2018solar, yaglikci2022understanding, mathew2017using, lee2017design, chandrasekar2014exploiting, das2018vrldram}}\yctcomment{8}{yes, solardram is included} that show that several timing parameters (e.g., $t_{RCD}$, $t_{WR}$, $t_{RP}$, \gls{tras}) in commodity DRAM chips implement large guardbands. 
\om{8}{To improve robustness and to account for dynamic variability\yctcomment{8}{we couldn't parse "variability" or something else?}, e.g., aging}, \X{} can be combined with error correction mechanisms\yct{7}{~\cite{zhang2021quantifying, nair2016xed, hamming1950error, hocquenghem1959codes, bose1960class, qureshi2021rethinking, fakhrzadehgan2022safeguard, chen2014memguard, khan2014efficacy, kim2015bamboo, reed1960polynomial, yoon2010virtualized, qureshi2015avatar, lee2017design, patel2017reaper}}\yctcomment{9}{added REAPER}. We leave the exploration of such systems for future work. 
}}

\section{Related Work}
\label{sec:related_work}
{{To our knowledge, this} is the first work that 
{\om{7}{i)~}experimentally studies how reducing \agy{0}{charge restoration latency} affects \om{7}{modern} DRAM chips'} RowHammer vulnerability\om{7}{, ii)~}\agy{0}{investigate\om{7}{s} implications on state-of-the-art RowHammer mitigation mechanisms\om{7}{, and iii)~proposes a new technique}\yct{7}{, called \X{}}\om{8}{, which} reduces the performance overheads of existing \om{7}{RowHammer} mitigation mechanisms by reducing the time spent on preventive refreshes. \secref{sec:evaluation} already quantitatively evaluates \X{}'s benefits \om{8}{when used with} five state-of-the-art RowHammer mitigation mechanism\om{9}{s:} PARA~\cite{kim2014flipping}, RFM~\cite{jedec2020ddr5}, PRAC~\cite{jedec2024ddr5}, Hydra~\cite{qureshi2022hydra}, and Graphene~\cite{park2020graphene}.} 
\yct{7}{In this section, \iey{7}{we discuss related works in three \om{8}{related} directions.}}
}

\yctcomment{7}{I added the missing works}
\head{DRAM operation under reduced latency}
Many works characterize DRAM under reduced latency\yct{7}{~\cite{chang2017understanding, lee2017design, olgun2021quac, kim2019drange, yaglikci2022hira, kim2018dram, qureshi2015avatar, gao2019computedram, gao2022frac, lee2015adaptive, orosa2021codic, kim2018solar, chandrasekar2014exploiting, talukder2019prelatpuf, yuksel2024functionally, yuksel2024pulsar, chang2016understanding, yuksel2024simultaneous, das2018vrldram, hassan2016chargecache, shin2015dram, wang2018reducing}}\yctcomment{8}{added ambit, vivek 2015 and rowclone} and develop techniques that \yct{7}{reduce} the latency of DRAM operations to improve system throughput and energy efficiency\yct{7}{~\cite{wang2018caldram, qureshi2015avatar, hassan2016chargecache, yaglikci2022hira, kim2018solar, hassan2019crow, luo2020clrdram, mathew2017using, koppula2019eden, orosa2023approximate, chang2017understanding, lee2015adaptive, chandrasekar2014exploiting, das2018vrldram, qin2023cdardram, lee2013tiered, lee2017design, zhang2016restore, shin2014nuat, wang2018reducing, shin2015dram}}, generate \iey{7}{true} random numbers or physical unclonable functions using DRAM chips\yct{7}{~\cite{olgun2021quac, kim2019drange, olgun2022pidram, bostanci2022drstrange, talukder2019prelatpuf, orosa2021codic, kim2018dram, mutlu2024memory}}, and perform \om{8}{computation using DRAM} techniques\yct{7}{~\cite{gao2019computedram, gao2022frac, yuksel2024functionally, yuksel2024pulsar, yuksel2024simultaneous, olgun2022pidram, olgun2023drambender, mutlu2024memory, hajinazar2021simdram, oliveira2024mimdram, seshadri2017ambit, seshadri2013rowclone, vivek2015fastbulk}}. {\om{8}{None} of these works study the effect of reduced charge restoration latency on RowHammer vulnerability.}


\head{\iey{7}{Experimental RowHammer characterization}}
{Prior work{s} extensively characterize the RowHammer vulnerability in real DRAM chips\understandingRowHammerAllCitations{}. These works {demonstrate {(}using real DDR3, DDR4, LPDDR4, and HBM2 DRAM chips{)} how a DRAM chip's RowHammer vulnerability varies with}
i)~DRAM refresh rate~\cite{hassan2021utrr,frigo2020trrespass,kim2014flipping}, ii)~the physical distance between aggressor and victim rows\yct{7}{~\cite{kim2014flipping,kim2020revisiting,lang2023blaster}}, iii)~DRAM generation and technology node~\cite{orosa2021deeper,kim2014flipping,kim2020revisiting,hassan2021utrr}, iv)~temperature~\cite{orosa2021deeper,park2016experiments}, v)~the time the aggressor row stays active\yct{7}{~\cite{orosa2021deeper,park2016experiments, luo2023rowpress, luo2024rowpress, luo2024experimental, olgun2024read}}, vi)~physical location of the victim cell\yct{7}{~\cite{orosa2021deeper, olgun2023hbm, olgun2024read, yaglikci2024spatial}}, \agy{0}{vii)~reduced wordline voltage~\cite{yaglikci2022understanding}\om{7}{, and viii)~time~\cite{olgun2025variable}}. \om{8}{None} of these works investigate the effect of charge restoration latency on} RowHammer vulnerability in real DRAM chips. Our characterization study \agy{0}{advances} the analyses in these works by uncovering new insights into RowHammer \om{7}{under reduced refresh latency}.}

\head{RowHammer attacks and mitigation mechanisms}
{Many prior works~\exploitingRowHammerAllCitations{} show that RowHammer can be exploited to mount system-level attacks to compromise system security {and safety} (e.g., to acquire root privileges {or leak private data}).
To protect against these attacks,
{many prior works~\mitigatingRowHammerAllCitations{} propose RowHammer mitigation mechanisms that prevent bitflips from compromising a system.}
The observations {we make in this work} can be leveraged to reduce \agy{0}{the performance overhead of existing mitigation mechanisms by reducing the execution time \om{7}{and energy} spent on preventive refreshes\om{7}{,} as we \om{7}{quantitatively} demonstrate in~\secref{sec:evaluation}.}}

\section{Conclusion}
\label{sec:conclusion}

\om{8}{We} present the first rigorous experimental study on the \om{1}{effects of reduced refresh latency on} RowHammer vulnerability \om{5}{of modern DRAM chips}. \yct{8}{\om{8}{Our} analysis of \param{\nCHIPS{}} real DDR4 DRAM chips from three major manufacturers \yct{7}{demonstrate that charge restoration latency of \om{8}{RowHammer-}preventive refreshes can be significantly reduced}
at the expense of requiring slightly more preventive refreshes.}
\yct{7}{We propose a new mechanism, \emph{\Xlong{}} (\emph{\X{}}), which \om{8}{robustly} reduces the preventive refresh latency in existing RowHammer mitigation mechanisms. Our evaluation demonstrates that \X{} significantly enhances system performance and energy efficiency \om{8}{when used with} five state-of-the-art RowHammer mitigation mechanisms while introducing small additional area overhead. By doing so, \X{} enables new trade-offs in RowHammer mitigation.}
\yct{7}{We hope and expect that the understanding we develop via our rigorous experimental characterization and the resulting \X{} mechanism will inspire DRAM manufacturers and system designers to efficiently and scalably enable robust \om{8}{and efficient} operation as DRAM technology node scaling exacerbates read disturbance.}

\section*{Acknowledgments} {
We thank the anonymous reviewers of \om{8}{HPCA 2025 (both main submission and artifact evaluation), MICRO 2024, and ISCA 2024} for the encouraging feedback.
We thank the SAFARI Research Group members for valuable feedback and the stimulating scientific and intellectual environment.
We acknowledge the generous gift funding provided by our industrial partners (especially Google, Huawei, Intel, Microsoft, VMware), which has been instrumental in enabling the research we have been conducting on read disturbance in DRAM since 2011~\cite{kim2023flipping}.
This work was also in part supported by the Google Security and Privacy Research Award, the Microsoft Swiss Joint Research Center, \yct{3}{and the ETH Future Computing Laboratory (EFCL)}\yctcomment{5}{Oguz Hoca requested to include EFCL}.
}
\yctcomment{8}{Double-checked all references}

\bibliographystyle{IEEEtran}
\bibliography{refs}

\clearpage

%
%
%
%
%


\appendix
\section{Artifact Appendix}

\subsection{Abstract}

Our artifact contains the data, source code, and scripts needed to reproduce our results.
We provide: 
1)~the scripts to parse and plot our DRAM Bender results, 
2)~the source code of our simulation infrastructure based on Ramulator 2.0, 
3)~all evaluated memory access traces and all major evaluation results, and
4)~the data and scripts to reproduce all key figures in the paper.
We identify the following as key results:
\begin{itemize}
    \item Effect of partial charge restoration on RowHammer vulnerability in terms of RowHammer \gls{nrh} and \gls{ber}
    \item Effect of repeated partial charge restoration on RowHammer vulnerability in terms of RowHammer \gls{nrh}
    \item \X's impact on system performance
    \item \X's impact on DRAM energy consumption
\end{itemize}

\subsection{Artifact Check-list (meta-information)}
\begin{table}[ht]
\centering
\footnotesize
\label{table:ae_metainfo}
\begin{tabular}{lm{4.8cm}}
\textbf{Parameter} & \textbf{Value} \\ \hline
Program & C++ programs\newline 
            Python3 scripts\newline
            Shell scripts \\ \hline
Compilation & C++ compiler with c++20 features \\ \hline
Run-time environment & Ubuntu 20.04 (or similar) Linux\newline
                        C++20 build toolchain (tested with GCC 10)\newline
                        Python 3.10+\newline
                        Git \\ \hline
Metrics & RowHammer \gls{nrh} and \gls{ber}\newline
            Instruction-per-cycles (IPC)\newline
            Weighted speedup\newline
            DRAM energy \\ \hline
Experiment customization & Possible. See \secref{sec:exp_custom} \\ \hline
Disk space requirement & $\approx$ 30 GiB \\ \hline
Experiment completion time & $\approx$ 2 days (on a compute cluster with 250 cores) \\ \hline
Publicly available? & DRAM Bender results (\url{https://zenodo.org/records/14343791})\newline
                        Benchmarks (\url{https://zenodo.org/records/14345886})\newline
                        Zenodo (\url{https://zenodo.org/records/14878588})\newline
                        GitHub (\url{https://github.com/CMU-SAFARI/PaCRAM}) \\ \hline
Code licences & MIT \\ \hline
\end{tabular}
\end{table}

\subsection{Description}
 
\noindent\emph{We highly recommend using Slurm with a cluster that can run experiments in bulk.}

\subsubsection{How to access}

~Source code and scripts are available at \url{https://github.com/CMU-SAFARI/PaCRAM}.

\subsubsection{Hardware dependencies}

~We recommend using a PC with 32 GiB of main memory.
Approximately 30 GiB of disk space is needed to store intermediate and final evaluation results.

\subsubsection{Software dependencies}

\begin{itemize}
    \item GNU Make, CMake 3.20+
    \item C++20 build toolchain (tested with GCC 10)
    \item Python 3.9+ (tested with Python 3.10.6)
    \item pip packages: matplotlib, seaborn, pandas, pyyaml
    \item Ubuntu 22.04
    \item (Optional) Slurm 20+
\end{itemize}

\subsubsection{Benchmarks}

~We use workload memory traces collected from SPEC2006, SPEC2017, TPC, MediaBench, and YCSB benchmark suites.
These traces are available at \url{https://zenodo.org/records/14345886}.
Install scripts will download and extract the traces.

\subsection{Installation}

\lstset{
    backgroundcolor=\color{gray!20}, 
    basicstyle=\ttfamily\bfseries\footnotesize,
    columns=fullflexible,
    frame=single,
    breaklines=true,
    postbreak=\mbox{\textcolor{red}{$\hookrightarrow$}\space},
    showstringspaces=false,
    numbersep=5pt,
    xleftmargin=0pt,
    numbers=none,
    keywordstyle=\color{black},  
    identifierstyle=\color{black},  
    commentstyle=\color{black},  
    stringstyle=\color{black}  
}

The following steps will download and prepare the repository for the main experiments:
\begin{enumerate}
    \item Clone the git repository.
    \begin{lstlisting}[language=bash]
$ git clone https://github.com/CMU-SAFARI/PaCRAM
    \end{lstlisting}
    \item Install Python dependencies, compile Ramulator 2.0, download DRAM Bender experiment results, and download workload traces.
    \begin{lstlisting}[language=bash]
$ ./setup_all.sh
    \end{lstlisting}
\end{enumerate}

\subsection{Evaluation and Expected Results}
We categorize our evaluation into two parts: i) Off-the-shelf DDR4 DRAM characterization and ii) System evaluation of \X{}.

\subsubsection{Off-the-shelf DDR4 DRAM characterization}
~We used DRAM Bender~\cite{safari-drambender} to characterize real DDR4 DRAM chips.
The data we gather from \nCHIPS{} DDR4 DRAM chips are publicly available at \url{https://zenodo.org/records/14343791}. We make the following claims.

\head{Claim 1.1 (C1.1)} Charge restoration latency (\gls{tras}) can be reduced 1) with no effect on or 2) at the cost of increasing the RowHammer vulnerability (i.e., lower RowHammer \gls{nrh} and higher RowHammer \gls{ber}), while reducing charge restoration latency beyond a safe minimum value can cause data retention failures. This property is proven by measuring the RowHammer vulnerability of real DDR4 DRAM chips with reduced charge restoration latency as described in \secref{subsec:testing_methodology} whose results are illustrated in Figs.~\ref{fig:nRH_single_pcr} and~\ref{fig:ber_single_pcr}. 

\head{Claim 1.2 (C1.2)} It is not safe to reduce charge restoration latency (\gls{tras}) for all refresh operations as performing partial charge restoration many times can cause data retention failures. This property is proven by measuring the RowHammer vulnerability of real DDR4 DRAM chips with repeated reduced charge restoration latency as described in \secref{subsec:testing_methodology} whose results are illustrated in \figref{fig:nrh_multiple_pcr}. 

To parse the results and plot the key figures of our experimental results:
\begin{enumerate}
    \item Parse and plot the results for RowHammer \gls{nrh} and \gls{ber}.
    \begin{lstlisting}[language=bash]
$ ./plot_db_figures.sh
(or ./plot_db_figures_slurm.sh if Slurm is available)
\end{lstlisting}
\end{enumerate}

\subsubsection{System evaluation of \X{}}
~We used Ramulator 2.0~\cite{ramulator2github} to demonstrate the potential benefits of \X{} on system performance and energy efficiency. We make the following claims.

\head{Claim 2.1 (C2.1)} \X{} significantly improves system performance for both single-core and multi-programmed workloads. This property is proven by evaluating system performance of \X{} when used with five state-of-the-art RowHammer mitigation mechanisms as described in \secref{sec:eval_methodology} whose results are illustrated in Figs.~\ref{fig:lrr_analysis} and~\ref{fig:perf_imp}.


\head{Claim 2.2 (C2.2)} \X{} significantly improves energy efficiency of the system. This property is proven by evaluating DRAM energy consumption of \X{} when used with five state-of-the-art RowHammer mitigation mechanisms as described in \secref{sec:eval_methodology} whose results are illustrated in \figref{fig:edp}.

To run simulations, parse the results, and plot the key figures of \X's evaluation:
\begin{enumerate}
    \item Prepare Ramulator 2.0 simulation warmup checkpoints.
    \begin{lstlisting}[language=bash]
$ ./prepare_warmups.sh
(or ./prepare_warmups_slurm.sh if Slurm is available)
    \end{lstlisting}
    \item Wait for the warmup runs to end. The following displays the status of the warmup runs. You can restart failed runs simply by rerunning step 1.
    \begin{lstlisting}[language=bash]
$ python3 check_warmup_status.py
    \end{lstlisting}
    \item Launch Ramulator 2.0 simulation runs.
    \begin{lstlisting}[language=bash]
$ ./run_ramulator_all.sh
(or ./run_ramulator_all_slurm.sh if Slurm is available)
    \end{lstlisting}
    \item Wait for the simulations to end. The following displays the status of the runs. You can restart failed runs simply by rerunning step 3.
    \begin{lstlisting}[language=bash]
$ python3 check_run_status.py
    \end{lstlisting}
    \item Parse simulation results and collect statistics.
    \begin{lstlisting}[language=bash]
$ ./parse_ram_results.sh
    \end{lstlisting}
    \item Generate all figures that support C2.1 and C2.2.
    \begin{lstlisting}[language=bash]
$ ./plot_ram_figures.sh
    \end{lstlisting}
\end{enumerate}

\subsection{Experiment Customization}
\label{sec:exp_custom}
Our scripts provide easy configuration of the 1)~evaluated RowHammer mitigation mechanisms, 2)~tested RowHammer thresholds, 3)~preventive refresh latency, 4)~periodic refresh latency, and 5)~simulated workload combinations. The run parameters are configurable in \mbox{\texttt{Ram\_scripts/utils\_runs.py}} with 1)~\mbox{\textit{MITIGATION\_LIST}} and 2)~\mbox{\textit{NRH\_VALUES}}. 3)~The preventive refresh latency and 4)~periodic refresh latency can be configured using \mbox{\texttt{latency\_factor\_vrr}} and \mbox{\texttt{latency\_factor\_rfc}} parameters in a configuration file, respectively. 5)~Simulated workload combinations can be updated in \mbox{\texttt{{Ram\_scripts/mixes\_multicore.txt}}}.

\subsection{Methodology}

Submission, reviewing, and badging methodology:\yctcomment{8}{I will fix this margin issue.}

\begin{itemize}
  \item \url{https://www.acm.org/publications/policies/artifact-review-and-badging-current}
  \item \url{https://cTuning.org/ae}
\end{itemize}

\clearpage

\section{\yct{7}{\X{} \agy{7}{for Other \om{10}{Types of}\\Charge Restoration}}}
\cql{GC1.3}{}\agy{1}{\X{} can be extended to \om{7}{also} reduce the \om{7}{charge} restoration latency for \om{11}{\emph{periodic refreshes and dynamic accesses}}\om{7}{, in addition to preventive refreshes (as we \om{10}{have done} so far).}}
To \om{10}{exemplify} the potential benefits of \agy{1}{such extensions}, 
\yct{7}{we analyze the impact of reducing charge restoration latency for periodic refreshes, which restore every cell's charge in the module every \gls{trefw} to prevent data retention failures, similarly to prior work~\cite{das2018vrldram}.
\yct{10}{To do so, we modify \X{} such the latency of \emph{all} periodic refreshes is reduced since our experimental characterization demonstrates that charge restoration latency can be reduced significantly without causing any data retention failures (\takeref{take:retention}).}
\yct{7}{We use a configuration that does \emph{not} employ any RowHammer mitigation mechanism (i.e., no preventive refreshes are issued) and sweep the \agy{7}{charge restoration latency for periodic refreshes.}}}

\begin{figure}[ht]
\centering
\includegraphics[width=0.9\linewidth]{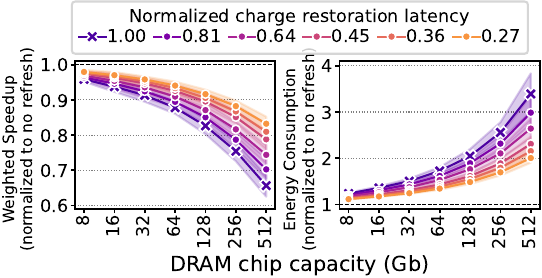}
\caption{\yct{7}{System performance (left) and energy consumption (right) versus different DRAM chip capacities for different periodic refresh latencies}}
\label{fig:periodic_ref}
\end{figure}

\srev{\figref{fig:periodic_ref} demonstrates \yct{7}{multi-core system} performance \sql{R3}(left subplot) and DRAM energy consumption (right subplot) of \X{} with reduced periodic refreshes \yct{7}{and the baseline system with nominal periodic refreshes} normalized to \yct{7}{a \agy{7}{hypothetical} system \agy{7}{which}
does \emph{not} perform any periodic refreshes} (y-axis) for different DRAM chip densities} (x-axis). \yct{7}{As DRAM chip density increases, the number of rows per bank increases. Hence, the number of rows refreshed with a single periodic refresh and the periodic refresh latency increase~\dramStandardCitations{}.} Six curves represent different \yct{7}{periodic} refresh latencies normalized to the nominal \yct{7}{periodic} refresh latency, the baseline system with nominal refresh latency is where \yct{7}{periodic} refresh latency is 1.00 \yct{7}{(marked with $\bm{\times}$)}. \yct{7}{The shades show the variation across tested workloads.}

\srev{We make \param{three} observations from \figref{fig:periodic_ref}. 
\yct{7}{
First, \yct{7}{\X{} significantly improves system performance and energy efficiency for all DRAM chip densities and reduced periodic refresh latencies (i.e., $<1.00$) 
\agy{7}{compared to} \yct{7}{the baseline system that uses nominal charge restoration latency\yct{7}{, across all tested workloads.}} 
For example, reducing periodic refresh latency by 64\% (i.e., \Xh's best-observed
charge restoration latency) improves multi-core system performance and energy efficiency by 23.31\%, and 36.49\% of a \param{512Gb} DRAM chip, respectively. 
Second, performance improvement and energy efficiency increase as periodic refresh latency decreases. For example, system performance increases by 8.83\% when periodic refresh latency decreases from 64\% to 36\%.}}
\yct{7}{Third, for all periodic refresh latency values, periodic refresh overheads increase as DRAM chip capacity increases.}
Based on these observations, we conclude that reducing the latency of periodic refreshes \agy{1}{improves \X{}'s performance benefits.}}

\yct{10}{To robustly employ this extension \om{11}{of \X{} to periodic refreshes}, \X{} needs to guarantee that a DRAM row does \emph{not} receive \gls{thpcr} consecutive periodic refreshes using reduced charge restoration latency. As each DRAM row is refreshed with periodic refresh once every refresh window (\gls{trefw}), \X{} can safely use reduced charge restoration latency for \gls{thpcr} refresh windows. After periodically refreshing each DRAM row using reduced charge restoration latency for \gls{thpcr} times, \X{} uses nominal charge restoration latency for the next refresh window to fully restore each cell's charge. To do so, \X{} implements a single counter that counts the number of refresh windows using reduced charge restoration latency and increments the counter every \gls{trefw}. When the counter reaches \gls{thpcr}, \X{} uses nominal charge restoration latency for that refresh window and resets the counter to zero.}

\clearpage

\secondbibtrue  
\begin{landscape}

\section{Summary Tables of All Tested DRAM Modules}    
\label{sec:ext_tables}

\newcolumntype{?}{!{\vrule width 2.5pt}}
\begin{table}[ht!]
\renewcommand{\arraystretch}{1.2}
\setlength{\tabcolsep}{2pt}
\caption{Summary of all tested DDR4 DRAM modules\protect\footnotemark\ and their lowest observed RowHammer thresholds ($\bm{N}_{\bm{RH}}$) when the victim rows are refreshed using different charge restoration latencies. Red cells denote $\bm{N}_{\bm{RH}}$ of zero (i.e., victim rows exhibit bitflips without hammering due to partial charge restoration)}
\label{tab:pcrsummary}
\resizebox{\columnwidth}{!}{%
\begin{tabular}{|c|ccccccc?ccccccc|}
\hline
 &  &  &  &  &  &  &  & \multicolumn{7}{c|}{\textbf{\begin{tabular}[c]{@{}c@{}}Lowest observed $\bm{N}_{\bm{RH}}$ of DRAM module when the victim rows are refreshed\\ using $\bm{t}_{\bm{RAS(Red)}}\bm{=Mt}_{\bm{RAS(Nom)}}$, $\bm{N}_{\bm{RH}}$ (normalized to $\bm{N}_{\bm{RH}}$ when $\bm{M=1.00}$)\end{tabular}}} \\ \cline{9-15} 
\multirow{-2}{*}{\textbf{Mfr.}} & \multirow{-2}{*}{\textbf{DRAM Part}} & \multirow{-2}{*}{\textbf{\begin{tabular}[c]{@{}c@{}}Form\\ Factor\end{tabular}}} & \multirow{-2}{*}{\textbf{\begin{tabular}[c]{@{}c@{}}Die\\ Rev.\end{tabular}}} & \multirow{-2}{*}{\textbf{\begin{tabular}[c]{@{}c@{}}Chip\\ Density\end{tabular}}} & \multirow{-2}{*}{\textbf{DQ}} & \multirow{-2}{*}{\textbf{\begin{tabular}[c]{@{}c@{}}Date\\ Code\end{tabular}}} & \multirow{-2}{*}{\textbf{ID}} & \multicolumn{1}{c|}{\textbf{\begin{tabular}[c]{@{}c@{}}$\bm{M=1.00}$\\ $\bm{(33ns)}$\end{tabular}}} & \multicolumn{1}{c|}{\textbf{\begin{tabular}[c]{@{}c@{}}$\bm{M=0.81}$\\ $\bm{(27ns)}$\end{tabular}}} & \multicolumn{1}{c|}{\textbf{\begin{tabular}[c]{@{}c@{}}$\bm{M=0.64}$\\ $\bm{(21ns)}$\end{tabular}}} & \multicolumn{1}{c|}{\textbf{\begin{tabular}[c]{@{}c@{}}$\bm{M=0.45}$\\ $\bm{(15ns)}$\end{tabular}}} & \multicolumn{1}{c|}{\textbf{\begin{tabular}[c]{@{}c@{}}$\bm{M=0.36}$\\ $\bm{(12ns)}$\end{tabular}}} & \multicolumn{1}{c|}{\textbf{\begin{tabular}[c]{@{}c@{}}$\bm{M=0.27}$\\ $\bm{(9ns)}$\end{tabular}}} & \textbf{\begin{tabular}[c]{@{}c@{}}$\bm{M=0.18}$\\ $\bm{(6ns)}$\end{tabular}} \\ \specialrule{2.5pt}{1pt}{1pt}
 & H5AN4G8NMFR-TFC [m1] & SO-DIMM & M & 4 Gb & x8 & N/A & H0 & \multicolumn{1}{c|}{No bitflips} & \multicolumn{1}{c|}{No bitflips} & \multicolumn{1}{c|}{No bitflips} & \multicolumn{1}{c|}{No bitflips} & \multicolumn{1}{c|}{No bitflips} & \multicolumn{1}{c|}{No bitflips} & No bitflips \\
 & \cellcolor[HTML]{DAE8FC}Unknown & \cellcolor[HTML]{DAE8FC}SO-DIMM & \cellcolor[HTML]{DAE8FC}X & \cellcolor[HTML]{DAE8FC}4 Gb & \cellcolor[HTML]{DAE8FC}x8 & \cellcolor[HTML]{DAE8FC}N/A & \cellcolor[HTML]{DAE8FC}H1 & \multicolumn{1}{c|}{\cellcolor[HTML]{DAE8FC}56.2K (1.00)} & \multicolumn{1}{c|}{\cellcolor[HTML]{DAE8FC}53.1K (0.94)} & \multicolumn{1}{c|}{\cellcolor[HTML]{DAE8FC}55.5K (0.99)} & \multicolumn{1}{c|}{\cellcolor[HTML]{DAE8FC}56.2K (1.00)} & \multicolumn{1}{c|}{\cellcolor[HTML]{DAE8FC}55.5K (0.99)} & \multicolumn{1}{c|}{\cellcolor[HTML]{DAE8FC}45.3K (0.81)} & \cellcolor[HTML]{DAE8FC}44.1K (0.78) \\
 & H5AN4G8N/AFR-TFC [m2] & SO-DIMM & A & 4 Gb & x8 & N/A & H2 & \multicolumn{1}{c|}{39.1K (1.00)} & \multicolumn{1}{c|}{40.6K (1.04)} & \multicolumn{1}{c|}{40.6K (1.04)} & \multicolumn{1}{c|}{39.1K (1.00)} & \multicolumn{1}{c|}{39.1K (1.00)} & \multicolumn{1}{c|}{39.1K (1.00)} & 37.9K (0.97) \\
 & \cellcolor[HTML]{DAE8FC}H5AN8G4NMFR-UKC [m3] & \cellcolor[HTML]{DAE8FC}R-DIMM & \cellcolor[HTML]{DAE8FC}M & \cellcolor[HTML]{DAE8FC}8 Gb & \cellcolor[HTML]{DAE8FC}x4 & \cellcolor[HTML]{DAE8FC}N/A & \cellcolor[HTML]{DAE8FC}H3 & \multicolumn{1}{c|}{\cellcolor[HTML]{DAE8FC}59.8K (1.00)} & \multicolumn{1}{c|}{\cellcolor[HTML]{DAE8FC}59.8K (1.00)} & \multicolumn{1}{c|}{\cellcolor[HTML]{DAE8FC}59.8K (1.00)} & \multicolumn{1}{c|}{\cellcolor[HTML]{DAE8FC}59.4K (0.99)} & \multicolumn{1}{c|}{\cellcolor[HTML]{DAE8FC}56.2K (0.94)} & \multicolumn{1}{c|}{\cellcolor[HTML]{DAE8FC}56.2K (0.94)} & \cellcolor[HTML]{DAE8FC}55.9K (0.93) \\
 & H5AN8G8NDJR-XNC [m4] & R-DIMM & D & 8 Gb & x8 & 2048 & H4 & \multicolumn{1}{c|}{11.7K (1.00)} & \multicolumn{1}{c|}{11.7K (1.00)} & \multicolumn{1}{c|}{11.7K (1.00)} & \multicolumn{1}{c|}{11.7K (1.00)} & \multicolumn{1}{c|}{11.7K (1.00)} & \multicolumn{1}{c|}{10.2K (0.87)} & \cellcolor[HTML]{FFCCC9}0 (0) \\
 & \cellcolor[HTML]{DAE8FC}H5AN8G8NDJR-XNC [m4] & \cellcolor[HTML]{DAE8FC}R-DIMM & \cellcolor[HTML]{DAE8FC}D & \cellcolor[HTML]{DAE8FC}8 Gb & \cellcolor[HTML]{DAE8FC}x8 & \cellcolor[HTML]{DAE8FC}2048 & \cellcolor[HTML]{DAE8FC}H5 & \multicolumn{1}{c|}{\cellcolor[HTML]{DAE8FC}10.2K (1.00)} & \multicolumn{1}{c|}{\cellcolor[HTML]{DAE8FC}10.9K (1.08)} & \multicolumn{1}{c|}{\cellcolor[HTML]{DAE8FC}10.2K (1.00)} & \multicolumn{1}{c|}{\cellcolor[HTML]{DAE8FC}10.9K (1.08)} & \multicolumn{1}{c|}{\cellcolor[HTML]{DAE8FC}10.2K (1.00)} & \multicolumn{1}{c|}{\cellcolor[HTML]{DAE8FC}10.2K (1.00)} & \cellcolor[HTML]{FFCCC9}0 (0) \\
 & H5AN8G4N/AFR-VKC [m5] & R-DIMM & A & 8 Gb & x4 & N/A & H6 & \multicolumn{1}{c|}{23.8K (1.00)} & \multicolumn{1}{c|}{23.8K (1.00)} & \multicolumn{1}{c|}{23.8K (1.00)} & \multicolumn{1}{c|}{23.4K (0.98)} & \multicolumn{1}{c|}{22.3K (0.93)} & \multicolumn{1}{c|}{22.3K (0.93)} & 18.0K (0.75) \\
 & \cellcolor[HTML]{DAE8FC}H5AN/AG8NCJR-XNC [m6] & \cellcolor[HTML]{DAE8FC}U-DIMM & \cellcolor[HTML]{DAE8FC}C & \cellcolor[HTML]{DAE8FC}16 Gb & \cellcolor[HTML]{DAE8FC}x8 & \cellcolor[HTML]{DAE8FC}2136 & \cellcolor[HTML]{DAE8FC}H7 & \multicolumn{1}{c|}{\cellcolor[HTML]{DAE8FC}8.6K (1.00)} & \multicolumn{1}{c|}{\cellcolor[HTML]{DAE8FC}8.6K (1.00)} & \multicolumn{1}{c|}{\cellcolor[HTML]{DAE8FC}7.8K (0.91)} & \multicolumn{1}{c|}{\cellcolor[HTML]{DAE8FC}8.6K (1.00)} & \multicolumn{1}{c|}{\cellcolor[HTML]{DAE8FC}8.6K (1.00)} & \multicolumn{1}{c|}{\cellcolor[HTML]{DAE8FC}7.0K (0.82)} & \cellcolor[HTML]{FFCCC9}0 (0) \\
\multirow{-9}{*}{\begin{tabular}[c]{@{}c@{}}Mfr. H\\ (SK Hynix)\end{tabular}} & H5AN/AG8NCJR-XNC [m6] & U-DIMM & C & 16 Gb & x8 & 2136 & H8 & \multicolumn{1}{c|}{10.5K (1.00)} & \multicolumn{1}{c|}{10.5K (1.00)} & \multicolumn{1}{c|}{10.2K (0.96)} & \multicolumn{1}{c|}{8.6K (0.81)} & \multicolumn{1}{c|}{8.6K (0.81)} & \multicolumn{1}{c|}{7.8K (0.74)} & \cellcolor[HTML]{FFCCC9}0 (0) \\ \hline
 & \cellcolor[HTML]{DAE8FC}MT40A2G4WE-083E:B [m7] & \cellcolor[HTML]{DAE8FC}R-DIMM & \cellcolor[HTML]{DAE8FC}B & \cellcolor[HTML]{DAE8FC}8 Gb & \cellcolor[HTML]{DAE8FC}x4 & \cellcolor[HTML]{DAE8FC}N/A & \cellcolor[HTML]{DAE8FC}M0 & \multicolumn{1}{c|}{\cellcolor[HTML]{DAE8FC}43.8K (1.00)} & \multicolumn{1}{c|}{\cellcolor[HTML]{DAE8FC}44.5K (1.02)} & \multicolumn{1}{c|}{\cellcolor[HTML]{DAE8FC}44.5K (1.02)} & \multicolumn{1}{c|}{\cellcolor[HTML]{DAE8FC}44.5K (1.02)} & \multicolumn{1}{c|}{\cellcolor[HTML]{DAE8FC}44.5K (1.02)} & \multicolumn{1}{c|}{\cellcolor[HTML]{DAE8FC}44.5K (1.02)} & \cellcolor[HTML]{DAE8FC}44.5K (1.02) \\
 & MT40A2G4WE-083E:B [m7] & R-DIMM & B & 8 Gb & x4 & N/A & M1 & \multicolumn{1}{c|}{37.1K (1.00)} & \multicolumn{1}{c|}{37.9K (1.02)} & \multicolumn{1}{c|}{37.9K (1.02)} & \multicolumn{1}{c|}{37.9K (1.02)} & \multicolumn{1}{c|}{37.9K (1.02)} & \multicolumn{1}{c|}{37.9K (1.02)} & 37.9K (1.02) \\
 & \cellcolor[HTML]{DAE8FC}MT40A2G4WE-083E:B [m7] & \cellcolor[HTML]{DAE8FC}R-DIMM & \cellcolor[HTML]{DAE8FC}B & \cellcolor[HTML]{DAE8FC}8 Gb & \cellcolor[HTML]{DAE8FC}x4 & \cellcolor[HTML]{DAE8FC}N/A & \cellcolor[HTML]{DAE8FC}M2 & \multicolumn{1}{c|}{\cellcolor[HTML]{DAE8FC}42.6K (1.00)} & \multicolumn{1}{c|}{\cellcolor[HTML]{DAE8FC}43.8K (1.03)} & \multicolumn{1}{c|}{\cellcolor[HTML]{DAE8FC}44.1K (1.04)} & \multicolumn{1}{c|}{\cellcolor[HTML]{DAE8FC}44.1K (1.04)} & \multicolumn{1}{c|}{\cellcolor[HTML]{DAE8FC}44.1K (1.04)} & \multicolumn{1}{c|}{\cellcolor[HTML]{DAE8FC}44.1K (1.04)} & \cellcolor[HTML]{DAE8FC}44.1K (1.04) \\
 & MT40A2G8SA-062E:F [m8] & SO-DIMM & F & 16 Gb & x8 & 2237 & M3 & \multicolumn{1}{c|}{6.2K (1.00)} & \multicolumn{1}{c|}{6.2K (1.00)} & \multicolumn{1}{c|}{6.2K (1.00)} & \multicolumn{1}{c|}{6.2K (1.00)} & \multicolumn{1}{c|}{6.2K (1.00)} & \multicolumn{1}{c|}{6.2K (1.00)} & 6.2K (1.00) \\
 & \cellcolor[HTML]{DAE8FC}MT40A1G16KD-062E:E [m9] & \cellcolor[HTML]{DAE8FC}SO-DIMM & \cellcolor[HTML]{DAE8FC}E & \cellcolor[HTML]{DAE8FC}16 Gb & \cellcolor[HTML]{DAE8FC}x16 & \cellcolor[HTML]{DAE8FC}2046 & \cellcolor[HTML]{DAE8FC}M4 & \multicolumn{1}{c|}{\cellcolor[HTML]{DAE8FC}5.1K (1.00)} & \multicolumn{1}{c|}{\cellcolor[HTML]{DAE8FC}5.1K (1.00)} & \multicolumn{1}{c|}{\cellcolor[HTML]{DAE8FC}5.1K (1.00)} & \multicolumn{1}{c|}{\cellcolor[HTML]{DAE8FC}5.1K (1.00)} & \multicolumn{1}{c|}{\cellcolor[HTML]{DAE8FC}5.1K (1.00)} & \multicolumn{1}{c|}{\cellcolor[HTML]{DAE8FC}5.1K (1.00)} & \cellcolor[HTML]{DAE8FC}5.1K (1.00) \\
 & MT40A4G4JC-062E:E [m10] & R-DIMM & E & 16 Gb & x4 & 2014 & M5 & \multicolumn{1}{c|}{5.9K (1.00)} & \multicolumn{1}{c|}{5.9K (1.00)} & \multicolumn{1}{c|}{5.9K (1.00)} & \multicolumn{1}{c|}{5.9K (1.00)} & \multicolumn{1}{c|}{5.9K (1.00)} & \multicolumn{1}{c|}{5.9K (1.00)} & 5.5K (0.93) \\
\multirow{-7}{*}{\begin{tabular}[c]{@{}c@{}}Mfr. M\\ (Micron)\end{tabular}} & \cellcolor[HTML]{DAE8FC}MT40A1G16RC-062E:B [m11] & \cellcolor[HTML]{DAE8FC}SO-DIMM & \cellcolor[HTML]{DAE8FC}B & \cellcolor[HTML]{DAE8FC}16 Gb & \cellcolor[HTML]{DAE8FC}x16 & \cellcolor[HTML]{DAE8FC}2126 & \cellcolor[HTML]{DAE8FC}M6 & \multicolumn{1}{c|}{\cellcolor[HTML]{DAE8FC}13.3K (1.00)} & \multicolumn{1}{c|}{\cellcolor[HTML]{DAE8FC}13.3K (1.00)} & \multicolumn{1}{c|}{\cellcolor[HTML]{DAE8FC}13.3K (1.00)} & \multicolumn{1}{c|}{\cellcolor[HTML]{DAE8FC}13.3K (1.00)} & \multicolumn{1}{c|}{\cellcolor[HTML]{DAE8FC}13.3K (1.00)} & \multicolumn{1}{c|}{\cellcolor[HTML]{DAE8FC}13.3K (1.00)} & \cellcolor[HTML]{DAE8FC}13.3K (1.00) \\ \hline
 & K4A4G085WF-BCTD [m12] & U-DIMM & F & 4 Gb & x8 & N/A & S0 & \multicolumn{1}{c|}{12.5K (1.00)} & \multicolumn{1}{c|}{11.7K (0.94)} & \multicolumn{1}{c|}{12.5K (1.00)} & \multicolumn{1}{c|}{11.7K (0.94)} & \multicolumn{1}{c|}{10.2K (0.81)} & \multicolumn{1}{c|}{6.2K (0.50)} & \cellcolor[HTML]{FFCCC9}0 (0) \\
 & \cellcolor[HTML]{DAE8FC}K4A4G085WF-BCTD [m12] & \cellcolor[HTML]{DAE8FC}U-DIMM & \cellcolor[HTML]{DAE8FC}F & \cellcolor[HTML]{DAE8FC}4 Gb & \cellcolor[HTML]{DAE8FC}x8 & \cellcolor[HTML]{DAE8FC}N/A & \cellcolor[HTML]{DAE8FC}S1 & \multicolumn{1}{c|}{\cellcolor[HTML]{DAE8FC}14.1K (1.00)} & \multicolumn{1}{c|}{\cellcolor[HTML]{DAE8FC}14.1K (1.00)} & \multicolumn{1}{c|}{\cellcolor[HTML]{DAE8FC}12.9K (0.92)} & \multicolumn{1}{c|}{\cellcolor[HTML]{DAE8FC}10.9K (0.78)} & \multicolumn{1}{c|}{\cellcolor[HTML]{DAE8FC}9.8K (0.69)} & \multicolumn{1}{c|}{\cellcolor[HTML]{DAE8FC}7.0K (0.50)} & \cellcolor[HTML]{FFCCC9}0 (0) \\
 & K4A4G085WE-BCPB [m13] & SO-DIMM & E & 4 Gb & x8 & 1708 & S2 & \multicolumn{1}{c|}{25.8K (1.00)} & \multicolumn{1}{c|}{26.2K (1.02)} & \multicolumn{1}{c|}{25.0K (0.97)} & \multicolumn{1}{c|}{24.2K (0.94)} & \multicolumn{1}{c|}{22.7K (0.88)} & \multicolumn{1}{c|}{19.9K (0.77)} & 5.1K (0.20) \\
 & \cellcolor[HTML]{DAE8FC}K4A4G085WE-BCPB [m13] & \cellcolor[HTML]{DAE8FC}SO-DIMM & \cellcolor[HTML]{DAE8FC}E & \cellcolor[HTML]{DAE8FC}4 Gb & \cellcolor[HTML]{DAE8FC}x8 & \cellcolor[HTML]{DAE8FC}1708 & \cellcolor[HTML]{DAE8FC}S3 & \multicolumn{1}{c|}{\cellcolor[HTML]{DAE8FC}21.9K (1.00)} & \multicolumn{1}{c|}{\cellcolor[HTML]{DAE8FC}21.9K (1.00)} & \multicolumn{1}{c|}{\cellcolor[HTML]{DAE8FC}21.9K (1.00)} & \multicolumn{1}{c|}{\cellcolor[HTML]{DAE8FC}20.3K (0.93)} & \multicolumn{1}{c|}{\cellcolor[HTML]{DAE8FC}19.5K (0.89)} & \multicolumn{1}{c|}{\cellcolor[HTML]{DAE8FC}17.6K (0.80)} & \cellcolor[HTML]{FFCCC9}0 (0) \\
 & K4A4G085WE-BCPB [m13] & SO-DIMM & E & 4 Gb & x8 & 1708 & S4 & \multicolumn{1}{c|}{25.0K (1.00)} & \multicolumn{1}{c|}{25.0K (1.00)} & \multicolumn{1}{c|}{25.0K (1.00)} & \multicolumn{1}{c|}{24.6K (0.98)} & \multicolumn{1}{c|}{21.5K (0.86)} & \multicolumn{1}{c|}{\cellcolor[HTML]{FFCCC9}0 (0)} & \cellcolor[HTML]{FFCCC9}0 (0) \\
 & \cellcolor[HTML]{DAE8FC}Unknown & \cellcolor[HTML]{DAE8FC}SO-DIMM & \cellcolor[HTML]{DAE8FC}C & \cellcolor[HTML]{DAE8FC}4 Gb & \cellcolor[HTML]{DAE8FC}x16 & \cellcolor[HTML]{DAE8FC}N/A & \cellcolor[HTML]{DAE8FC}S5 & \multicolumn{1}{c|}{\cellcolor[HTML]{DAE8FC}11.3K (1.00)} & \multicolumn{1}{c|}{\cellcolor[HTML]{DAE8FC}10.2K (0.90)} & \multicolumn{1}{c|}{\cellcolor[HTML]{DAE8FC}10.5K (0.93)} & \multicolumn{1}{c|}{\cellcolor[HTML]{DAE8FC}10.2K (0.90)} & \multicolumn{1}{c|}{\cellcolor[HTML]{DAE8FC}9.8K (0.86)} & \multicolumn{1}{c|}{\cellcolor[HTML]{DAE8FC}9.0K (0.79)} & \cellcolor[HTML]{FFCCC9}0 (0) \\
 & K4A8G085WD-BCTD [m14] & U-DIMM & D & 8 Gb & x8 & 2110 & S6 & \multicolumn{1}{c|}{7.8K (1.00)} & \multicolumn{1}{c|}{7.0K (0.90)} & \multicolumn{1}{c|}{7.0K (0.90)} & \multicolumn{1}{c|}{7.0K (0.90)} & \multicolumn{1}{c|}{6.2K (0.80)} & \multicolumn{1}{c|}{3.9K (0.50)} & \cellcolor[HTML]{FFCCC9}0 (0) \\
 & \cellcolor[HTML]{DAE8FC}K4A8G085WD-BCTD [m14] & \cellcolor[HTML]{DAE8FC}U-DIMM & \cellcolor[HTML]{DAE8FC}D & \cellcolor[HTML]{DAE8FC}8 Gb & \cellcolor[HTML]{DAE8FC}x8 & \cellcolor[HTML]{DAE8FC}2110 & \cellcolor[HTML]{DAE8FC}S7 & \multicolumn{1}{c|}{\cellcolor[HTML]{DAE8FC}7.8K (1.00)} & \multicolumn{1}{c|}{\cellcolor[HTML]{DAE8FC}7.8K (1.00)} & \multicolumn{1}{c|}{\cellcolor[HTML]{DAE8FC}7.0K (0.90)} & \multicolumn{1}{c|}{\cellcolor[HTML]{DAE8FC}6.2K (0.80)} & \multicolumn{1}{c|}{\cellcolor[HTML]{DAE8FC}5.5K (0.70)} & \multicolumn{1}{c|}{\cellcolor[HTML]{DAE8FC}3.9K (0.50)} & \cellcolor[HTML]{FFCCC9}0 (0) \\
 & K4A8G085WD-BCTD [m14] & U-DIMM & D & 8 Gb & x8 & 2110 & S8 & \multicolumn{1}{c|}{7.8K (1.00)} & \multicolumn{1}{c|}{6.6K (0.85)} & \multicolumn{1}{c|}{7.8K (1.00)} & \multicolumn{1}{c|}{6.2K (0.80)} & \multicolumn{1}{c|}{5.1K (0.65)} & \multicolumn{1}{c|}{3.9K (0.50)} & \cellcolor[HTML]{FFCCC9}0 (0) \\
 & \cellcolor[HTML]{DAE8FC}K4A8G085WD-BCTD [m14] & \cellcolor[HTML]{DAE8FC}U-DIMM & \cellcolor[HTML]{DAE8FC}D & \cellcolor[HTML]{DAE8FC}8 Gb & \cellcolor[HTML]{DAE8FC}x8 & \cellcolor[HTML]{DAE8FC}2110 & \cellcolor[HTML]{DAE8FC}S9 & \multicolumn{1}{c|}{\cellcolor[HTML]{DAE8FC}7.8K (1.00)} & \multicolumn{1}{c|}{\cellcolor[HTML]{DAE8FC}7.8K (1.00)} & \multicolumn{1}{c|}{\cellcolor[HTML]{DAE8FC}7.8K (1.00)} & \multicolumn{1}{c|}{\cellcolor[HTML]{DAE8FC}6.6K (0.85)} & \multicolumn{1}{c|}{\cellcolor[HTML]{DAE8FC}6.2K (0.80)} & \multicolumn{1}{c|}{\cellcolor[HTML]{DAE8FC}3.9K (0.50)} & \cellcolor[HTML]{FFCCC9}0 (0) \\
 & K4A8G085WC-BCRC [m15] & R-DIMM & C & 8 Gb & x8 & 1809 & S10 & \multicolumn{1}{c|}{14.1K (1.00)} & \multicolumn{1}{c|}{14.1K (1.00)} & \multicolumn{1}{c|}{14.1K (1.00)} & \multicolumn{1}{c|}{13.3K (0.94)} & \multicolumn{1}{c|}{12.5K (0.89)} & \multicolumn{1}{c|}{10.2K (0.72)} & \cellcolor[HTML]{FFCCC9}0 (0) \\
 & \cellcolor[HTML]{DAE8FC}K4A8G085WB-BCTD [m16] & \cellcolor[HTML]{DAE8FC}R-DIMM & \cellcolor[HTML]{DAE8FC}B & \cellcolor[HTML]{DAE8FC}8 Gb & \cellcolor[HTML]{DAE8FC}x8 & \cellcolor[HTML]{DAE8FC}2053 & \cellcolor[HTML]{DAE8FC}S11 & \multicolumn{1}{c|}{\cellcolor[HTML]{DAE8FC}28.1K (1.00)} & \multicolumn{1}{c|}{\cellcolor[HTML]{DAE8FC}28.9K (1.03)} & \multicolumn{1}{c|}{\cellcolor[HTML]{DAE8FC}28.1K (1.00)} & \multicolumn{1}{c|}{\cellcolor[HTML]{DAE8FC}26.6K (0.94)} & \multicolumn{1}{c|}{\cellcolor[HTML]{DAE8FC}27.3K (0.97)} & \multicolumn{1}{c|}{\cellcolor[HTML]{FFCCC9}0 (0)} & \cellcolor[HTML]{FFCCC9}0 (0) \\
 & K4AAG085WA-BCWE [m17] & U-DIMM & A & 8 Gb & x8 & 2212 & S12 & \multicolumn{1}{c|}{9.0K (1.00)} & \multicolumn{1}{c|}{8.2K (0.91)} & \multicolumn{1}{c|}{7.8K (0.87)} & \multicolumn{1}{c|}{9.0K (1.00)} & \multicolumn{1}{c|}{7.0K (0.78)} & \multicolumn{1}{c|}{\cellcolor[HTML]{FFCCC9}0 (0)} & \cellcolor[HTML]{FFCCC9}0 (0) \\
\multirow{-14}{*}{\begin{tabular}[c]{@{}c@{}}Mfr. S\\ (Samsung)\end{tabular}} & \cellcolor[HTML]{DAE8FC}Unknown & \cellcolor[HTML]{DAE8FC}U-DIMM & \cellcolor[HTML]{DAE8FC}B & \cellcolor[HTML]{DAE8FC}16 Gb & \cellcolor[HTML]{DAE8FC}x8 & \cellcolor[HTML]{DAE8FC}2315 & \cellcolor[HTML]{DAE8FC}S13 & \multicolumn{1}{c|}{\cellcolor[HTML]{DAE8FC}7.0K (1.00)} & \multicolumn{1}{c|}{\cellcolor[HTML]{DAE8FC}7.8K (1.11)} & \multicolumn{1}{c|}{\cellcolor[HTML]{DAE8FC}7.0K (1.00)} & \multicolumn{1}{c|}{\cellcolor[HTML]{DAE8FC}6.6K (0.94)} & \multicolumn{1}{c|}{\cellcolor[HTML]{DAE8FC}7.0K (1.00)} & \multicolumn{1}{c|}{\cellcolor[HTML]{DAE8FC}5.9K (0.83)} & \cellcolor[HTML]{FFCCC9}0 (0) \\ \hline
\end{tabular}%
}
\end{table}

\footnotetext{We report the die revision marked on the DRAM chip package (if available). A die revision of “X” means the original markings on the DRAM chips\om{11}{'} package are removed by the DRAM module vendor, and thus the die revision could not be identified. In most cases, we report the date code of a DRAM module in the WWYY format (i.e., 2048 means the module is manufactured in the 48th week of year 2020) as marked on the label of the module. We report “N/A” if no date is marked on the label of a module.\label{fn:dierev}}

\clearpage

\yctcomment{11}{The reason why NAs in tab4 and 0(0) in fig3 was not the same was I had filtered low NPCR values but I fixed it now and they match.}

\begin{table}[ht!]
\renewcommand{\arraystretch}{1.2}
\setlength{\tabcolsep}{2pt}
\caption{Summary of all tested DDR4 DRAM modules\fnref{fn:dierev} and their \X{} configuration parameters (RowHammer threshold, $\bm{N}_{\bm{RH}}$; the maximum number of consecutive partial charge restorations, $\bm{N}_{\bm{PCR}}$; full charge restoration interval, $\bm{t}_{\bm{FCRI}}$) for different charge restoration latencies\yct{11}{. Red cells denote the configurations where \X{} is not applicable due to i) no bitflips observed in the module (e.g., H0) or ii) reducing charge restoration latency causes bitflips without hammering (i.e., $\bm{N}_{\bm{RH}}\bm{=0}$)}}\label{tab:pacramparams}
\resizebox{\columnwidth}{!}{%
\begin{tabular}{|c|ccccccc?cccccccccccccccccc|}
\hline
 &  &  &  &  &  &  &  & \multicolumn{18}{c|}{\textbf{PaCRAM parameters with $\bm{t}_{\bm{RAS(Red)}}\bm{=Mt}_{\bm{RAS(Nom)}}$}} \\ \cline{9-26} 
 &  &  &  &  &  &  &  & \multicolumn{3}{c|}{\textbf{$\bm{M=0.81}$ $\bm{(27ns)}$}} & \multicolumn{3}{c|}{\textbf{$\bm{M=0.64}$ $\bm{(21ns)}$}} & \multicolumn{3}{c|}{\textbf{$\bm{M=0.45}$ $\bm{(15ns)}$}} & \multicolumn{3}{c|}{\textbf{$\bm{M=0.36}$ $\bm{(12ns)}$}} & \multicolumn{3}{c|}{\textbf{$\bm{M=0.27}$ $\bm{(9ns)}$}} & \multicolumn{3}{c|}{\textbf{$\bm{M=0.18}$ $\bm{(6ns)}$}} \\ \cline{9-26} 
\multirow{-3}{*}{\textbf{Mfr.}} & \multirow{-3}{*}{\textbf{DRAM Part}} & \multirow{-3}{*}{\textbf{\begin{tabular}[c]{@{}c@{}}Form\\ Factor\end{tabular}}} & \multirow{-3}{*}{\textbf{\begin{tabular}[c]{@{}c@{}}Die\\ Rev.\end{tabular}}} & \multirow{-3}{*}{\textbf{\begin{tabular}[c]{@{}c@{}}Chip\\ Density\end{tabular}}} & \multirow{-3}{*}{\textbf{DQ}} & \multirow{-3}{*}{\textbf{\begin{tabular}[c]{@{}c@{}}Date\\ Code\end{tabular}}} & \multirow{-3}{*}{\textbf{ID}} & $\bm{N}_{\bm{RH}}$ & $\bm{N}_{\bm{PCR}}$ & \multicolumn{1}{c|}{$\bm{t}_{\bm{FCRI}}$} & $\bm{N}_{\bm{RH}}$ & $\bm{N}_{\bm{PCR}}$ & \multicolumn{1}{c|}{$\bm{t}_{\bm{FCRI}}$} & $\bm{N}_{\bm{RH}}$ & $\bm{N}_{\bm{PCR}}$ & \multicolumn{1}{c|}{$\bm{t}_{\bm{FCRI}}$} & $\bm{N}_{\bm{RH}}$ & $\bm{N}_{\bm{PCR}}$ & \multicolumn{1}{c|}{$\bm{t}_{\bm{FCRI}}$} & $\bm{N}_{\bm{RH}}$ & $\bm{N}_{\bm{PCR}}$ & \multicolumn{1}{c|}{$\bm{t}_{\bm{FCRI}}$} & $\bm{N}_{\bm{RH}}$ & $\bm{N}_{\bm{PCR}}$ & $\bm{t}_{\bm{FCRI}}$ \\ \specialrule{2.5pt}{1pt}{1pt}
 & H5AN4G8NMFR-TFC [m1] & SO-DIMM & M & 4 Gb & x8 & N/A & H0 & \cellcolor[HTML]{FFCCC9} & \cellcolor[HTML]{FFCCC9}$N/A$ & \multicolumn{1}{c|}{\cellcolor[HTML]{FFCCC9}} & \cellcolor[HTML]{FFCCC9} & \cellcolor[HTML]{FFCCC9}$N/A$ & \multicolumn{1}{c|}{\cellcolor[HTML]{FFCCC9}} & \cellcolor[HTML]{FFCCC9} & \cellcolor[HTML]{FFCCC9}$N/A$ & \multicolumn{1}{c|}{\cellcolor[HTML]{FFCCC9}} & \cellcolor[HTML]{FFCCC9} & \cellcolor[HTML]{FFCCC9}$N/A$ & \multicolumn{1}{c|}{\cellcolor[HTML]{FFCCC9}} & \cellcolor[HTML]{FFCCC9} & \cellcolor[HTML]{FFCCC9}$N/A$ & \multicolumn{1}{c|}{\cellcolor[HTML]{FFCCC9}} & \cellcolor[HTML]{FFCCC9} & \cellcolor[HTML]{FFCCC9}$N/A$ & \cellcolor[HTML]{FFCCC9} \\
 & \cellcolor[HTML]{DAE8FC}Unknown & \cellcolor[HTML]{DAE8FC}SO-DIMM & \cellcolor[HTML]{DAE8FC}X & \cellcolor[HTML]{DAE8FC}4 Gb & \cellcolor[HTML]{DAE8FC}x8 & \cellcolor[HTML]{DAE8FC}N/A & \cellcolor[HTML]{DAE8FC}H1 & \cellcolor[HTML]{DAE8FC}$50.0K$ & \cellcolor[HTML]{DAE8FC}$15.0K$ & \multicolumn{1}{c|}{\cellcolor[HTML]{DAE8FC}$36.0s$} & \cellcolor[HTML]{DAE8FC}$49.6K$ & \cellcolor[HTML]{DAE8FC}$15.0K$ & \multicolumn{1}{c|}{\cellcolor[HTML]{DAE8FC}$35.7s$} & \cellcolor[HTML]{DAE8FC}$50.0K$ & \cellcolor[HTML]{DAE8FC}$15.0K$ & \multicolumn{1}{c|}{\cellcolor[HTML]{DAE8FC}$36.0s$} & \cellcolor[HTML]{DAE8FC}$50.0K$ & \cellcolor[HTML]{DAE8FC}$15.0K$ & \multicolumn{1}{c|}{\cellcolor[HTML]{DAE8FC}$36.0s$} & \cellcolor[HTML]{DAE8FC}$47.7K$ & \cellcolor[HTML]{DAE8FC}$15.0K$ & \multicolumn{1}{c|}{\cellcolor[HTML]{DAE8FC}$34.3s$} & \cellcolor[HTML]{DAE8FC}$44.1K$ & \cellcolor[HTML]{DAE8FC}$1$ & \cellcolor[HTML]{DAE8FC}$2ms$ \\
 & H5AN4G8NAFR-TFC [m2] & SO-DIMM & A & 4 Gb & x8 & N/A & H2 & $34.8K$ & $15.0K$ & \multicolumn{1}{c|}{$25.0s$} & $34.8K$ & $15.0K$ & \multicolumn{1}{c|}{$25.0s$} & $34.8K$ & $15.0K$ & \multicolumn{1}{c|}{$25.0s$} & $34.8K$ & $15.0K$ & \multicolumn{1}{c|}{$25.0s$} & $34.4K$ & $15.0K$ & \multicolumn{1}{c|}{$24.8s$} & \cellcolor[HTML]{FFFFFF}$37.9K$ & \cellcolor[HTML]{FFFFFF}$1$ & \cellcolor[HTML]{FFFFFF}$1ms$ \\
 & \cellcolor[HTML]{DAE8FC}H5AN8G4NMFR-UKC [m3] & \cellcolor[HTML]{DAE8FC}R-DIMM & \cellcolor[HTML]{DAE8FC}M & \cellcolor[HTML]{DAE8FC}8 Gb & \cellcolor[HTML]{DAE8FC}x4 & \cellcolor[HTML]{DAE8FC}N/A & \cellcolor[HTML]{DAE8FC}H3 & \cellcolor[HTML]{DAE8FC}$56.2K$ & \cellcolor[HTML]{DAE8FC}$15.0K$ & \multicolumn{1}{c|}{\cellcolor[HTML]{DAE8FC}$40.5s$} & \cellcolor[HTML]{DAE8FC}$57.0K$ & \cellcolor[HTML]{DAE8FC}$15.0K$ & \multicolumn{1}{c|}{\cellcolor[HTML]{DAE8FC}$41.1s$} & \cellcolor[HTML]{DAE8FC}$56.2K$ & \cellcolor[HTML]{DAE8FC}$15.0K$ & \multicolumn{1}{c|}{\cellcolor[HTML]{DAE8FC}$40.5s$} & \cellcolor[HTML]{DAE8FC}$56.2K$ & \cellcolor[HTML]{DAE8FC}$15.0K$ & \multicolumn{1}{c|}{\cellcolor[HTML]{DAE8FC}$40.5s$} & \cellcolor[HTML]{DAE8FC}$56.2K$ & \cellcolor[HTML]{DAE8FC}$15.0K$ & \multicolumn{1}{c|}{\cellcolor[HTML]{DAE8FC}$40.5s$} & \cellcolor[HTML]{DAE8FC}$55.9K$ & \cellcolor[HTML]{DAE8FC}$1$ & \cellcolor[HTML]{DAE8FC}$2ms$ \\
 & H5AN8G8NDJR-XNC [m4] & R-DIMM & D & 8 Gb & x8 & 2048 & H4 & $10.9K$ & $15.0K$ & \multicolumn{1}{c|}{$7.9s$} & $10.9K$ & $15.0K$ & \multicolumn{1}{c|}{$7.9s$} & $10.9K$ & $15.0K$ & \multicolumn{1}{c|}{$7.9s$} & $10.9K$ & $15.0K$ & \multicolumn{1}{c|}{$7.9s$} & \cellcolor[HTML]{FFFFFF}$10.2K$ & \cellcolor[HTML]{FFFFFF}$1$ & \multicolumn{1}{c|}{\cellcolor[HTML]{FFFFFF}$489\mu s$} & \cellcolor[HTML]{FFCCC9} & \cellcolor[HTML]{FFCCC9}$N/A$ & \cellcolor[HTML]{FFCCC9} \\
 & \cellcolor[HTML]{DAE8FC}H5AN8G8NDJR-XNC [m4] & \cellcolor[HTML]{DAE8FC}R-DIMM & \cellcolor[HTML]{DAE8FC}D & \cellcolor[HTML]{DAE8FC}8 Gb & \cellcolor[HTML]{DAE8FC}x8 & \cellcolor[HTML]{DAE8FC}2048 & \cellcolor[HTML]{DAE8FC}H5 & \cellcolor[HTML]{DAE8FC}$10.2K$ & \cellcolor[HTML]{DAE8FC}$15.0K$ & \multicolumn{1}{c|}{\cellcolor[HTML]{DAE8FC}$7.3s$} & \cellcolor[HTML]{DAE8FC}$10.2K$ & \cellcolor[HTML]{DAE8FC}$15.0K$ & \multicolumn{1}{c|}{\cellcolor[HTML]{DAE8FC}$7.3s$} & \cellcolor[HTML]{DAE8FC}$10.2K$ & \cellcolor[HTML]{DAE8FC}$15.0K$ & \multicolumn{1}{c|}{\cellcolor[HTML]{DAE8FC}$7.3s$} & \cellcolor[HTML]{DAE8FC}$10.2K$ & \cellcolor[HTML]{DAE8FC}$15.0K$ & \multicolumn{1}{c|}{\cellcolor[HTML]{DAE8FC}$7.3s$} & \cellcolor[HTML]{DAE8FC}$9.4K$ & \cellcolor[HTML]{DAE8FC}$300$ & \multicolumn{1}{c|}{\cellcolor[HTML]{DAE8FC}$135ms$} & \cellcolor[HTML]{FFCCC9} & \cellcolor[HTML]{FFCCC9}$N/A$ & \cellcolor[HTML]{FFCCC9} \\
 & H5AN8G4NAFR-VKC [m5] & R-DIMM & A & 8 Gb & x4 & N/A & H6 & $22.7K$ & $15.0K$ & \multicolumn{1}{c|}{$16.3s$} & $22.7K$ & $15.0K$ & \multicolumn{1}{c|}{$16.3s$} & $22.7K$ & $15.0K$ & \multicolumn{1}{c|}{$16.3s$} & $22.3K$ & $15.0K$ & \multicolumn{1}{c|}{$16.0s$} & $22.3K$ & $15.0K$ & \multicolumn{1}{c|}{$16.0s$} & \cellcolor[HTML]{FFFFFF}$18.0K$ & \cellcolor[HTML]{FFFFFF}$1$ & \cellcolor[HTML]{FFFFFF}$864\mu s$ \\
 & \cellcolor[HTML]{DAE8FC}H5ANAG8NCJR-XNC [m6] & \cellcolor[HTML]{DAE8FC}U-DIMM & \cellcolor[HTML]{DAE8FC}C & \cellcolor[HTML]{DAE8FC}16 Gb & \cellcolor[HTML]{DAE8FC}x8 & \cellcolor[HTML]{DAE8FC}2136 & \cellcolor[HTML]{DAE8FC}H7 & \cellcolor[HTML]{DAE8FC}$8.6K$ & \cellcolor[HTML]{DAE8FC}$15.0K$ & \multicolumn{1}{c|}{\cellcolor[HTML]{DAE8FC}$6.2s$} & \cellcolor[HTML]{DAE8FC}$7.8K$ & \cellcolor[HTML]{DAE8FC}$15.0K$ & \multicolumn{1}{c|}{\cellcolor[HTML]{DAE8FC}$5.6s$} & \cellcolor[HTML]{DAE8FC}$7.8K$ & \cellcolor[HTML]{DAE8FC}$15.0K$ & \multicolumn{1}{c|}{\cellcolor[HTML]{DAE8FC}$5.6s$} & \cellcolor[HTML]{DAE8FC}$7.8K$ & \cellcolor[HTML]{DAE8FC}$15.0K$ & \multicolumn{1}{c|}{\cellcolor[HTML]{DAE8FC}$5.6s$} & \cellcolor[HTML]{DAE8FC}$6.2K$ & \cellcolor[HTML]{DAE8FC}$15.0K$ & \multicolumn{1}{c|}{\cellcolor[HTML]{DAE8FC}$4.5s$} & \cellcolor[HTML]{FFCCC9} & \cellcolor[HTML]{FFCCC9}$N/A$ & \cellcolor[HTML]{FFCCC9} \\
\multirow{-9}{*}{\begin{tabular}[c]{@{}c@{}}Mfr. H\\ (SK Hynix)\end{tabular}} & H5ANAG8NCJR-XNC [m6] & U-DIMM & C & 16 Gb & x8 & 2136 & H8 & $7.8K$ & $15.0K$ & \multicolumn{1}{c|}{$5.6s$} & $7.8K$ & $15.0K$ & \multicolumn{1}{c|}{$5.6s$} & $7.8K$ & $15.0K$ & \multicolumn{1}{c|}{$5.6s$} & $7.8K$ & $15.0K$ & \multicolumn{1}{c|}{$5.6s$} & $6.2K$ & $15.0K$ & \multicolumn{1}{c|}{$4.5s$} & \cellcolor[HTML]{FFCCC9} & \cellcolor[HTML]{FFCCC9}$N/A$ & \cellcolor[HTML]{FFCCC9} \\ \hline
 & \cellcolor[HTML]{DAE8FC}MT40A2G4WE-083E:B [m7] & \cellcolor[HTML]{DAE8FC}R-DIMM & \cellcolor[HTML]{DAE8FC}B & \cellcolor[HTML]{DAE8FC}8 Gb & \cellcolor[HTML]{DAE8FC}x4 & \cellcolor[HTML]{DAE8FC}N/A & \cellcolor[HTML]{DAE8FC}M0 & \cellcolor[HTML]{DAE8FC}$43.8K$ & \cellcolor[HTML]{DAE8FC}$15.0K$ & \multicolumn{1}{c|}{\cellcolor[HTML]{DAE8FC}$31.5s$} & \cellcolor[HTML]{DAE8FC}$43.8K$ & \cellcolor[HTML]{DAE8FC}$15.0K$ & \multicolumn{1}{c|}{\cellcolor[HTML]{DAE8FC}$31.5s$} & \cellcolor[HTML]{DAE8FC}$43.8K$ & \cellcolor[HTML]{DAE8FC}$15.0K$ & \multicolumn{1}{c|}{\cellcolor[HTML]{DAE8FC}$31.5s$} & \cellcolor[HTML]{DAE8FC}$43.8K$ & \cellcolor[HTML]{DAE8FC}$15.0K$ & \multicolumn{1}{c|}{\cellcolor[HTML]{DAE8FC}$31.5s$} & \cellcolor[HTML]{DAE8FC}$43.8K$ & \cellcolor[HTML]{DAE8FC}$15.0K$ & \multicolumn{1}{c|}{\cellcolor[HTML]{DAE8FC}$31.5s$} & \cellcolor[HTML]{DAE8FC}$43.8K$ & \cellcolor[HTML]{DAE8FC}$15.0K$ & \cellcolor[HTML]{DAE8FC}$31.5s$ \\
 & MT40A2G4WE-083E:B [m7] & R-DIMM & B & 8 Gb & x4 & N/A & M1 & $43.4K$ & $15.0K$ & \multicolumn{1}{c|}{$31.2s$} & $40.6K$ & $15.0K$ & \multicolumn{1}{c|}{$29.3s$} & $39.5K$ & $15.0K$ & \multicolumn{1}{c|}{$28.4s$} & $39.1K$ & $15.0K$ & \multicolumn{1}{c|}{$28.1s$} & $40.6K$ & $15.0K$ & \multicolumn{1}{c|}{$29.3s$} & \cellcolor[HTML]{FFFFFF}$40.6K$ & \cellcolor[HTML]{FFFFFF}$15.0K$ & \cellcolor[HTML]{FFFFFF}$29.3s$ \\
 & \cellcolor[HTML]{DAE8FC}MT40A2G4WE-083E:B [m7] & \cellcolor[HTML]{DAE8FC}R-DIMM & \cellcolor[HTML]{DAE8FC}B & \cellcolor[HTML]{DAE8FC}8 Gb & \cellcolor[HTML]{DAE8FC}x4 & \cellcolor[HTML]{DAE8FC}N/A & \cellcolor[HTML]{DAE8FC}M2 & \cellcolor[HTML]{DAE8FC}$37.1K$ & \cellcolor[HTML]{DAE8FC}$15.0K$ & \multicolumn{1}{c|}{\cellcolor[HTML]{DAE8FC}$26.7s$} & \cellcolor[HTML]{DAE8FC}$37.1K$ & \cellcolor[HTML]{DAE8FC}$15.0K$ & \multicolumn{1}{c|}{\cellcolor[HTML]{DAE8FC}$26.7s$} & \cellcolor[HTML]{DAE8FC}$37.1K$ & \cellcolor[HTML]{DAE8FC}$15.0K$ & \multicolumn{1}{c|}{\cellcolor[HTML]{DAE8FC}$26.7s$} & \cellcolor[HTML]{DAE8FC}$37.1K$ & \cellcolor[HTML]{DAE8FC}$15.0K$ & \multicolumn{1}{c|}{\cellcolor[HTML]{DAE8FC}$26.7s$} & \cellcolor[HTML]{DAE8FC}$37.1K$ & \cellcolor[HTML]{DAE8FC}$15.0K$ & \multicolumn{1}{c|}{\cellcolor[HTML]{DAE8FC}$26.7s$} & \cellcolor[HTML]{DAE8FC}$37.1K$ & \cellcolor[HTML]{DAE8FC}$15.0K$ & \cellcolor[HTML]{DAE8FC}$26.7s$ \\
 & MT40A2G8SA-062E:F [m8] & SO-DIMM & F & 16 Gb & x8 & 2237 & M3 & $5.5K$ & $15.0K$ & \multicolumn{1}{c|}{$3.9s$} & $5.5K$ & $15.0K$ & \multicolumn{1}{c|}{$3.9s$} & $5.5K$ & $15.0K$ & \multicolumn{1}{c|}{$3.9s$} & $5.5K$ & $15.0K$ & \multicolumn{1}{c|}{$3.9s$} & $5.5K$ & $15.0K$ & \multicolumn{1}{c|}{$3.9s$} & \cellcolor[HTML]{FFFFFF}$5.5K$ & \cellcolor[HTML]{FFFFFF}$15.0K$ & \cellcolor[HTML]{FFFFFF}$3.9s$ \\
 & \cellcolor[HTML]{DAE8FC}MT40A1G16KD-062E:E [m9] & \cellcolor[HTML]{DAE8FC}SO-DIMM & \cellcolor[HTML]{DAE8FC}E & \cellcolor[HTML]{DAE8FC}16 Gb & \cellcolor[HTML]{DAE8FC}x16 & \cellcolor[HTML]{DAE8FC}2046 & \cellcolor[HTML]{DAE8FC}M4 & \cellcolor[HTML]{DAE8FC}$5.9K$ & \cellcolor[HTML]{DAE8FC}$15.0K$ & \multicolumn{1}{c|}{\cellcolor[HTML]{DAE8FC}$4.2s$} & \cellcolor[HTML]{DAE8FC}$5.5K$ & \cellcolor[HTML]{DAE8FC}$15.0K$ & \multicolumn{1}{c|}{\cellcolor[HTML]{DAE8FC}$3.9s$} & \cellcolor[HTML]{DAE8FC}$5.5K$ & \cellcolor[HTML]{DAE8FC}$15.0K$ & \multicolumn{1}{c|}{\cellcolor[HTML]{DAE8FC}$3.9s$} & \cellcolor[HTML]{DAE8FC}$5.5K$ & \cellcolor[HTML]{DAE8FC}$15.0K$ & \multicolumn{1}{c|}{\cellcolor[HTML]{DAE8FC}$3.9s$} & \cellcolor[HTML]{DAE8FC}$5.5K$ & \cellcolor[HTML]{DAE8FC}$15.0K$ & \multicolumn{1}{c|}{\cellcolor[HTML]{DAE8FC}$3.9s$} & \cellcolor[HTML]{DAE8FC}$5.5K$ & \cellcolor[HTML]{DAE8FC}$15.0K$ & \cellcolor[HTML]{DAE8FC}$3.9s$ \\
 & MT40A4G4JC-062E:E [m10] & R-DIMM & E & 16 Gb & x4 & 2014 & M5 & $6.6K$ & $15.0K$ & \multicolumn{1}{c|}{$4.8s$} & $6.2K$ & $15.0K$ & \multicolumn{1}{c|}{$4.5s$} & $6.2K$ & $15.0K$ & \multicolumn{1}{c|}{$4.5s$} & $6.2K$ & $15.0K$ & \multicolumn{1}{c|}{$4.5s$} & $6.2K$ & $15.0K$ & \multicolumn{1}{c|}{$4.5s$} & \cellcolor[HTML]{FFFFFF}$6.2K$ & \cellcolor[HTML]{FFFFFF}$15.0K$ & \cellcolor[HTML]{FFFFFF}$4.5s$ \\
\multirow{-7}{*}{\begin{tabular}[c]{@{}c@{}}Mfr. M\\ (Micron)\end{tabular}} & \cellcolor[HTML]{DAE8FC}MT40A1G16RC-062E:B [m11] & \cellcolor[HTML]{DAE8FC}SO-DIMM & \cellcolor[HTML]{DAE8FC}B & \cellcolor[HTML]{DAE8FC}16 Gb & \cellcolor[HTML]{DAE8FC}x16 & \cellcolor[HTML]{DAE8FC}2126 & \cellcolor[HTML]{DAE8FC}M6 & \cellcolor[HTML]{DAE8FC}$13.3K$ & \cellcolor[HTML]{DAE8FC}$15.0K$ & \multicolumn{1}{c|}{\cellcolor[HTML]{DAE8FC}$9.6s$} & \cellcolor[HTML]{DAE8FC}$13.3K$ & \cellcolor[HTML]{DAE8FC}$15.0K$ & \multicolumn{1}{c|}{\cellcolor[HTML]{DAE8FC}$9.6s$} & \cellcolor[HTML]{DAE8FC}$13.3K$ & \cellcolor[HTML]{DAE8FC}$15.0K$ & \multicolumn{1}{c|}{\cellcolor[HTML]{DAE8FC}$9.6s$} & \cellcolor[HTML]{DAE8FC}$13.3K$ & \cellcolor[HTML]{DAE8FC}$15.0K$ & \multicolumn{1}{c|}{\cellcolor[HTML]{DAE8FC}$9.6s$} & \cellcolor[HTML]{DAE8FC}$13.3K$ & \cellcolor[HTML]{DAE8FC}$15.0K$ & \multicolumn{1}{c|}{\cellcolor[HTML]{DAE8FC}$9.6s$} & \cellcolor[HTML]{DAE8FC}$13.3K$ & \cellcolor[HTML]{DAE8FC}$15.0K$ & \cellcolor[HTML]{DAE8FC}$9.6s$ \\ \hline
 & K4A4G085WF-BCTD [m12] & U-DIMM & F & 4 Gb & x8 & N/A & S0 & $11.7K$ & $15.0K$ & \multicolumn{1}{c|}{$8.4s$} & $11.7K$ & $15.0K$ & \multicolumn{1}{c|}{$8.4s$} & $10.9K$ & $15.0K$ & \multicolumn{1}{c|}{$7.9s$} & $9.4K$ & $10.0K$ & \multicolumn{1}{c|}{$4.5s$} & \cellcolor[HTML]{FFFFFF}$6.2K$ & \cellcolor[HTML]{FFFFFF}$1$ & \multicolumn{1}{c|}{\cellcolor[HTML]{FFFFFF}$300\mu s$} & \cellcolor[HTML]{FFCCC9} & \cellcolor[HTML]{FFCCC9}$N/A$ & \cellcolor[HTML]{FFCCC9} \\
 & \cellcolor[HTML]{DAE8FC}K4A4G085WF-BCTD [m12] & \cellcolor[HTML]{DAE8FC}U-DIMM & \cellcolor[HTML]{DAE8FC}F & \cellcolor[HTML]{DAE8FC}4 Gb & \cellcolor[HTML]{DAE8FC}x8 & \cellcolor[HTML]{DAE8FC}N/A & \cellcolor[HTML]{DAE8FC}S1 & \cellcolor[HTML]{DAE8FC}$14.1K$ & \cellcolor[HTML]{DAE8FC}$15.0K$ & \multicolumn{1}{c|}{\cellcolor[HTML]{DAE8FC}$10.1s$} & \cellcolor[HTML]{DAE8FC}$13.3K$ & \cellcolor[HTML]{DAE8FC}$15.0K$ & \multicolumn{1}{c|}{\cellcolor[HTML]{DAE8FC}$9.6s$} & \cellcolor[HTML]{DAE8FC}$12.1K$ & \cellcolor[HTML]{DAE8FC}$15.0K$ & \multicolumn{1}{c|}{\cellcolor[HTML]{DAE8FC}$8.7s$} & \cellcolor[HTML]{DAE8FC}$9.8K$ & \cellcolor[HTML]{DAE8FC}$15.0K$ & \multicolumn{1}{c|}{\cellcolor[HTML]{DAE8FC}$7.0s$} & \cellcolor[HTML]{DAE8FC}$5.1K$ & \cellcolor[HTML]{DAE8FC}$2$ & \multicolumn{1}{c|}{\cellcolor[HTML]{DAE8FC}$487\mu s$} & \cellcolor[HTML]{FFCCC9} & \cellcolor[HTML]{FFCCC9}$N/A$ & \cellcolor[HTML]{FFCCC9} \\
 & K4A4G085WE-BCPB [m13] & SO-DIMM & E & 4 Gb & x8 & 1708 & S2 & $23.8K$ & $15.0K$ & \multicolumn{1}{c|}{$17.2s$} & $23.4K$ & $15.0K$ & \multicolumn{1}{c|}{$16.9s$} & $22.3K$ & $15.0K$ & \multicolumn{1}{c|}{$16.0s$} & $20.7K$ & $15.0K$ & \multicolumn{1}{c|}{$14.9s$} & \cellcolor[HTML]{FFFFFF}$19.9K$ & \cellcolor[HTML]{FFFFFF}$1$ & \multicolumn{1}{c|}{\cellcolor[HTML]{FFFFFF}$955\mu s$} & \cellcolor[HTML]{FFFFFF}$5.1K$ & \cellcolor[HTML]{FFFFFF}$1$ & \cellcolor[HTML]{FFFFFF}$244\mu s$ \\
 & \cellcolor[HTML]{DAE8FC}K4A4G085WE-BCPB [m13] & \cellcolor[HTML]{DAE8FC}SO-DIMM & \cellcolor[HTML]{DAE8FC}E & \cellcolor[HTML]{DAE8FC}4 Gb & \cellcolor[HTML]{DAE8FC}x8 & \cellcolor[HTML]{DAE8FC}1708 & \cellcolor[HTML]{DAE8FC}S3 & \cellcolor[HTML]{DAE8FC}$19.9K$ & \cellcolor[HTML]{DAE8FC}$15.0K$ & \multicolumn{1}{c|}{\cellcolor[HTML]{DAE8FC}$14.3s$} & \cellcolor[HTML]{DAE8FC}$19.5K$ & \cellcolor[HTML]{DAE8FC}$15.0K$ & \multicolumn{1}{c|}{\cellcolor[HTML]{DAE8FC}$14.1s$} & \cellcolor[HTML]{DAE8FC}$18.8K$ & \cellcolor[HTML]{DAE8FC}$15.0K$ & \multicolumn{1}{c|}{\cellcolor[HTML]{DAE8FC}$13.5s$} & \cellcolor[HTML]{DAE8FC}$17.2K$ & \cellcolor[HTML]{DAE8FC}$15.0K$ & \multicolumn{1}{c|}{\cellcolor[HTML]{DAE8FC}$12.4s$} & \cellcolor[HTML]{DAE8FC}$17.6K$ & \cellcolor[HTML]{DAE8FC}$1$ & \multicolumn{1}{c|}{\cellcolor[HTML]{DAE8FC}$844\mu s$} & \cellcolor[HTML]{FFCCC9} & \cellcolor[HTML]{FFCCC9}$N/A$ & \cellcolor[HTML]{FFCCC9} \\
 & K4A4G085WE-BCPB [m13] & SO-DIMM & E & 4 Gb & x8 & 1708 & S4 & $20.3K$ & $15.0K$ & \multicolumn{1}{c|}{$14.6s$} & $20.3K$ & $15.0K$ & \multicolumn{1}{c|}{$14.6s$} & $19.1K$ & $15.0K$ & \multicolumn{1}{c|}{$13.8s$} & $18.0K$ & $15.0K$ & \multicolumn{1}{c|}{$12.9s$} & \cellcolor[HTML]{FFCCC9} & \cellcolor[HTML]{FFCCC9}$N/A$ & \multicolumn{1}{c|}{\cellcolor[HTML]{FFCCC9}} & \cellcolor[HTML]{FFCCC9} & \cellcolor[HTML]{FFCCC9}$N/A$ & \cellcolor[HTML]{FFCCC9} \\
 & \cellcolor[HTML]{DAE8FC}Unknown & \cellcolor[HTML]{DAE8FC}SO-DIMM & \cellcolor[HTML]{DAE8FC}C & \cellcolor[HTML]{DAE8FC}4 Gb & \cellcolor[HTML]{DAE8FC}x16 & \cellcolor[HTML]{DAE8FC}N/A & \cellcolor[HTML]{DAE8FC}S5 & \cellcolor[HTML]{DAE8FC}$12.1K$ & \cellcolor[HTML]{DAE8FC}$15.0K$ & \multicolumn{1}{c|}{\cellcolor[HTML]{DAE8FC}$8.7s$} & \cellcolor[HTML]{DAE8FC}$12.1K$ & \cellcolor[HTML]{DAE8FC}$15.0K$ & \multicolumn{1}{c|}{\cellcolor[HTML]{DAE8FC}$8.7s$} & \cellcolor[HTML]{DAE8FC}$11.7K$ & \cellcolor[HTML]{DAE8FC}$15.0K$ & \multicolumn{1}{c|}{\cellcolor[HTML]{DAE8FC}$8.4s$} & \cellcolor[HTML]{DAE8FC}$9.4K$ & \cellcolor[HTML]{DAE8FC}$15.0K$ & \multicolumn{1}{c|}{\cellcolor[HTML]{DAE8FC}$6.8s$} & \cellcolor[HTML]{DAE8FC}$5.1K$ & \cellcolor[HTML]{DAE8FC}$2$ & \multicolumn{1}{c|}{\cellcolor[HTML]{DAE8FC}$487\mu s$} & \cellcolor[HTML]{FFCCC9} & \cellcolor[HTML]{FFCCC9}$N/A$ & \cellcolor[HTML]{FFCCC9} \\
 & K4A8G085WD-BCTD [m14] & U-DIMM & D & 8 Gb & x8 & 2110 & S6 & $7.0K$ & $15.0K$ & \multicolumn{1}{c|}{$5.1s$} & $7.0K$ & $15.0K$ & \multicolumn{1}{c|}{$5.1s$} & $6.2K$ & $15.0K$ & \multicolumn{1}{c|}{$4.5s$} & $3.9K$ & $2.0K$ & \multicolumn{1}{c|}{$374ms$} & \cellcolor[HTML]{FFFFFF}$3.9K$ & \cellcolor[HTML]{FFFFFF}$1$ & \multicolumn{1}{c|}{\cellcolor[HTML]{FFFFFF}$187\mu s$} & \cellcolor[HTML]{FFCCC9} & \cellcolor[HTML]{FFCCC9}$N/A$ & \cellcolor[HTML]{FFCCC9} \\
 & \cellcolor[HTML]{DAE8FC}K4A8G085WD-BCTD [m14] & \cellcolor[HTML]{DAE8FC}U-DIMM & \cellcolor[HTML]{DAE8FC}D & \cellcolor[HTML]{DAE8FC}8 Gb & \cellcolor[HTML]{DAE8FC}x8 & \cellcolor[HTML]{DAE8FC}2110 & \cellcolor[HTML]{DAE8FC}S7 & \cellcolor[HTML]{DAE8FC}$7.8K$ & \cellcolor[HTML]{DAE8FC}$15.0K$ & \multicolumn{1}{c|}{\cellcolor[HTML]{DAE8FC}$5.6s$} & \cellcolor[HTML]{DAE8FC}$7.0K$ & \cellcolor[HTML]{DAE8FC}$15.0K$ & \multicolumn{1}{c|}{\cellcolor[HTML]{DAE8FC}$5.1s$} & \cellcolor[HTML]{DAE8FC}$5.5K$ & \cellcolor[HTML]{DAE8FC}$15.0K$ & \multicolumn{1}{c|}{\cellcolor[HTML]{DAE8FC}$3.9s$} & \cellcolor[HTML]{DAE8FC}$5.5K$ & \cellcolor[HTML]{DAE8FC}$1$ & \multicolumn{1}{c|}{\cellcolor[HTML]{DAE8FC}$262\mu s$} & \cellcolor[HTML]{DAE8FC}$3.9K$ & \cellcolor[HTML]{DAE8FC}$1$ & \multicolumn{1}{c|}{\cellcolor[HTML]{DAE8FC}$187\mu s$} & \cellcolor[HTML]{FFCCC9} & \cellcolor[HTML]{FFCCC9}$N/A$ & \cellcolor[HTML]{FFCCC9} \\
 & K4A8G085WD-BCTD [m14] & U-DIMM & D & 8 Gb & x8 & 2110 & S8 & $7.8K$ & $15.0K$ & \multicolumn{1}{c|}{$5.6s$} & $7.8K$ & $15.0K$ & \multicolumn{1}{c|}{$5.6s$} & $5.9K$ & $15.0K$ & \multicolumn{1}{c|}{$4.2s$} & $3.9K$ & $15.0K$ & \multicolumn{1}{c|}{$2.8s$} & \cellcolor[HTML]{FFFFFF}$3.9K$ & \cellcolor[HTML]{FFFFFF}$1$ & \multicolumn{1}{c|}{\cellcolor[HTML]{FFFFFF}$187\mu s$} & \cellcolor[HTML]{FFCCC9} & \cellcolor[HTML]{FFCCC9}$N/A$ & \cellcolor[HTML]{FFCCC9} \\
 & \cellcolor[HTML]{DAE8FC}K4A8G085WD-BCTD [m14] & \cellcolor[HTML]{DAE8FC}U-DIMM & \cellcolor[HTML]{DAE8FC}D & \cellcolor[HTML]{DAE8FC}8 Gb & \cellcolor[HTML]{DAE8FC}x8 & \cellcolor[HTML]{DAE8FC}2110 & \cellcolor[HTML]{DAE8FC}S9 & \cellcolor[HTML]{DAE8FC}$8.6K$ & \cellcolor[HTML]{DAE8FC}$15.0K$ & \multicolumn{1}{c|}{\cellcolor[HTML]{DAE8FC}$6.2s$} & \cellcolor[HTML]{DAE8FC}$8.6K$ & \cellcolor[HTML]{DAE8FC}$15.0K$ & \multicolumn{1}{c|}{\cellcolor[HTML]{DAE8FC}$6.2s$} & \cellcolor[HTML]{DAE8FC}$6.6K$ & \cellcolor[HTML]{DAE8FC}$15.0K$ & \multicolumn{1}{c|}{\cellcolor[HTML]{DAE8FC}$4.8s$} & \cellcolor[HTML]{DAE8FC}$4.7K$ & \cellcolor[HTML]{DAE8FC}$15.0K$ & \multicolumn{1}{c|}{\cellcolor[HTML]{DAE8FC}$3.4s$} & \cellcolor[HTML]{DAE8FC}$3.1K$ & \cellcolor[HTML]{DAE8FC}$2$ & \multicolumn{1}{c|}{\cellcolor[HTML]{DAE8FC}$300\mu s$} & \cellcolor[HTML]{FFCCC9} & \cellcolor[HTML]{FFCCC9}$N/A$ & \cellcolor[HTML]{FFCCC9} \\
 & K4A8G085WC-BCRC [m15] & R-DIMM & C & 8 Gb & x8 & 1809 & S10 & $13.3K$ & $15.0K$ & \multicolumn{1}{c|}{$9.6s$} & $12.5K$ & $15.0K$ & \multicolumn{1}{c|}{$9.0s$} & $12.5K$ & $15.0K$ & \multicolumn{1}{c|}{$9.0s$} & $10.2K$ & $15.0K$ & \multicolumn{1}{c|}{$7.3s$} & \cellcolor[HTML]{FFFFFF}$10.2K$ & \cellcolor[HTML]{FFFFFF}$1$ & \multicolumn{1}{c|}{\cellcolor[HTML]{FFFFFF}$489\mu s$} & \cellcolor[HTML]{FFCCC9} & \cellcolor[HTML]{FFCCC9}$N/A$ & \cellcolor[HTML]{FFCCC9} \\
 & \cellcolor[HTML]{DAE8FC}K4A8G085WB-BCTD [m16] & \cellcolor[HTML]{DAE8FC}R-DIMM & \cellcolor[HTML]{DAE8FC}B & \cellcolor[HTML]{DAE8FC}8 Gb & \cellcolor[HTML]{DAE8FC}x8 & \cellcolor[HTML]{DAE8FC}2053 & \cellcolor[HTML]{DAE8FC}S11 & \cellcolor[HTML]{DAE8FC}$26.6K$ & \cellcolor[HTML]{DAE8FC}$15.0K$ & \multicolumn{1}{c|}{\cellcolor[HTML]{DAE8FC}$19.1s$} & \cellcolor[HTML]{DAE8FC}$26.6K$ & \cellcolor[HTML]{DAE8FC}$15.0K$ & \multicolumn{1}{c|}{\cellcolor[HTML]{DAE8FC}$19.1s$} & \cellcolor[HTML]{DAE8FC}$25.8K$ & \cellcolor[HTML]{DAE8FC}$15.0K$ & \multicolumn{1}{c|}{\cellcolor[HTML]{DAE8FC}$18.6s$} & \cellcolor[HTML]{DAE8FC}$25.0K$ & \cellcolor[HTML]{DAE8FC}$15.0K$ & \multicolumn{1}{c|}{\cellcolor[HTML]{DAE8FC}$18.0s$} & \cellcolor[HTML]{FFCCC9} & \cellcolor[HTML]{FFCCC9}$N/A$ & \multicolumn{1}{c|}{\cellcolor[HTML]{FFCCC9}} & \cellcolor[HTML]{FFCCC9} & \cellcolor[HTML]{FFCCC9}$N/A$ & \cellcolor[HTML]{FFCCC9} \\
 & K4AAG085WA-BCWE [m17] & U-DIMM & A & 8 Gb & x8 & 2212 & S12 & $8.6K$ & $15.0K$ & \multicolumn{1}{c|}{$6.2s$} & $9.0K$ & $15.0K$ & \multicolumn{1}{c|}{$6.5s$} & $7.8K$ & $15.0K$ & \multicolumn{1}{c|}{$5.6s$} & $6.2K$ & $15.0K$ & \multicolumn{1}{c|}{$4.5s$} & \cellcolor[HTML]{FFCCC9} & \cellcolor[HTML]{FFCCC9}$N/A$ & \multicolumn{1}{c|}{\cellcolor[HTML]{FFCCC9}} & \cellcolor[HTML]{FFCCC9} & \cellcolor[HTML]{FFCCC9}$N/A$ & \cellcolor[HTML]{FFCCC9} \\
\multirow{-14}{*}{\begin{tabular}[c]{@{}c@{}}Mfr. S\\ (Samsung)\end{tabular}} & \cellcolor[HTML]{DAE8FC}Unknown & \cellcolor[HTML]{DAE8FC}U-DIMM & \cellcolor[HTML]{DAE8FC}B & \cellcolor[HTML]{DAE8FC}16 Gb & \cellcolor[HTML]{DAE8FC}x8 & \cellcolor[HTML]{DAE8FC}2315 & \cellcolor[HTML]{DAE8FC}S13 & \cellcolor[HTML]{DAE8FC}$7.4K$ & \cellcolor[HTML]{DAE8FC}$15.0K$ & \multicolumn{1}{c|}{\cellcolor[HTML]{DAE8FC}$5.3s$} & \cellcolor[HTML]{DAE8FC}$7.0K$ & \cellcolor[HTML]{DAE8FC}$15.0K$ & \multicolumn{1}{c|}{\cellcolor[HTML]{DAE8FC}$5.1s$} & \cellcolor[HTML]{DAE8FC}$6.6K$ & \cellcolor[HTML]{DAE8FC}$15.0K$ & \multicolumn{1}{c|}{\cellcolor[HTML]{DAE8FC}$4.8s$} & \cellcolor[HTML]{DAE8FC}$6.2K$ & \cellcolor[HTML]{DAE8FC}$15.0K$ & \multicolumn{1}{c|}{\cellcolor[HTML]{DAE8FC}$4.5s$} & \cellcolor[HTML]{DAE8FC}$3.9K$ & \cellcolor[HTML]{DAE8FC}$5$ & \multicolumn{1}{c|}{\cellcolor[HTML]{DAE8FC}$937\mu s$} & \cellcolor[HTML]{FFCCC9} & \cellcolor[HTML]{FFCCC9}$N/A$ & \cellcolor[HTML]{FFCCC9} \\ \hline
\end{tabular}%
}
\end{table}

\end{landscape}
\nocitemodule{*}
\bibliographystylemodule{IEEEtran}
\bibliographymodule{modulerefs}

\end{document}